\documentclass[]{interact}

\usepackage{epstopdf}
\usepackage[caption=false]{subfig}
\usepackage[colorlinks=true,citecolor=blue]{hyperref}%
\usepackage{verbatim}
\usepackage{soul}
\usepackage[normalem]{ulem}
\usepackage{color}

\usepackage[numbers,sort&compress]{natbib}
\bibpunct[, ]{[}{]}{,}{n}{,}{,}
\makeatletter
\def\NAT@def@citea{\def\@citea{\NAT@separator}}
\makeatother

\usepackage[utf8]{inputenc}
\usepackage[T1]{fontenc}
\usepackage{dsfont}

\newcommand{\hs}{\hspace}
\newcommand{\vs}{\vspace}
\newcommand{\n}{\textbf}
\usepackage{mathrsfs}

\usepackage{color}

\begin{document}

\title{Shear deformation in CuZr metallic glass: A statistical and complex network analysis}
\author{\name{Fernando Corvacho\textsuperscript{a}$^{,*}$\thanks{$^*$~Corresponding author. Email: fernando.corvacho@ug.uchile.cl}, Victor Mu\~noz\textsuperscript{a} , Matias Sepulveda--Macias\textsuperscript{b} and Gonzalo Gutierrez\textsuperscript{a}}
\affil{\textsuperscript{a} Facultad de Ciencias, Departamento de F\'isica, Facultad de Ciencias, Universidad de Chile, Casilla 653, Santiago, Chile; \textsuperscript{b} Univ. Lyon, INSA-Lyon, LaMCoS, CNRS, UMR5259, 69621 Villeurbanne, France}}

\maketitle

        \begin{abstract}
        We have implemented a complex network description for metallic glasses, able to predict the 
        elasto-plastic regime, the location of shear bands and the statistics that controls the plastic 
        events that originate in the material due to a deformation process. By means of molecular dynamics 
        simulations, we perform a shear deformation test, obtaining the stress-strain curve for CuZr 
        metallic glass samples. The atomic configurations of the metallic glass are mapped to a graph, 
        where a node represents an atom whose stress/strain is above a certain threshold, and edges are 
        connections between existing nodes at consecutive timesteps in the simulation. We made a statistical 
        study of some physical descriptors such as shear stress, shear strain, volumetric strain and non-affine 
        displacement to use them as construction tools for complex networks. We have calculated their 
        probability density functions, skewness, kurtosis and gini coefficient to analyze the inequality of 
        the distributions. We study the evolution of the resulting complex network, by computing topological 
        metrics such as degree, clustering coefficient, betweenness and closeness centrality as a function of 
        the strain. We have obtained correlations between the physical phenomena produced by the deformation 
        with the data recorded by these metrics. By means the visual representation of the networks, we have 
        also found direct correlations between metrics and the local atomic shear strain, so that they are able 
        to predict the location of shear bands, as well as the formation of highly connected and interacting 
        communities, which we interpret as shear transformation zones. Our results suggest that the complex network approach has interesting capabilities for the description of mechanical properties of metallic glasses.
        \end{abstract}

\begin{keywords}
Metallic glasses, complex network, topological metrics, degree, clustering coefficient, centrality mesaures.
\end{keywords}


        \section{Introduction}

        \noindent Metallic glasses (MGs) are non-crystalline solids with very interesting physical, chemical 
        \cite{Wang2009Bulk}, magnetic \cite{Tiberto2007Magnetic} and mechanical \cite{Greer1995Metallic} 
        properties for their technological applications \cite{Anantharaman1984Metallic,Ashby2006Metallic}. 
        After the synthesis of the first MG in 1960 \cite{Klement1960Non} and the development of large-scale 
        metallic glasses (BMGs) in 1993 \cite{Peker1993Highly,Inoue1993New}, the theoretical and experimental 
        studies of these materials have been an area of intense research \cite{Suryanarayana2017Bulk}.
        The disordered nature of MGs causes them to exhibit excellent mechanical properties, such as high 
        yield strength, large elastic strain limits, good wear resistance, among others~\cite{Trexler2010Mechanical}. 
        They are even able to deform elastically to a strain limit greater than 2\%, which is an order of 
        magnitude higher than in their crystalline metallic form. However, it has been observed that BMGs suffers 
        brittle fracture due to its limited ductility under mechanical tests 
        \cite{Chen2008Mechanical,Schuh2007Mechanical}, which has inhibited its direct use as a structural 
        material \cite{Plummer2015Metallic}.

        \vs{0.2cm}
        \noindent The mechanical behavior and plastic deformation in BMGs, is still a topic that keeps 
        scientists and engineers very fascinated. It is well known that when a load is applied to a MG sample, 
        after the elastic limit is reached, it immediately undergoes catastrophic failure. Thus, in contrast 
        to its crystalline counterpart, where dislocations are the main carriers of plasticity, MGs do not have 
        this property due to the absence of grain boundaries, causing them to be materials with high 
        mechanical resistance but brittle \cite{Greer1995Metallic,Suryanarayana2017Bulk}. So far, there is no 
        theory that explains the physics behind these events. It has been hypothesized that this behavior can 
        be explained by the location of structural and dynamical heterogeneities called shear transformation zones 
        (STZs), where a pair of atoms ($\approx$ 8--20) rearrange to adapt to the applied strain 
        \cite{Argon1979Plastic,Falk1998Dynamics}. These STZs are initially randomly distributed along the 
        sample, but gradually begin to correlate both spatially and temporally, coalescing to give rise to shear 
        bands (SBs) \cite{Shimizu2007Theory}. The localization, or the formation, of a dominant SB results in 
        catastrophic failure, and the dynamics of this mechanism are thought to be key to understanding the 
        mechanical behavior of MGs \cite{Greer2013Shear}. In fact, several experimental and theoretical works 
        have been devoted to the problem of avoiding the localization of SBs to generate a homogeneous plastic 
        deformation in the material, improve its mechanical properties, and prevent fracture 
        \cite{Rezaei2017Ductility,Sepulveda2018Tensile}.

        \vs{0.2cm}
        \noindent Usually, to study the physical properties of materials, statistical mechanics is used through 
        atomic simulations. However, new mathematical tools have emerged to be used to study complexity and 
        "discretizable" systems. We are referring to the concepts of complex networks (CNs) and graph theory 
        \cite{Albert2002Statistical,Newman2006Structure}, where their application stands out in different fields 
        of science \cite{Gheibi2017Solar,Gosak2018Network,Wasserman1994Social,Baiesi2005Complex}. In physics, 
        these tools have proven useful for describing complex systems, by providing a new perspective for their 
        study, and revealing features which would be otherwise difficult to find with more traditional methods. 
        For instance, researchers have used network analysis to examine the structure and behavior of granular materials \cite{Papadopoulos2018Network}, representing force-chain interactions via networks. Similarly, network analysis has been employed to study plastic deformation in metals, calculating the stress-strain curve as a time series and comparing it with various topological measures\cite{Kiv2023Complex}. Several authors have applied components of complex network analysis to study the rheological properties, rejuvenation behavior, and local structure of metallic glasses. For example, computing the clustering coefficient\cite{Kui2007Correlation,Lee2011Networked,Foroughi2018Medium}. Additionally, it has been used successfully in the study of earthquakes \cite{Baiesi2005Complex,Abe2011Universalities, Vogel2017Time}. By means of a CN 
        representation of the spatiotemporal evolution of seismic events, it has been possible to show universal 
        characteristics of seismicity in different geological zones \cite{Abe2011Universalities}, investigate the transition involved 
        in the occurence of a large earthquake \cite{Pasten2016Time} and the relationship between the $b$-value 
        and coupling in a seismic zone \cite{Martin2022Complex}. A similar approach for CN construction has been
        followed to study solar flares \cite{Gheibi2017Solar} and the evolution of solar activity, as measured by 
        sunspots appearance in the solar photosphere, along the 23rd solar cycle \cite{Munoz2022Complex}, thus 
        showing the versatility of this technique to extract valuable information about 
        energy release events in physical systems. Interestingly, a systematic research on complex system approaches 
        for MGs subjected to an external deformation, is still needed. Since glass theory is still an underdeveloped 
        field, CN analysis may provide a different and useful viewpoint to the mechanical response of amorphous solids.

        \vs{0.2cm}
        \noindent In this research, we develop a microscopic study of the plastic deformation of a MG through 
        computational simulations and complex networks techniques. Using classical molecular dynamics (MD) 
        simulations, a sample of Cu$_{50}$Zr$_{50}$ MG is subjected to shear deformation 
        \cite{Inoue1999Stabilization}. MD simulations allow us to monitor atomic level events that appear when 
        applying the shear deformation in the material. Based on these data, provided by MD simulations, a 
        graph formed by nodes and vertices is built, and the time series of these graphs form the complex 
        network of the system, being an abstract representation of the spatiotemporal evolution of stress. For 
        these networks, various topological measurements are calculated for the networks, and compared with the 
        physical process represented by atomistic simulations. In section II, we describe the details of the MD 
        simulations, the deformation process and the physical observables obtained. In section III we present a 
        statistical study of some physical descriptors that are used for CNs. We computed the spatial 
        distributions and probability densities of these descriptors, as well as their moments, Lorenz curves, 
        and Gini coefficient. In section IV we explain the construction process of the graphs and present the 
        topological metrics that are used to characterize the resulting networks. In Section V the results are 
        presented, and finally in Section VI the conclusions are drawn.


        \section{Simulations details}

        \noindent To study the microscopic phenomena involved in the plastic deformation of a MG, we perform 
        a volume preserving shear deformation. Since we are working with a MG based on copper and zirconium, 
        we have used an atomic interaction for the system that obeys the embedded atom model (EAM) potential 
        developed by Cheng \emph{et al.} \cite{Cheng2011Atomic}. To carry out the simulations, the calculations 
        of trajectories and properties are done by the software LAMMPS \cite{Plimpton1995Fast}. This 
        computational tool has helped us with all the task and has been a fundamental tool for the physical 
        representation of the system as well as a data generating instrument for the development and application 
        of CNs. The amorphous Cu$_{50}$Zr$_{50}$ system consists of $N=580800$ atoms, with dimensions 
        $905\times452\times24\text{ \AA}^3$, where the $z$-direction was chosen thinner than the others to 
        facilitate the visualization of SBs. We consider periodic boundary conditions, during the whole deformation process, in all three directions.

        \begin{figure}[ht!]
                \centering
                \includegraphics[scale=0.35]{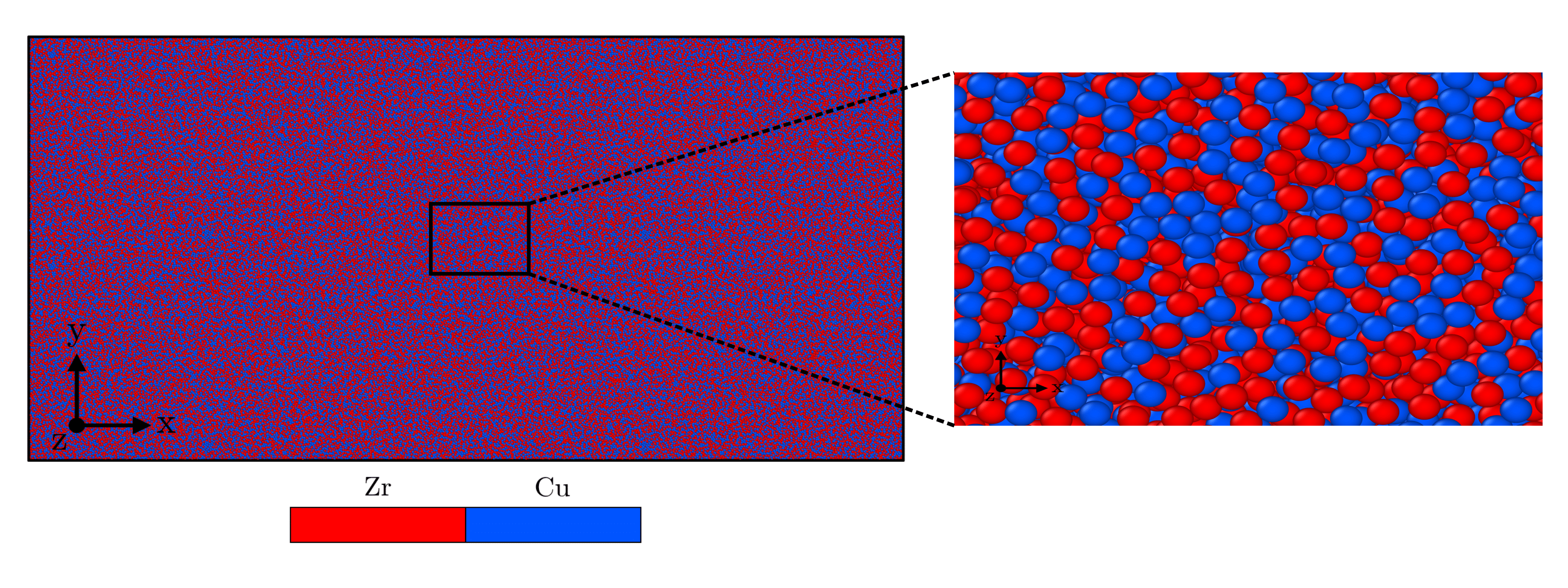}
                \caption{Visualization of the Cu$_{50}$Zr$_{50}$ MG system. Capture of the sample once its 
                preparation had finished, including a small region that allows checking the amorphous state 
                and the absence of crystallinity. This figure was obtained through the OVITO software 
                \cite{Stukowski2009Visualization}. }
                \label{fig:MG-sample}
        \end{figure}

        \vs{0.2cm}
        \noindent The glass 
        sample was prepared as follows: We start with a system of $145200$ Cu atoms with a FCC crystal structure, 
        where 50\% of these atoms are replaced by Zr atoms at random positions. Next, the system is heated from an 
        initial temperature of 300 K to 2200 K for 2 ns (with an integration step $\Delta\tau=1$ fs), in the NPT 
        ensemble, keeping the pressure constant at 0 GPa, thus obtaining a molten metal. The next step is to quench 
        the sample quickly enough to avoid crystallization and obtain a glassy state system. For this, we reduced the 
        temperature by 10 K followed by the procedure proposed by Wang \emph{et al.} \cite{Wang2012Structural}. The 
        $145200$ atoms in the glass are then replicated in the $x$ and $y$ direction, followed by a relaxation process 
        in the NPT ensemble to remove all the problems caused by replication. Finally, we let the system evolves in 
        the NVE ensemble for $100$ ps with a final minimization that ensures the interatomic forces are of the order 
        of $10^{-4}$ eV$\cdot$\AA$^{-1}$ and we thermalized it to $10$ K via Langevin thermostat, previous to the 
        deformation process, following the methodology described by Sepúlveda-Macías \emph{et al.} 
        \cite{Sepulveda2020Precursors,Sepulveda2016Onset}. 

        \begin{figure}[ht!]
                \centering
                \includegraphics[scale=0.5]{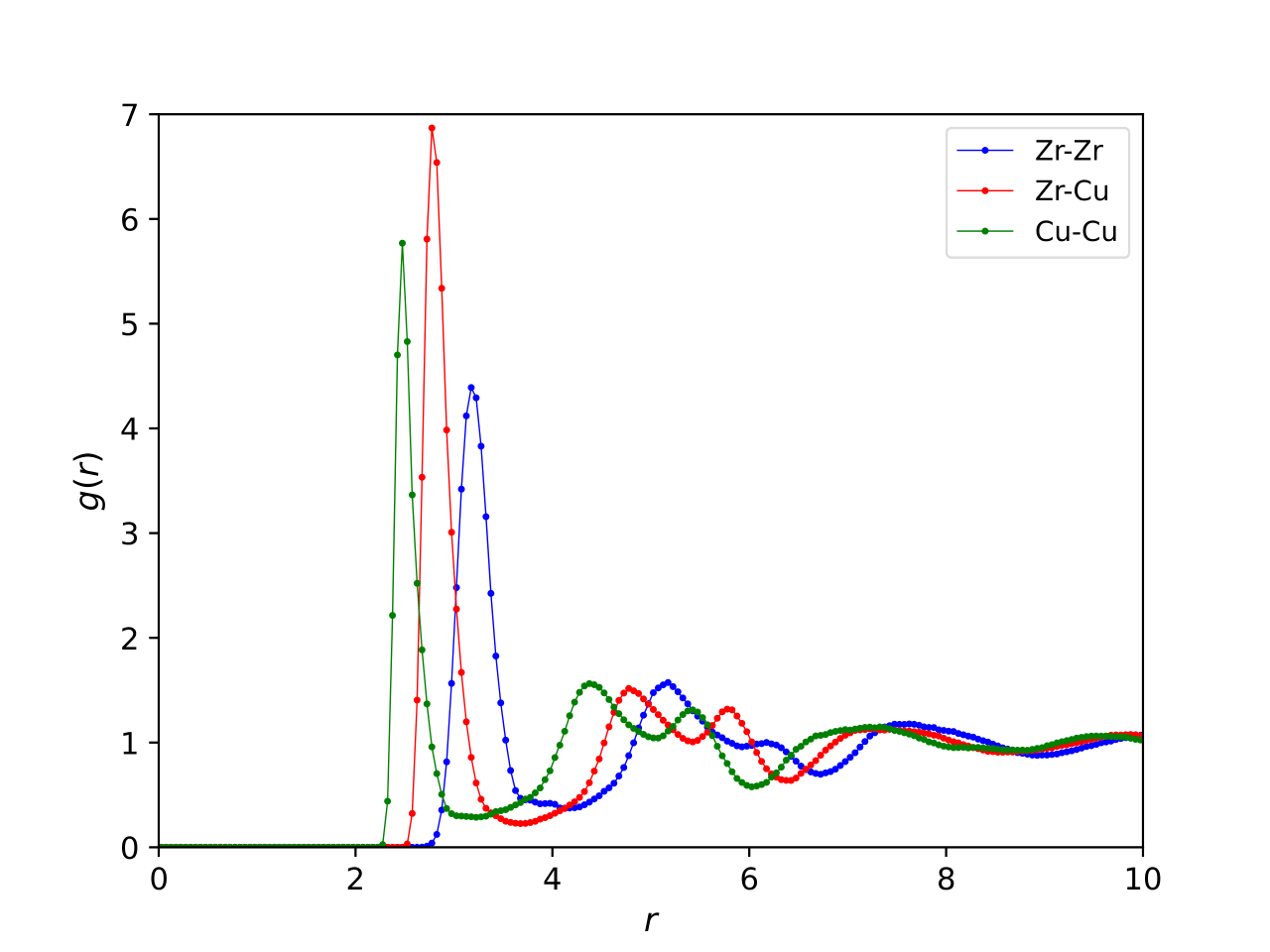}
                \caption{Partial radial distribution functions (RDF) for the MGs sample, computed at 10 K. Zr-Zr, Zr-Cu and Cu-Cu correlations are reviewed. The shape of these curves indicate a good agreement on the amorphous 
                state of the system.}
                \label{fig:rdf}
        \end{figure}

        \vs{0.2cm}
        \noindent The resulting MG sample is presented in Figure \ref{fig:MG-sample}. The partial radial distribution functions for the resulting Cu$_{50}$Zr$_{50}$ sample at 10 K are presented in Figure~\ref{fig:rdf}. As previously reported for this sample \cite{Cheng2011Atomic,Sepulveda2016Onset}, the location of the first peak and the 
        split of second peak are fingerprints of an amorphous state. It is important to note that if we inspect closely the local structure of the sample, in Fig. \ref{fig:MG-sample}, we noted the absence of any crystalline structure.

        \begin{figure}[ht!]
                \centering
                \includegraphics[scale=0.45]{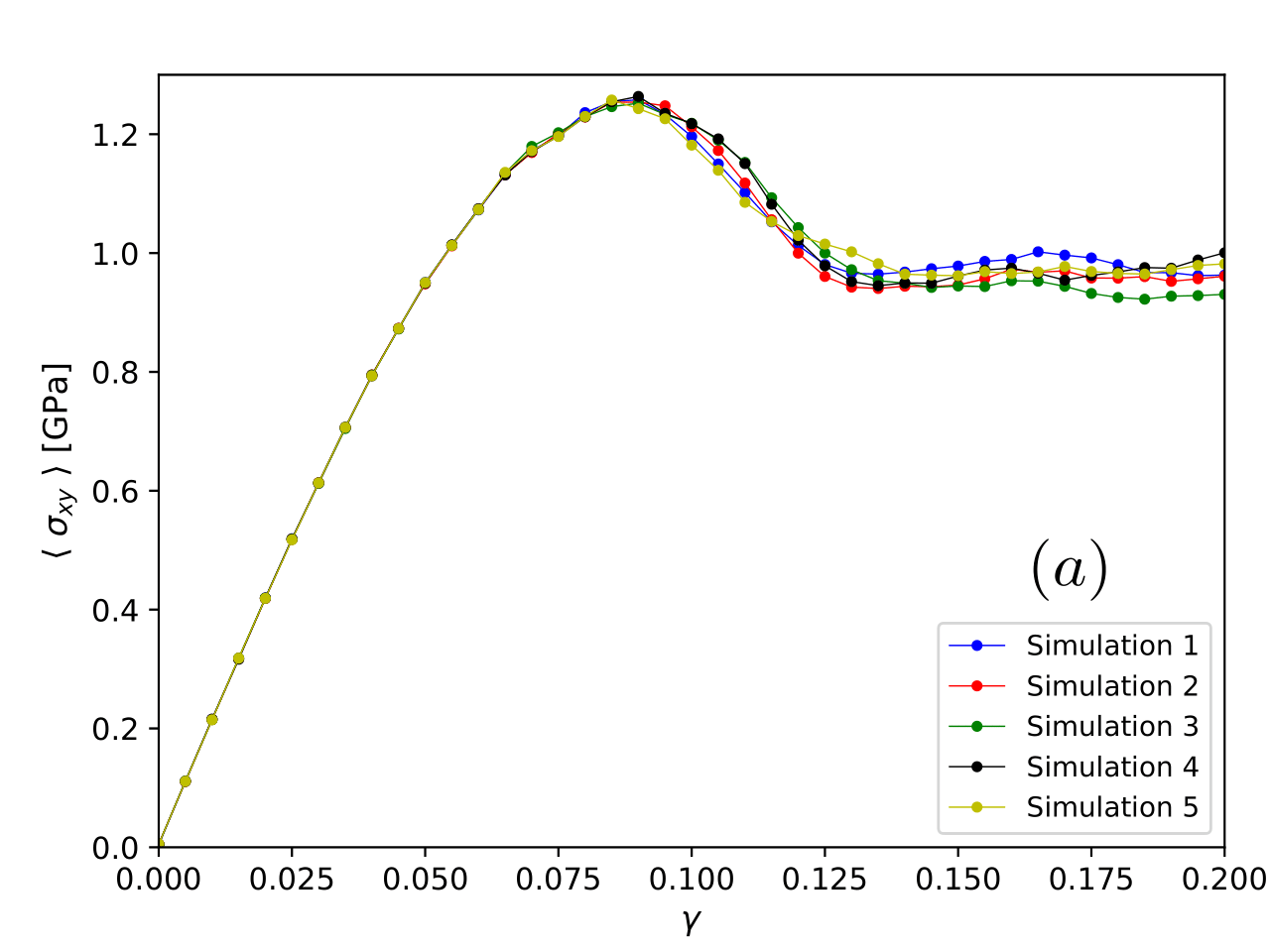}
                \includegraphics[scale=0.3]{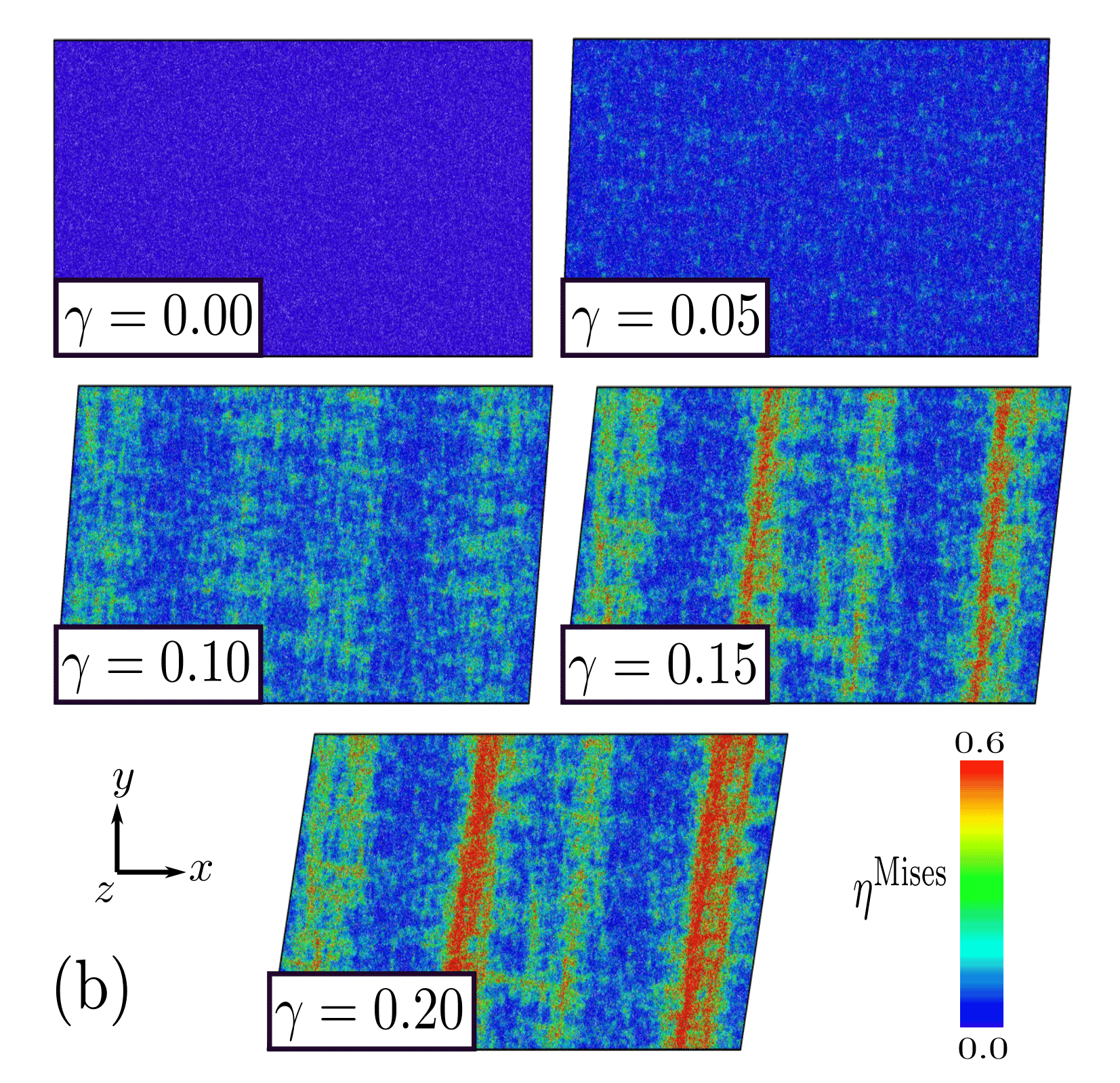}
                \caption{(a) Calculated stress-strain curves (S--S) for five simulations up to a global strain of $\gamma=0.20$. The stress $\langle\sigma_{xy}\rangle$ is calculated as the sum of the stress tensor 
                component $\sigma_{xy}$ of all the atoms averaged over the cell volume. (b) Shear strain localization in the $xy$ plane, here the atoms are colored according to their local atomic shear strain (von Mises strain, $\eta^{\text{Mises}}$), which indicates the deformation degree. This measurements were done and monitored by the OVITO software \cite{Stukowski2009Visualization}.} 
                \label{fig:stress_strain_lass}
        \end{figure}

        \vs{0.2cm}
        \noindent Once the preparation of the MG was finished, we proceed to apply a shear deformation with a 
        constant strain rate $\dot{\gamma}=5\times10^8$ s$^{-1}$, up to a maximum strain of $\gamma=0.20$ (duration 
        $0.4$ ns). We have obtained 5 different S--S curves for the elasto-plastic regime due to the thermalization.
        This process depends on an initial seed that generates the random force for the atoms in the theoretical model 
        of the Langevin thermostat. In the 5 simulations, we have occupied different seeds, so that at the microscopic 
        level each dynamic is different from one another, but at the macroscopic level, the material exhibits the same 
        mechanical state once the plastic events take place. 
        
        \vs{0.2cm}
        \noindent Figure \ref{fig:stress_strain_lass} shows the macroscopic stress-strain beaviour and the development  of the shear band during the deformation process. In Figure \ref{fig:stress_strain_lass}-(a) the S--S curves for five independent simulations are presented. Each of them exhibit two well-defined regimes: up to $\gamma\sim0.05$ the system responds elastically under shear stresses. Once this level of deformation is exceeded (yield strength), the sample begins to exhibit plastic and irreversible deformations, reaching its maximum load stress at $\gamma\sim0.09$. After reaching the maximum stress, the system undergoes a stress drop followed by a stress plateau beyond $\gamma\sim0.125$. At the atomic scale, it is possible to follow the evolution of the shear band using the local atomic shear strain, calculated by means of the von Mises strain $\eta^{\text{Mises}}$ and obtained through the strain tensor as proposed by Shimizu \emph{et al.} \cite{Shimizu2007Theory}. This is shown in Figure \ref{fig:stress_strain_lass}-(b). As can be seen, during the elastic regime, up to $\gamma=0.05$, there is no occurrence of plastic events. Thereafter, there is an increase in plastic events that are homogeneously distributed throughout the specimen. These plastic events, or STZs, begin to coalesce and align to give way to the formation of the shear band. This phenomenon is clearly seen in Figure \ref{fig:stress_strain_lass}-(b) at global $\gamma=0.15$ where the red colored atoms mark the location of the SBs, after this strain state two SBs are fully developed with an approximate width of $\sim 75$~\AA.     

        \vs{0.2cm}
        \noindent In the following section, we begin the statistical study of physical descriptors involved in the 
        deformation process, descriptors that are necessary for the application of our CN model. We compute stress tensors, strain tensors, and strain gradient tensors for each of the atoms in the cell to develop the statistics analysis.


        \section{Statistical study of physical descriptors and plastic events}

        \noindent Plasticity in MGs is a topic that still lacks a solid theoretical framework. 
        Until now, only models have been proposed that explain how the plastic deformation process 
        works, among them, structural heterogeneities known as shear transformation zones and 
        shear bands.

        \vs{0.2cm}
        \noindent In this work, we seek to give a microscopic characterization of the elasto--plastic 
        deformation, based on complex networks, that help us to describe the structural heterogeneities and the 
        properties of the Cu$_{50}$Zr$_{50}$ as a function of external shear deformation. We track the mechanical behavior of the MG up to the point where the system reaches the mechanical failure. That is, the point at which the material fractures into two or more parts.
        For this purpose, we use a methodology based on CN 
        where the atomic configurations of the system are mapped to a graph, thus obtaining 
        an abstract representation of the material, via interacting edges and vertices. For the 
        structure of these graphs, we have proposed a method that generates a growing network of 
        vertices and edges as a function of strain. The evolution of these graphs, which is the result 
        of mapping a time series of atomic configurations, results in a network with complex structural 
        and dynamical properties. Details about the construction of the network are presented in the next 
        section.

        \vs{0.2cm}
        \noindent One of the main ingredients for the construction of the CN is the selection of a 
        physical descriptor that we will use to designate the vertices. We know that the networks are 
        only a mathematical instrument to represent a system of discrete elements, but to make sense 
        of our problem, the physical information of the networks is contained in these descriptors. 
        Some descriptors that we review are: the \emph{Shear Stress}, the \emph{Shear Strain}, the 
        \emph{Volumetric Strain} and the \emph{Non-Affine Displacement}.

        \vs{0.2cm}
        \noindent The \emph{Shear Stress} or von Mises stress $\sigma^{\text{Mises}}$ is a quantity with 
        units of GPa$\cdot\text{\AA}^3/V$ that we compute for each atom $\ell$ using its stress tensor 
        through the expression

        \begin{equation}\label{shear_stress}
        \sigma^{\text{Mises}}_{\ell} = 
        \sqrt{ \sigma_{xy}^2 + \sigma_{xz}^2 + \sigma_{yz}^2 + 
                \frac{(\sigma_{xx}-\sigma_{yy})^2+(\sigma_{xx}-\sigma_{zz})^2+(\sigma_{yy}-\sigma_{zz})^2}{6} }.
        \end{equation}

        \vs{0.2cm}
        \noindent The \emph{Shear Strain} or von Mises strain $\eta^{\text{Mises}}$ is a dimensionless 
        quantity that we compute for each atom $\ell$ using its strain tensor through an expression 
        equivalent to \emph{Shear Stress}

        \begin{equation}\label{shear_strain}
        \eta^{\text{Mises}}_{\ell} = 
        \sqrt{ \eta_{xy}^2 + \eta_{xz}^2 + \eta_{yz}^2 + 
                \frac{(\eta_{xx}-\eta_{yy})^2+(\eta_{xx}-\eta_{zz})^2+(\eta_{yy}-\eta_{zz})^2}{6} }.
        \end{equation}

        \vs{0.2cm}
        \noindent The \emph{Volumetric Strain} $\eta^{\text{Vol}}$ is a dimensionless quantity that we 
        compute for each atom $\ell$ using the principal components of its strain tensor through the 
        expression

        \begin{equation}\label{volumetric_strain}
        \eta^{\text{Vol}}_{\ell} = \frac{1}{3}\Big(\eta_{xx}+\eta_{yy}+\eta_{zz}\Big).
        \end{equation}

        \vs{0.2cm}
        \noindent Finally, the \emph{Non-Affine Displacement} $\mathscr{D}^2$ is a dimensionless quantity 
        that corresponds to the least square error when trying to determine the best tensor transformation 
        $\bar{\bar{F}}_{\ell}$ which maps from an initial undeformed configuration $\{{\bf d}_{j\ell}^0\}$, to 
        a deformed configuration $\{{\bf d}_{j\ell}\}$. The index $j$ indicates the close neighbors to 
        $\ell$, with $N^0_{\ell}$ the number of neighbors of atom $\ell$ at the reference configuration. This 
        transformation $\bar{\bar{F}}$ is known as the deformation gradient tensor and is an element that 
        allows us to determine the strain tensor. In conclusion, this descriptor would be

        \begin{equation}\label{nonaffine_displacement}
        \mathscr{D}_{\ell}^2 = 
        \min\left\{\sum_{j\in N^0_{\ell}} \left|{\bf d}^0_{j\ell}\bar{\bar{F}}_{\ell} - {\bf d}_{j\ell}\right|^2\right\}.
        \end{equation}

        \begin{figure}[ht!]
                \centering
                \includegraphics[scale=0.7]{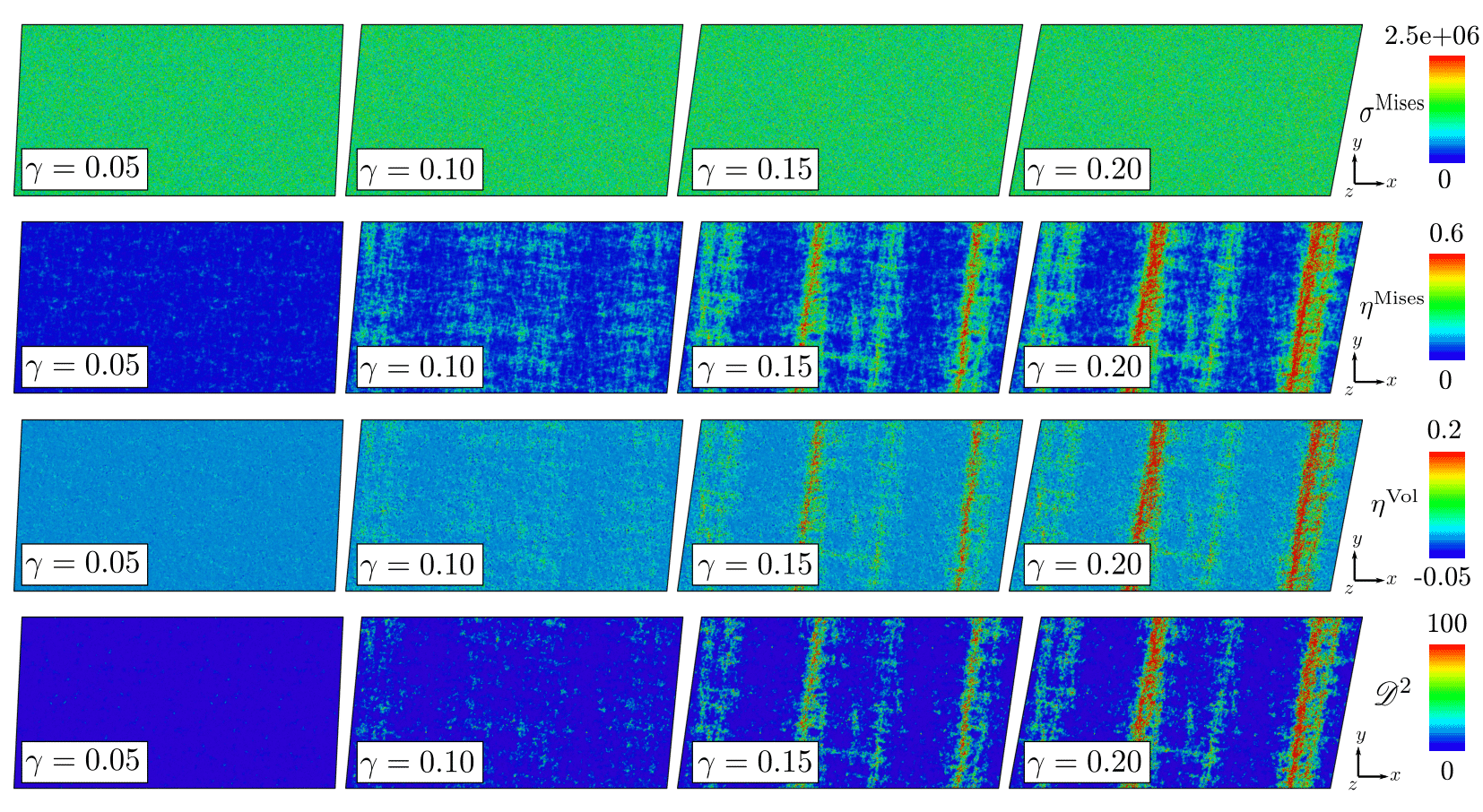}
                \caption{Spatial distribution, on the $xy$ plane, of descriptors $\alpha=\{\sigma^{\text{Mises}},\eta^{\text{Mises}}, \eta^{\text{Vol}},\mathscr{D}^ 2\}$. For the deformation states $\gamma=\{0.05, 0.10, 0.15, 0.20\}$, atoms are colored according to their descriptors values. When we compute the deformation gradient tensor $\bar{\bar{F}}$, we use a cutoff radius $r=5\text{ \AA}$ that defines the neighborhood $N_{\ell}^0$. The tensor $\bar{\bar{F}}$ allows to compute the strain tensor $\eta$ and all the descriptors.}
                \label{fig:spatial_distribution_descriptors}
        \end{figure}

        \vs{0.2cm}
        \noindent A summary of the defined descriptors, applied to our amorphous sample, is presented in Figure \ref{fig:spatial_distribution_descriptors}. The maps in Figure (\ref{fig:spatial_distribution_descriptors}) show the spatial distribution in the $xy$ plane of each descriptor at 4 different strain states.
        
       These results indicate a good agreement 
        with the phenomena of plastic deformation and shear band formation. For example, the spatial distribution of 
        $\eta^{\text{Mises}}$, $\eta^{\text{Vol}}$, and $\mathscr{D}^2$ shows that the system will eventually undergoes fracture at the place where the SBs are formed. 
        However, the spatial 
        distribution of $\sigma^{\text{Mises}}$ is distributed homogeneously along the cell, regardless 
        of whether the deformations are elastic or plastic. Although this result does not show any 
        signal about deformation, perhaps its application to network construction can reveal some information about plasticity.

        \begin{figure}[ht!]
                \centering
                \includegraphics[scale=0.4]{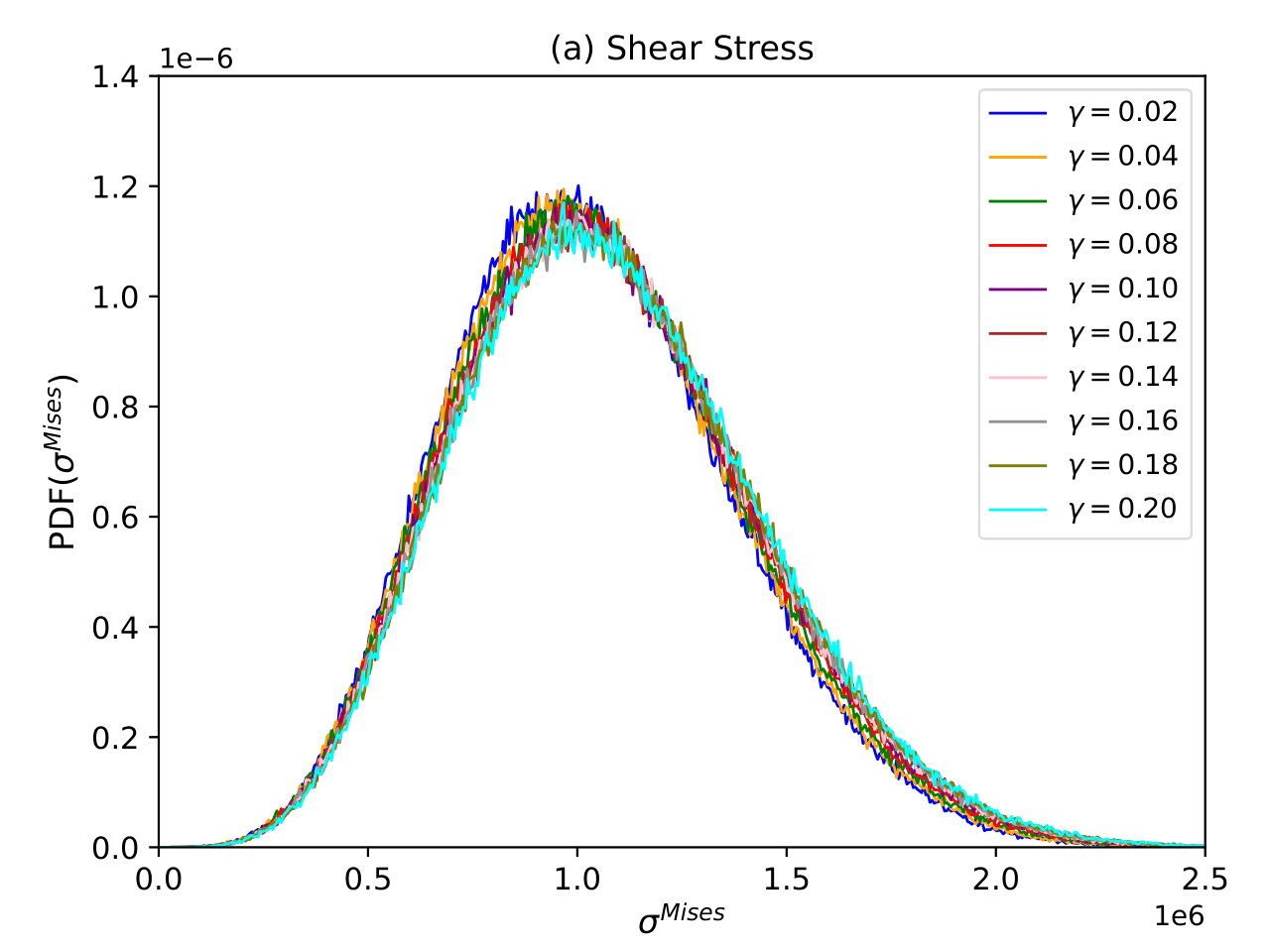}
                \includegraphics[scale=0.4]{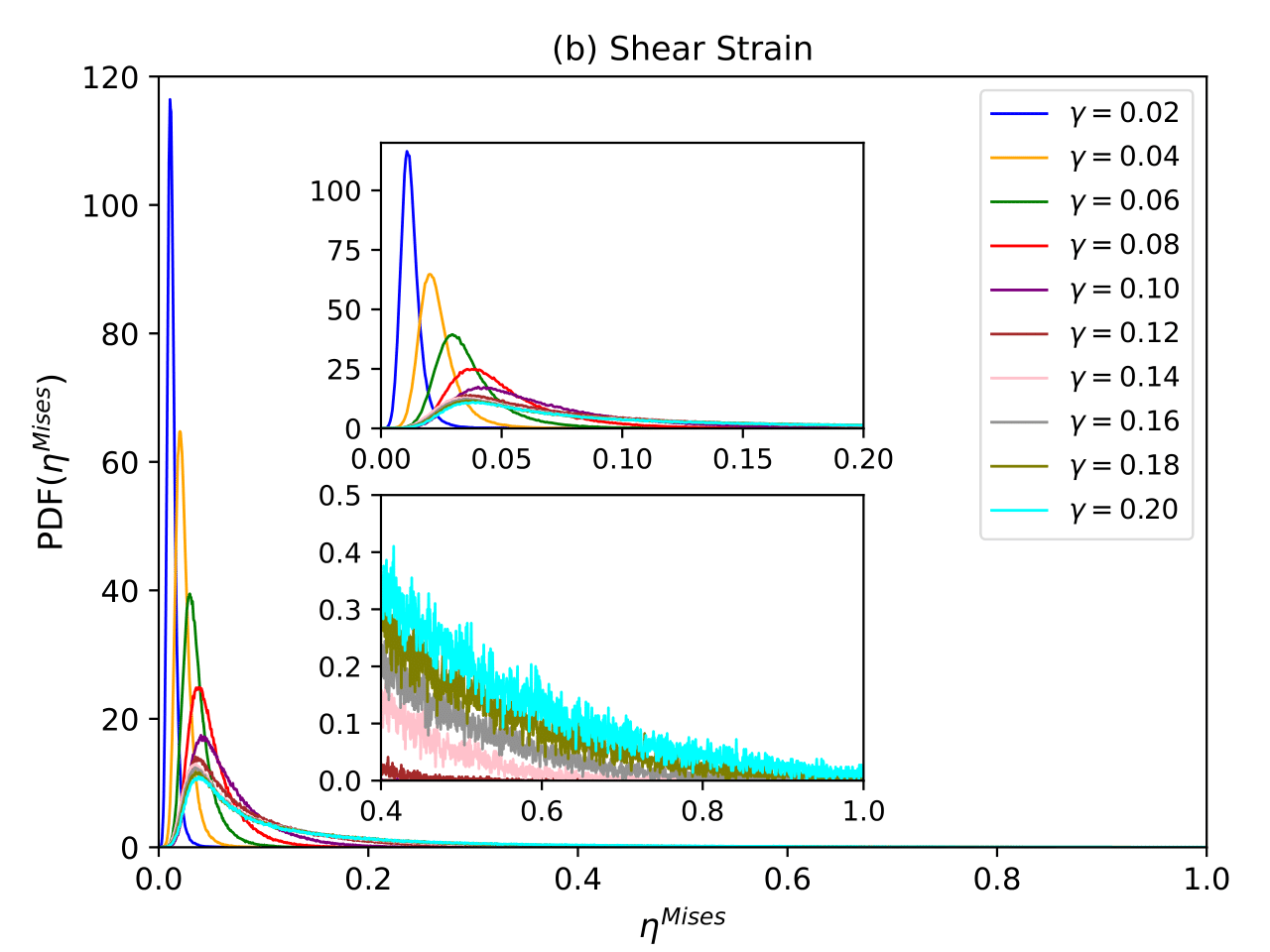}\\
                \includegraphics[scale=0.4]{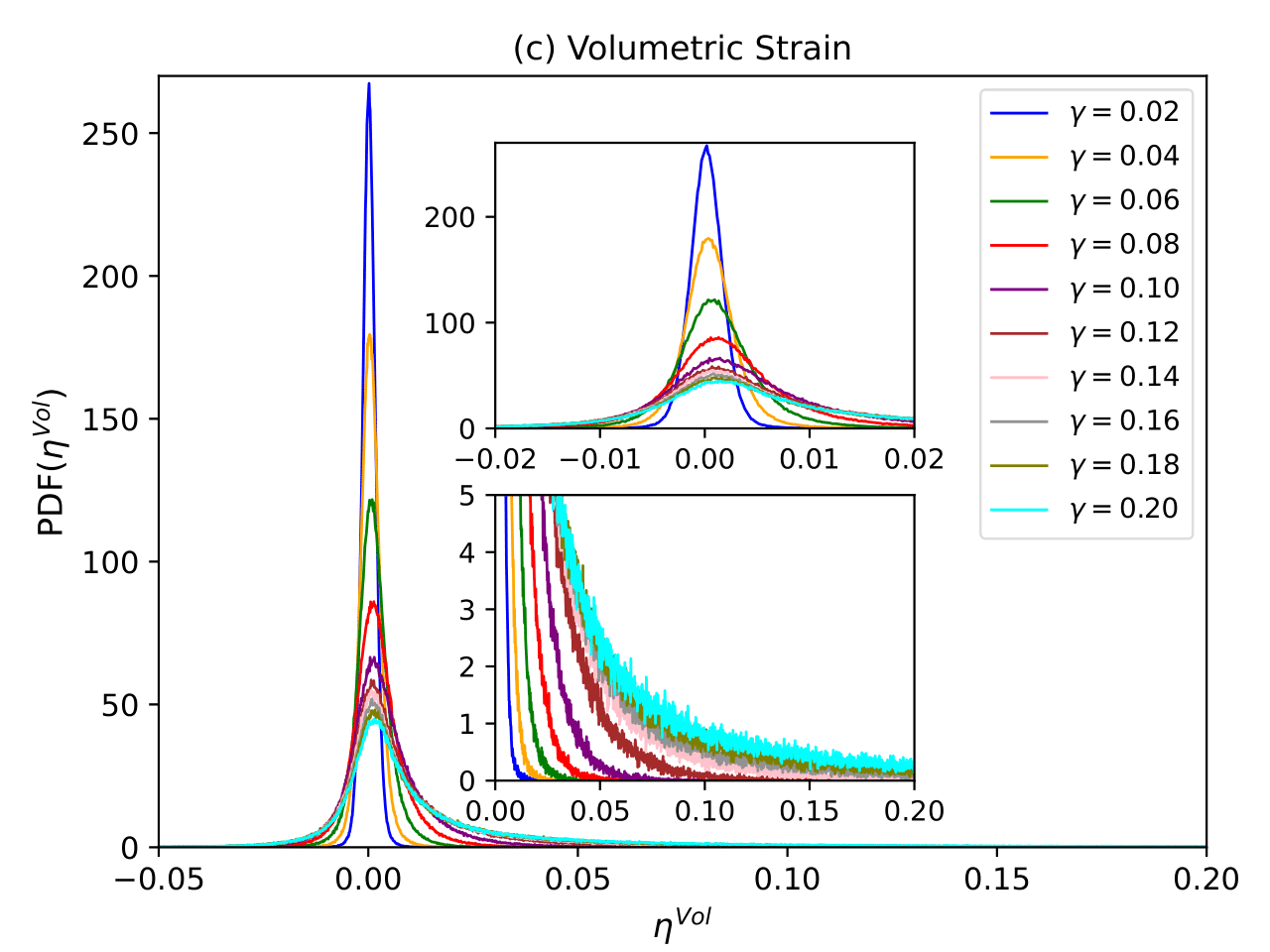}
                \includegraphics[scale=0.4]{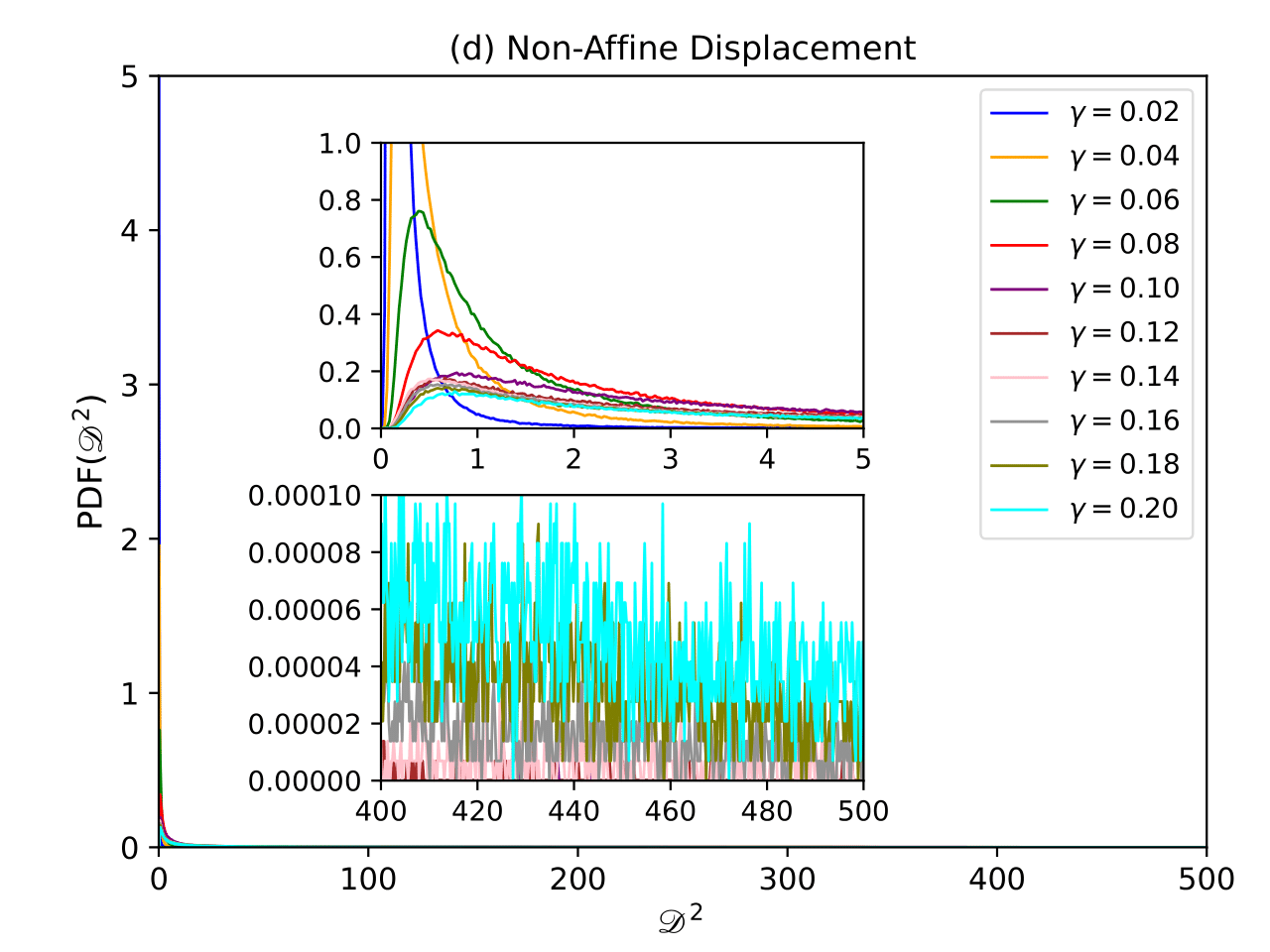}
                \caption{Probability density functions (value distributions) of (a) $\sigma^{\text{Mises}}$, 
                (b) $\eta^{\text{Mises}}$, (c) $\eta^{ \text{Vol}}$ and (d) $\mathscr{D}^2$, for different 
                deformation states $\gamma$. The inset in (b), (c), and (d) are zoomed curves of the same 
                respective distributions.}
                \label{fig:pdf-descriptors}
        \end{figure}

        \vs{0.2cm}
        \noindent We have also computed the numerical distribution via probability density that is 
        presented in Figure (\ref{fig:pdf-descriptors}), which shows us how the values descriptors are distributed among 
        each atom and for each deformation value. For the descriptor $\sigma^{\text{Mises}}$, we 
        see that its values approximately fit a Gaussian distribution, independent of the deformation regime, 
        unlike the others, which correlates with the homogeneous distribution seen in Figure 
        (\ref{fig:spatial_distribution_descriptors}). The $\eta^{\text{Mises}}$ distributions exhibit a dependency 
        on the deformation state, where for elastic strain, $\eta^{\text{Mises}}\in(0,0.1)$ , 
        and for plastic strains, $\eta^{\text{Mises}}\in(0,1)$. In this case, the standard deviation of the 
        data increases as well as the average $\langle\eta^{\text{Mises}}\rangle$. On the other hand, the 
        $\eta^{ \text{Vol}}$ distributions show a Gaussian trend for every value of $\gamma$ under the 
        restriction that as the strain increases, the standard deviation of the data increases, conserving their 
        averages $\langle\eta^{\text{Vol}}\rangle$ around $0$, which makes physical sense since the system 
        preserves its volume. Finally, the $\mathscr{D}^2$ distributions exhibit longer and longer tails as 
        the strain increases. These types of distributions are interesting to study since they apparently 
        have the form of a power law. Dozens of physical systems, such as earthquakes, solar flares, including 
        material deformation \cite{Baro2013Statistical}, have reported power laws to some quantity, giving a 
        better understanding of their behavior and complexity. It is interesting to study the tail of these 
        $\mathscr{D}^2$ distributions.

        \vs{0.2cm}
        \noindent In order to have a characterization of these distributions, we proceed to 
        calculate the time series of the moments, providing us with statistical information on the data. From 
        a statistical point of view, a distribution only provides probabilistic information about a random 
        variable. However, one way to characterize the sample space is by computing properties such as the mean, 
        variance, standard deviation, etc. These quantities can be calculated through the moments of a 
        distribution. For this analysis, we work with the standardized moment $\hat{\mu}$. In this context, 
        they are defined as

        \begin{equation}\label{standardized_moment}
        \hat{\mu}_k = \frac{\bar{\mu}_k}{\sigma^k} = \frac{\big\langle(\alpha-\mu)^k\big\rangle}{\sigma^k} 
                    = \frac{1}{\sigma^k}\int_{\alpha\in[a,b]}(\alpha-\mu)^k\rho(\alpha)d\alpha,
        \end{equation}

\begin{figure}[ht!]
                \centering
                \includegraphics[scale=0.21]{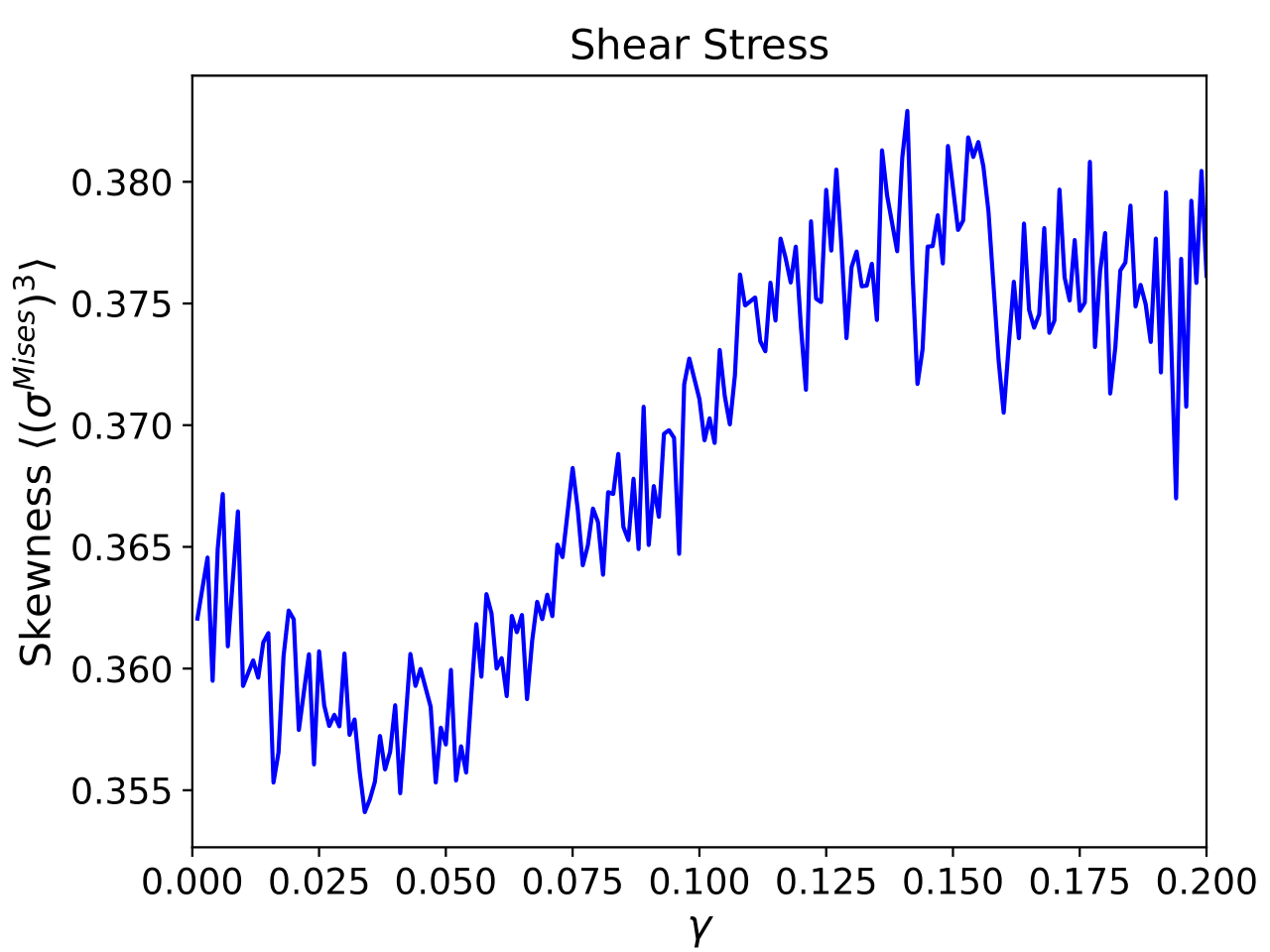}
                \includegraphics[scale=0.21]{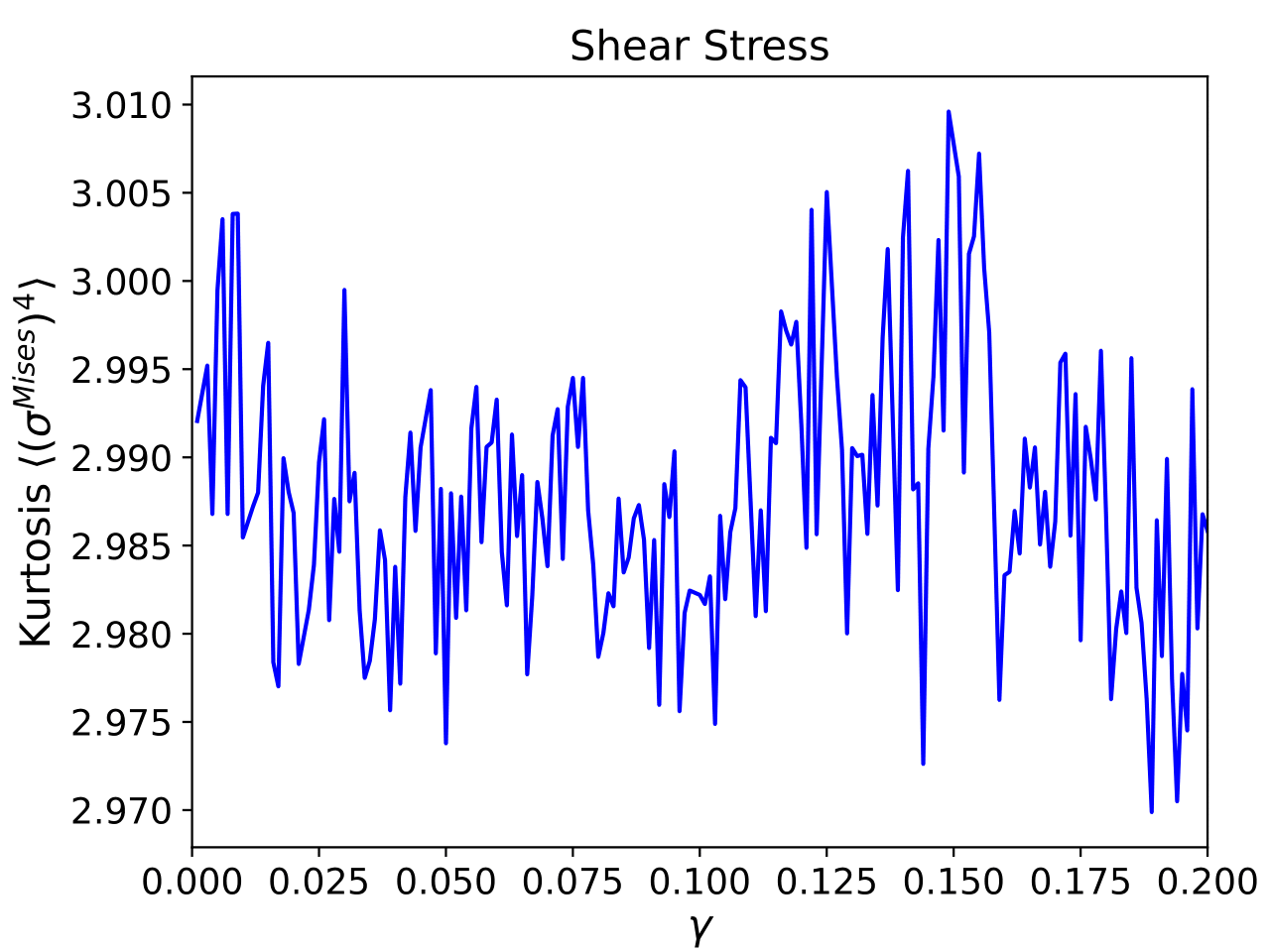}
                \includegraphics[scale=0.21]{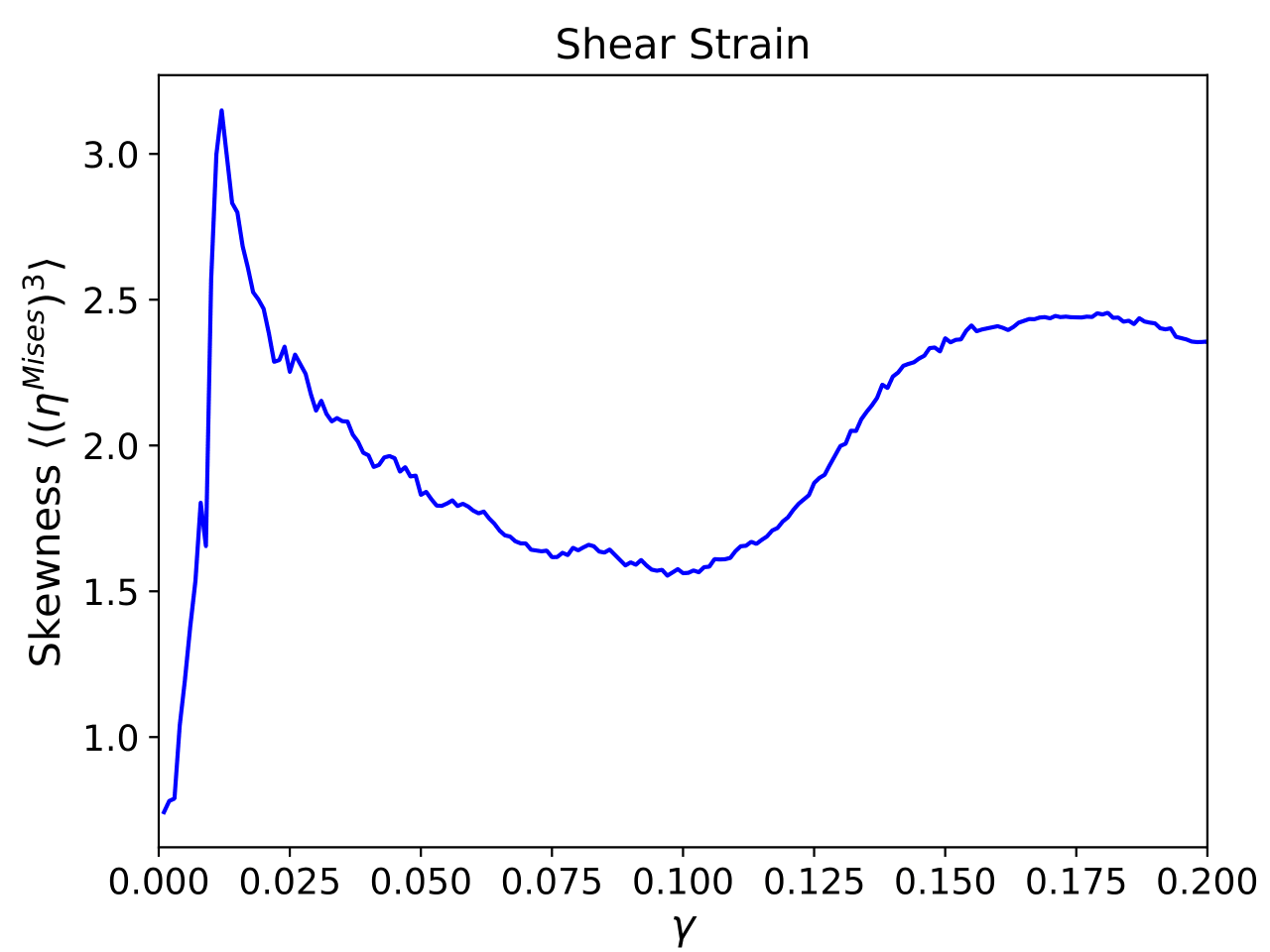}
                \includegraphics[scale=0.21]{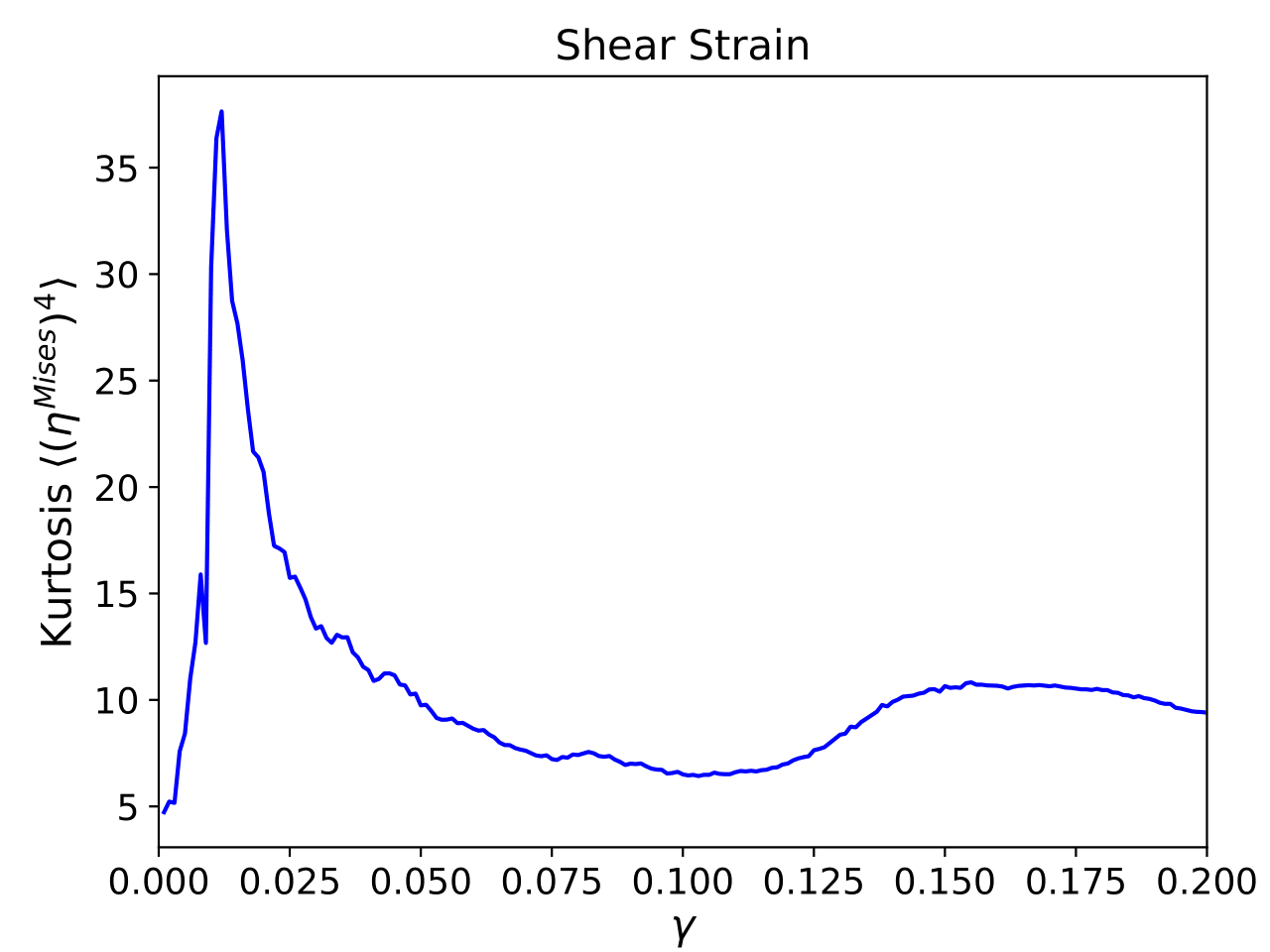}\\
                \includegraphics[scale=0.21]{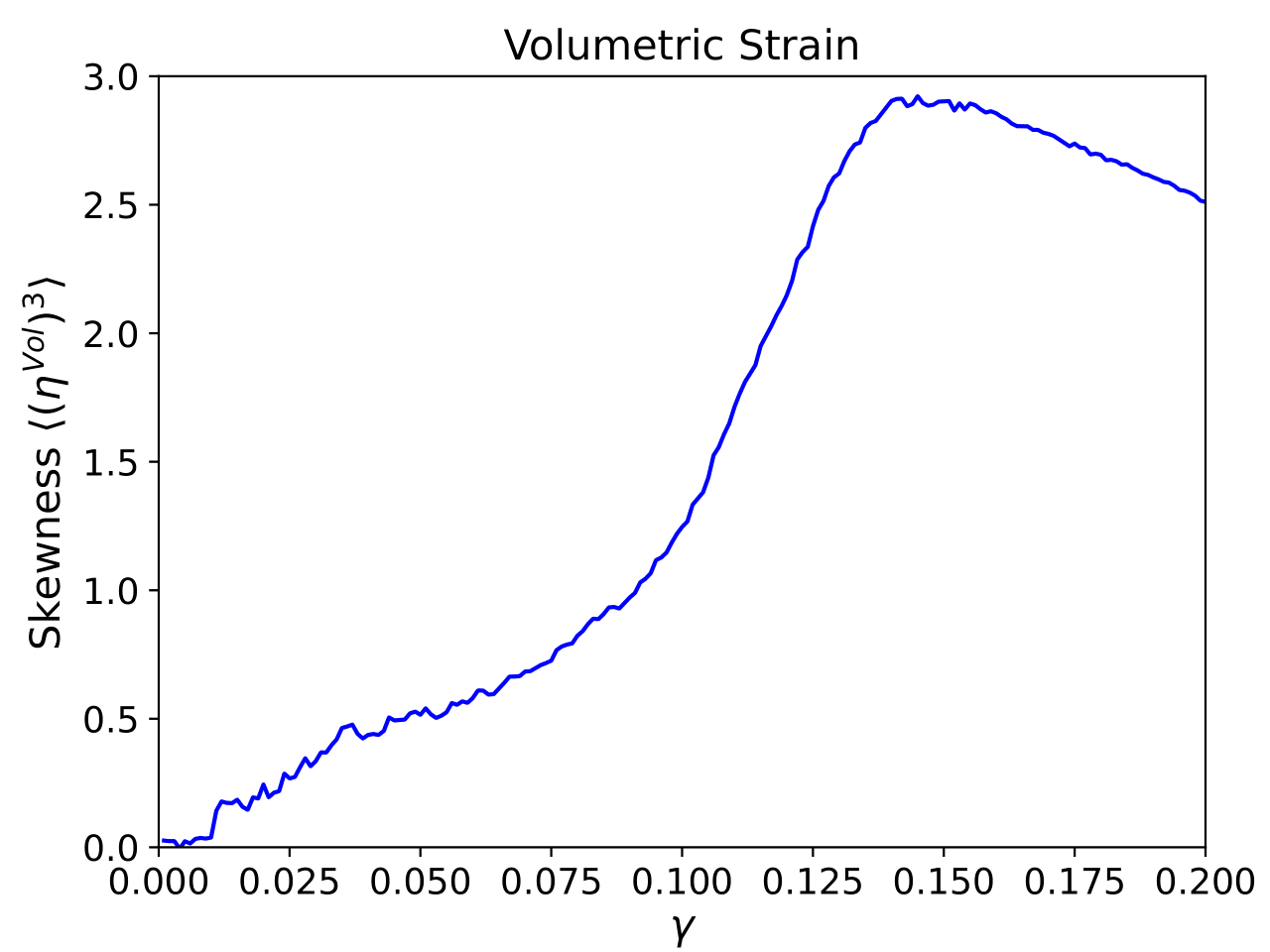}
                \includegraphics[scale=0.21]{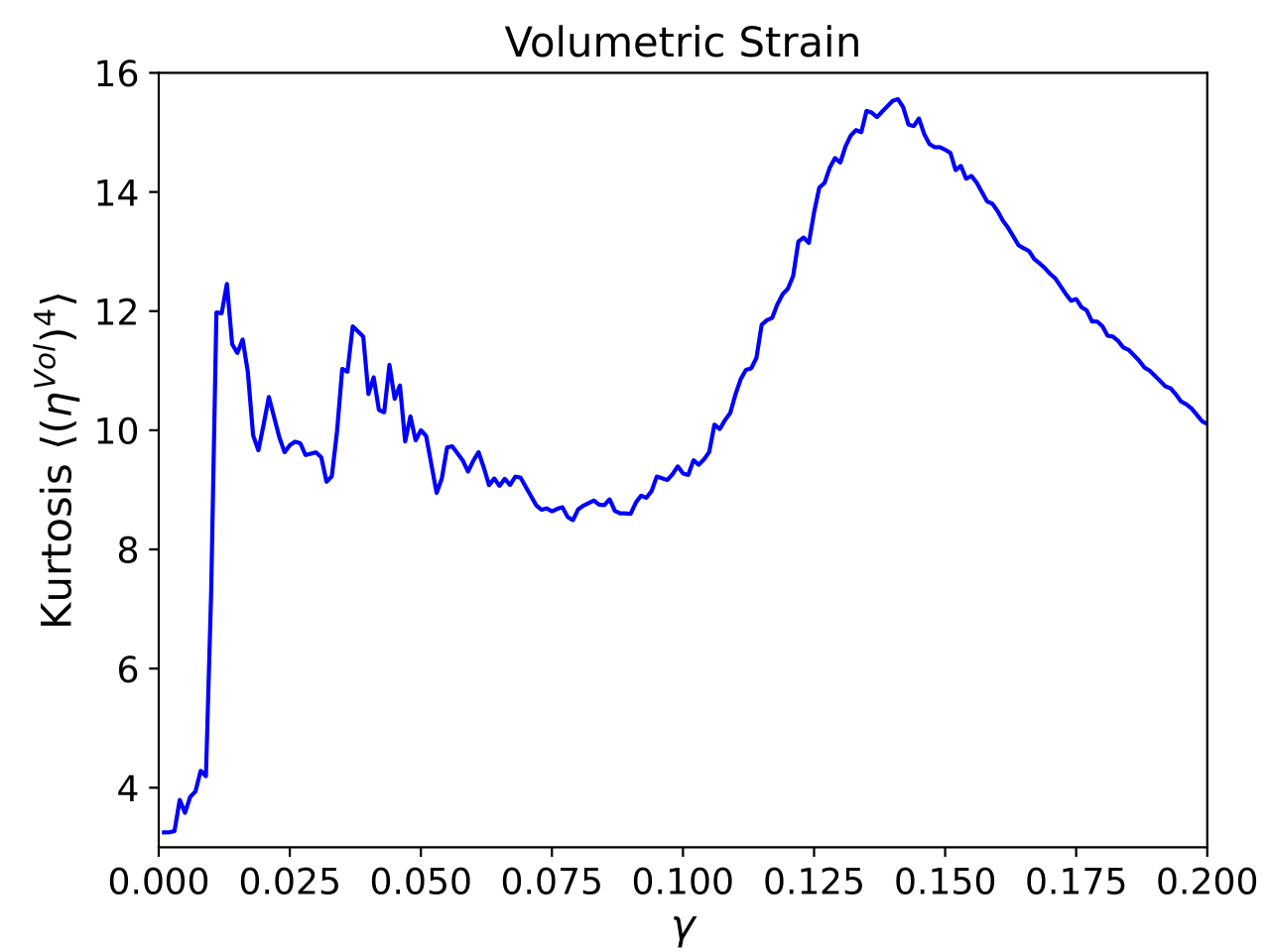}
                \includegraphics[scale=0.21]{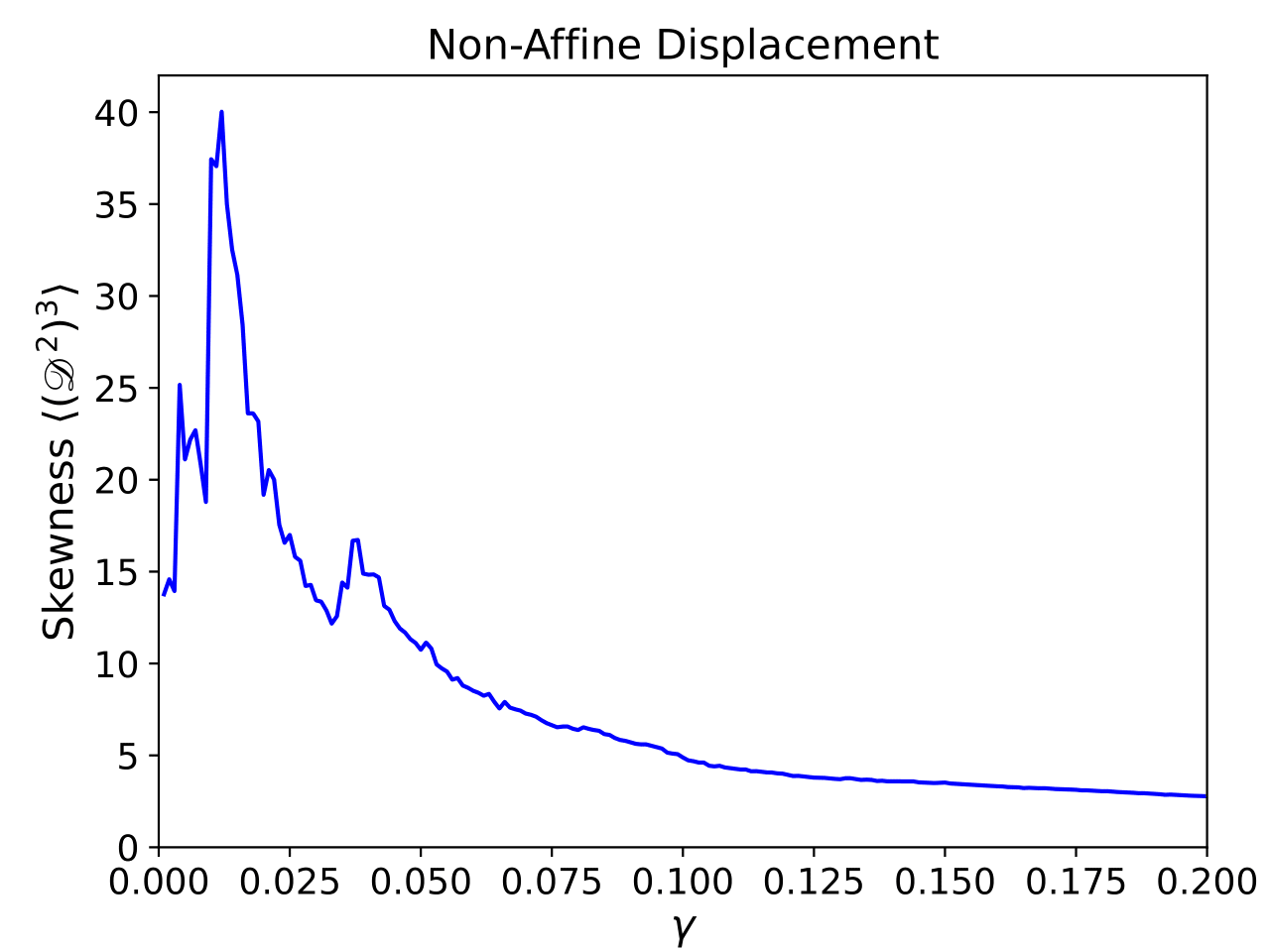}
                \includegraphics[scale=0.21]{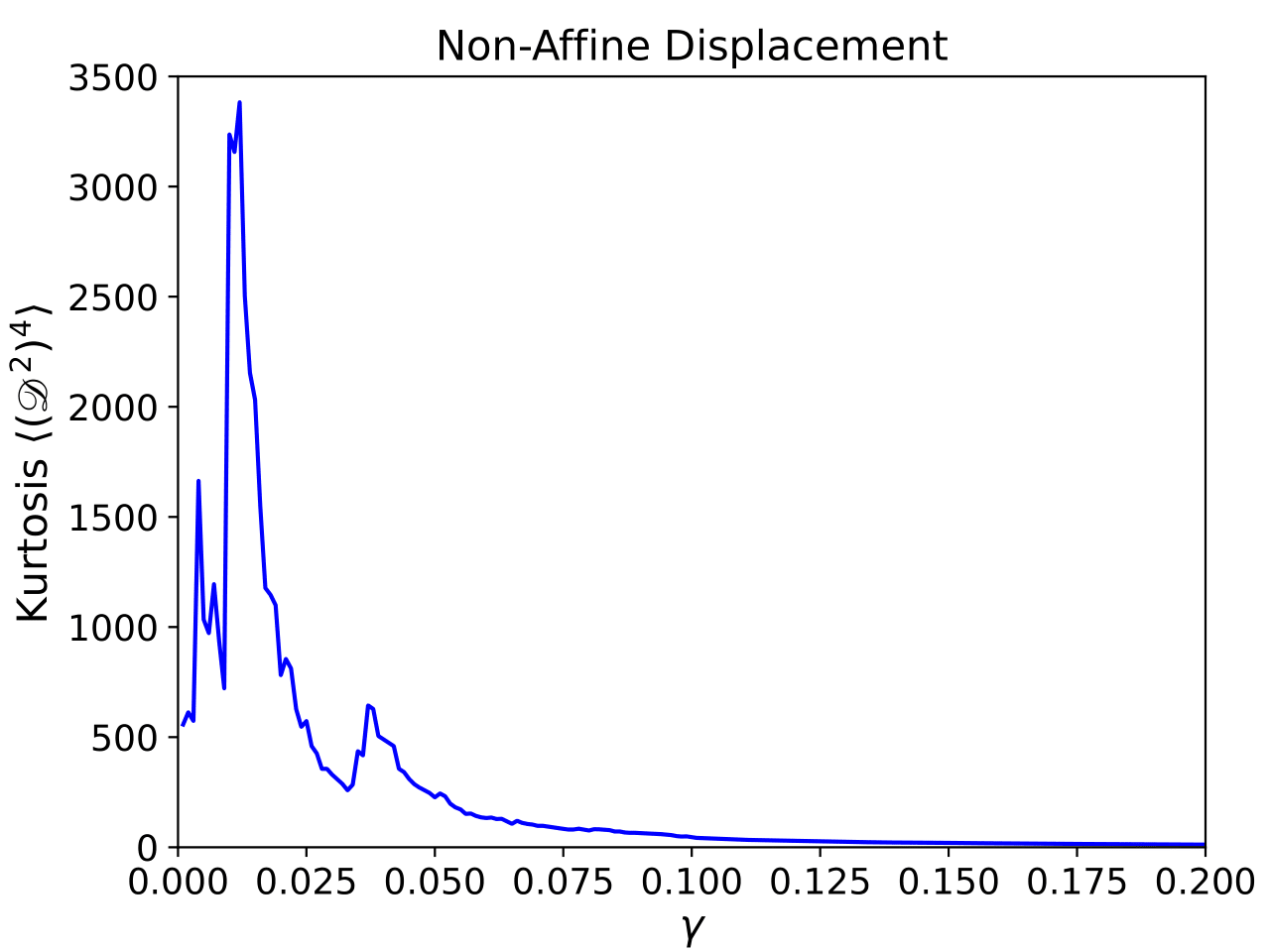}
                \caption{Third and fourth order standardized moments, from the distributions in Figure 
                (\ref{fig:pdf-descriptors}) as a function of strain and for descriptors 
                $\alpha=\{\sigma^{\text{Mises}} ,\eta^{\text{Mises}},\eta^{\text{Vol}},\mathscr{D}^2\}$. 
                These curves were calculated by means of the numerical integration that defines each 
                moment (acoording to equation (\ref{standardized_moment})) using the trapezoidal method.}
                \label{fig:mtos-descriptors}
        \end{figure}

        \vs{0.2cm}
        \noindent where $\alpha=\{\sigma^{\text{Mises}},\eta^{\text{Mises}},\sigma^{\text{Vol}},\mathscr{D}^2\}$ 
        are the physical descriptors and $\rho(\alpha)$ are the probability density functions of the Figure 
        (\ref{fig:pdf-descriptors}). For the expression (\ref{standardized_moment}), we work with the central 
        moment $\bar{\mu}_k$, centered on the average $\mu=\langle\alpha\rangle$, and normalized to the $k$th power of the 
        standard deviation $\sigma=\sqrt {\langle(\alpha-\mu)^2\rangle}$. In Figure (\ref{fig:mtos-descriptors}) 
        we present the results obtained for the third and fourth standardized moments of each one of the descriptors. 
        These results are basically the time series of the skewness and the kurtosis of the distributions. The 
        first and second standardized moments are not calculated since it is clear that they are $0$ and $1$ 
        respectively for the entire range of deformation.

        \vs{0.2cm}
        \noindent The third-order moment (skewness) $\hat{\mu}_3$ is a value that quantifies the asymmetry of a 
        distribution. The results presented in Figure (\ref{fig:mtos-descriptors}) show that for all distributions 
        there is always a positive skewness (heavy tails to the right) independent of the deformation state $\gamma$, 
        but that is interesting to note are the variations that this property exhibits. For example, for 
        $\sigma^{\text{Mises}}$ distributions, its skewness $\hat{\mu}_3\approx0.3$, that is, it is almost a Gaussian 
        with a slight asymmetry to the right. Interestingly, such asymmetry shows changes at points such as 
        elasto-plastic transition and fracture of the material. That the distribution becomes more asymmetric to the 
        right in the plastic regime and in the process of SBs formation, physically tells us that atoms with a high 
        stress appear as a result of the dissipation of elastic energy, an interesting result, all from one statistical 
        point of view. Now, the other descriptors show interesting variations, for example for the $\eta^{\text{Mises}}$ 
        it decreases in the plastic regime, for the $\eta^{\text{Vol}}$ it increases drastically before the failure and 
        for the $\mathscr{D}^2$ begins very asymmetric and as the strain increases, it attenuates.

        \vs{0.2cm}
        \noindent On the other hand we have the fourth-order moment (kurtosis) $\hat{\mu}_4$. This statistical measure 
        quantifies the degree of slope of the maximum and the tails of a distribution with respect to a Gaussian 
        distribution. For our results in Figure (\ref{fig:mtos-descriptors}), the kurtosis of each descriptor is positive 
        for all strain values, indicating that there are atypical data in the tails and a steeper maximum. These atypical 
        values are nothing more than the presence of highly stressed and/or deformed atoms. For the 
        $\sigma^{\text{Mises}}$ it is relatively constant showing that the difference to a Gaussian does not change, and 
        for the other descriptors, it indicates changes in relation to the deformation regime. This characterization is 
        just another complementary result to our physical interpretation of the phenomena that occur in the material.
        
        \vs{0.2cm}
        \noindent The next metric to study is a coefficient that allows to evaluate the level of inequality in a series 
        of data. This metric is known as the Gini coefficient, a measure of statistical dispersion which is used in 
        economics to study wealth inequality in a population. In our case, we are interested to compute how different 
        (numerically) the data for each $\alpha$ descriptor is as a function of strain. To evaluate this coefficient, we 
        need to calculate what is known as the Lorenz curve, which in this context would represent the fraction of the 
        accumulated descriptor (equivalent to the fraction of wealth of a population) as a function of the fraction of 
        accumulated atoms (equivalent to the population fraction).

        \begin{figure}[ht!]
                \centering
                \includegraphics[scale=0.4]{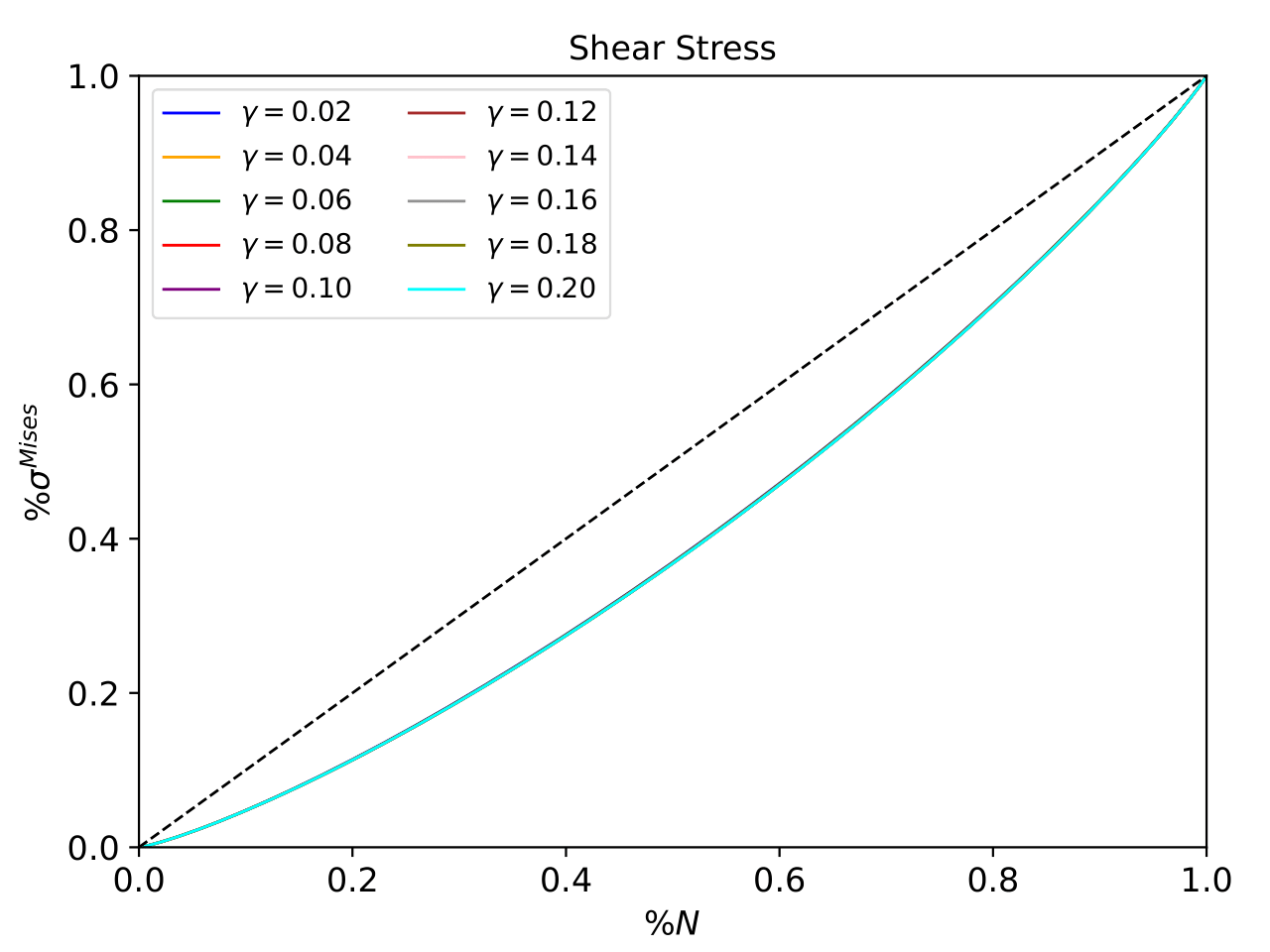}
                \includegraphics[scale=0.4]{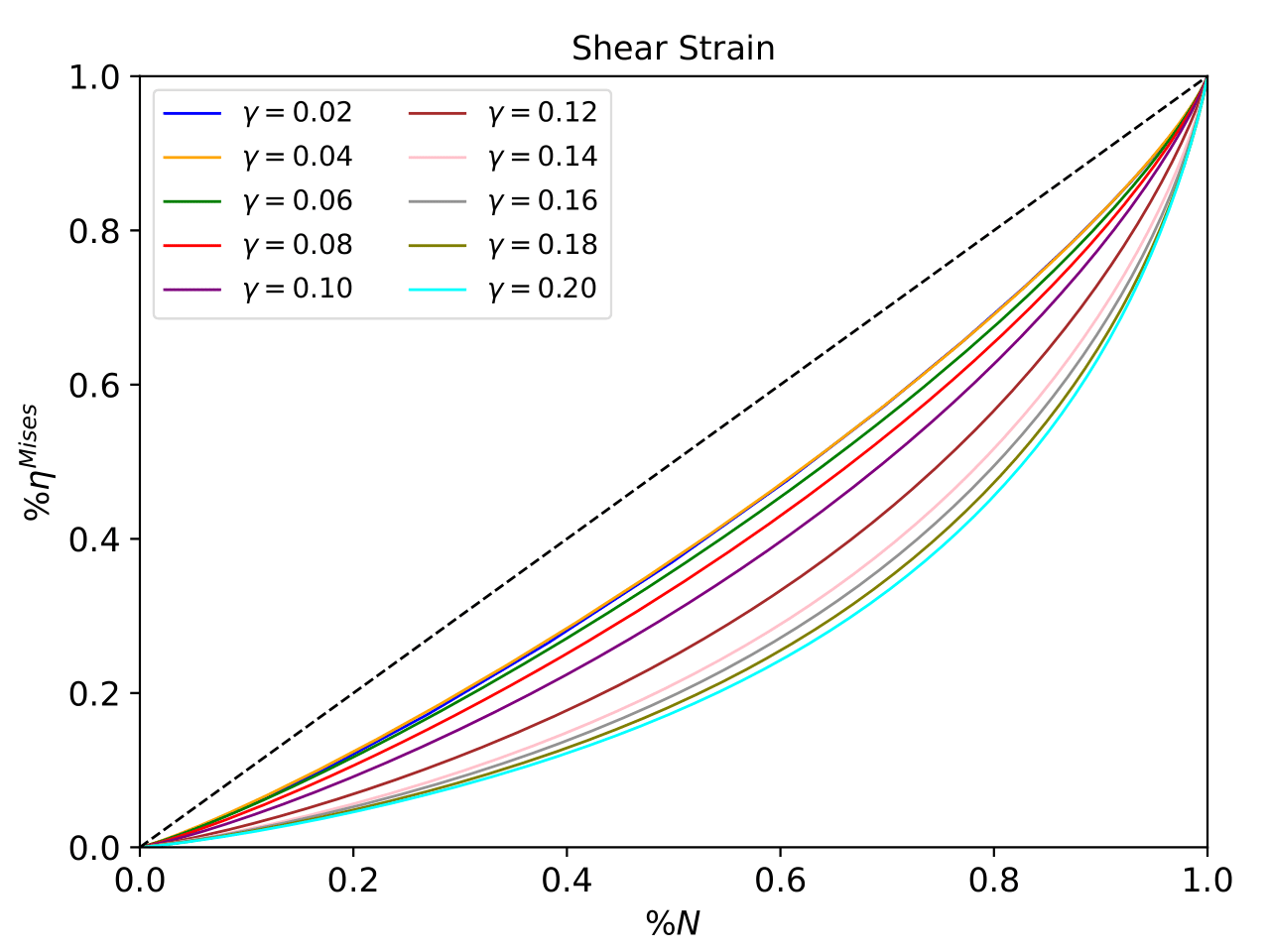}\\
                \includegraphics[scale=0.4]{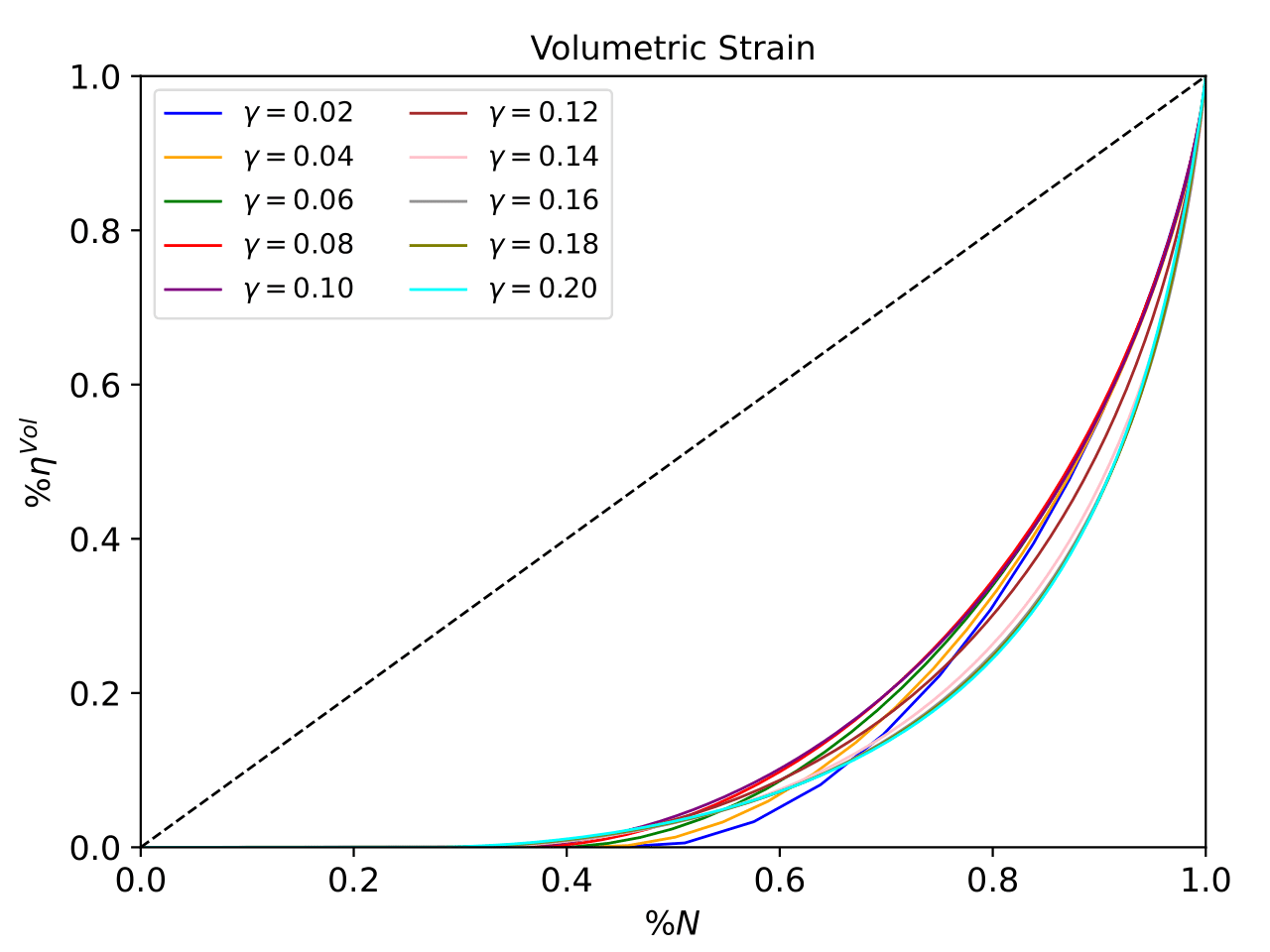}
                \includegraphics[scale=0.4]{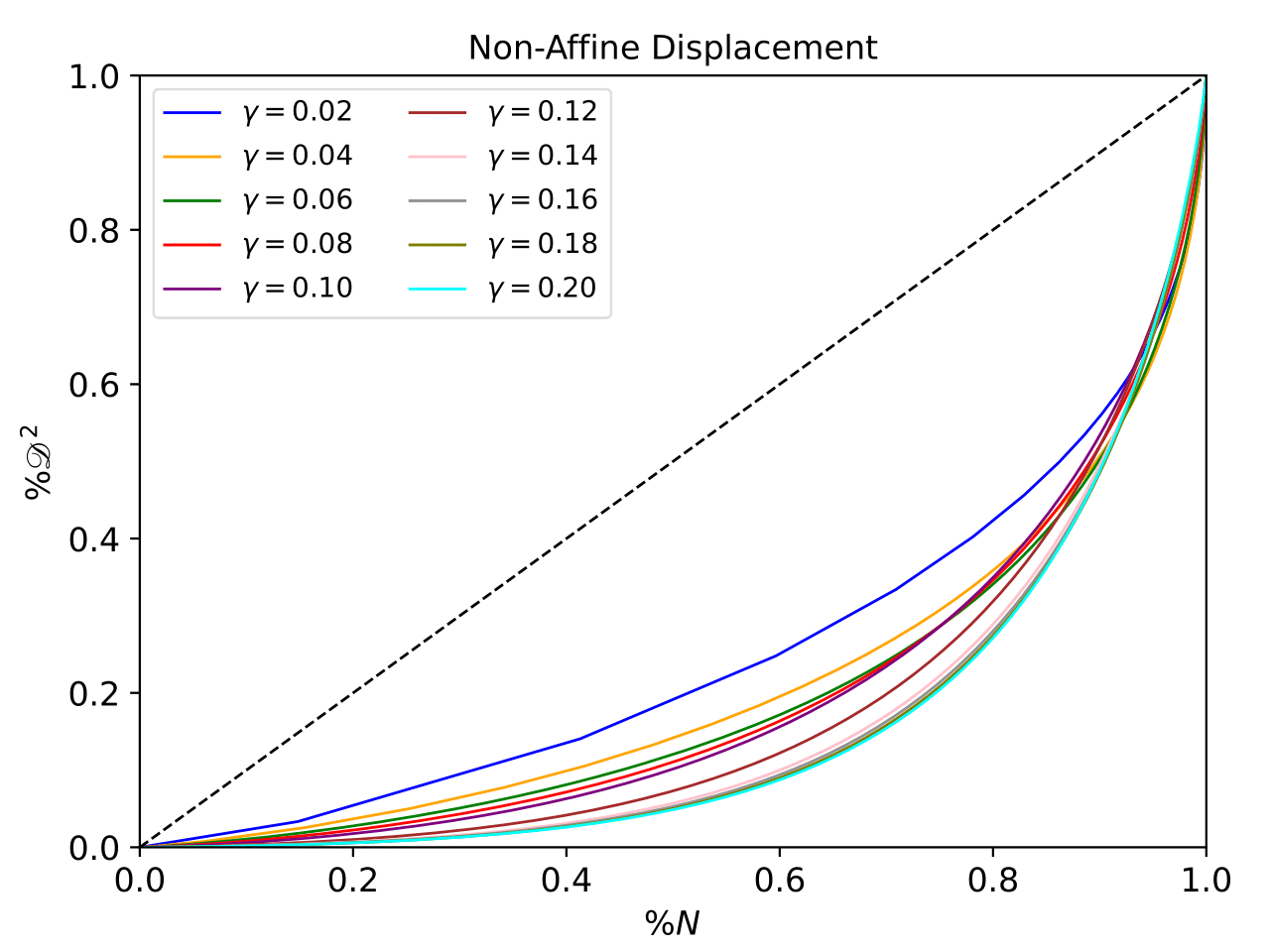}
                \caption{Lorenz curves of each of the distributions in the Figure (\ref{fig:pdf-descriptors}) 
                and for different strain values $\gamma$. The dashed line represents the maximum equality that 
                can be had in a data distribution. This means that for the same percentage of the population 
                there is the same percentage of wealth distributed among them. All the curves below the dashed 
                line account for a degree of inequality that the system exhibits.}
                \label{fig:lorenz-descriptors}
        \end{figure}

        \vs{0.2cm}
        \noindent In economics or in whatever context it is applied, Lorenz curves are a visual representation of how 
        unequal is the data distribution. These curves only take values between $0$ and $1$ since they evaluate the 
        fraction of the total wealth that a fraction of the complete population owns. If the curve approximates to 
        linear trend, the data distribution is more equality, since a given fraction of the population has the same 
        amount of wealth. However, if the curve takes the form of a semi-parabola with a minimum at zero, it indicates 
        that there is a degree of inequality in the data depending on the characteristics of said semi-parabola. 
        Reviewing Figure (\ref{fig:lorenz-descriptors}), we see that the Lorenz curves of the $\sigma^{\text{Mises}}$ 
        approximate a linear equation independent of the deformation regime, that is , the data is distributed almost 
        equally over the system. The above makes sense since the $\sigma^{\text{Mises}}$ is distributed homogeneously 
        according to the map in Figure (\ref{fig:spatial_distribution_descriptors}). On the other hand, the Lorenz 
        curves of the $\eta^{\text{Mises}}$ and the $\mathscr{D}^2$ take the form of a semi-parabola increasing its 
        convexity as the strain increases. At greater convexity, greater the level of inequality. In the case of 
        $\eta^{\text{Vol}}$, at least 40\% of the population of atoms presents a descriptor value that is very different 
        from the average, increasing the inequality considerably. Although the probability density of 
        $\sigma^{\text{Mises}}$ and $\eta^{\text{Vol}}$ look like a Gaussian, they will not necessarily have the same 
        properties.

        \vs{0.2cm}
        \noindent Now that we have an idea of how unequal is the distribution of each descriptor via Lorenz curves, 
        we proceed to quantify said inequality by calculating the Gini coefficient, which can be calculated by two 
        different ways:

        \begin{equation*}
        G = \frac{A}{A+B} = \bigg| 1 - \sum_{k=1}^{n-1}(X_{k+1}-X_k)(Y_{k+1}+Y_k) \bigg|,
        \end{equation*}

        \vs{0.2cm}
        \noindent where $A$ is the area between the line of perfect equality and the Lorenz curve, and $B$ is the 
        area under the Lorenz curve. Since the area under the perfect equality curve is $0.5$, it is clear to see 
        that $A+B=0.5$, so an alternative expression for this coefficient would be $G=1-2B$. On the other hand, we 
        have Brown's formula, an expression that uses the \emph{cumulative proportion of the population variable} 
        $X_k$ and the \emph{cumulative proportion of the descriptor variable} $Y_k$, to quantify this coefficient. 
        In order to have a more complete study, both results are presented for the entire time series of the Gini 
        coefficient of the all descriptors.

        \begin{figure}[ht!]
                \centering
                \includegraphics[scale=0.4]{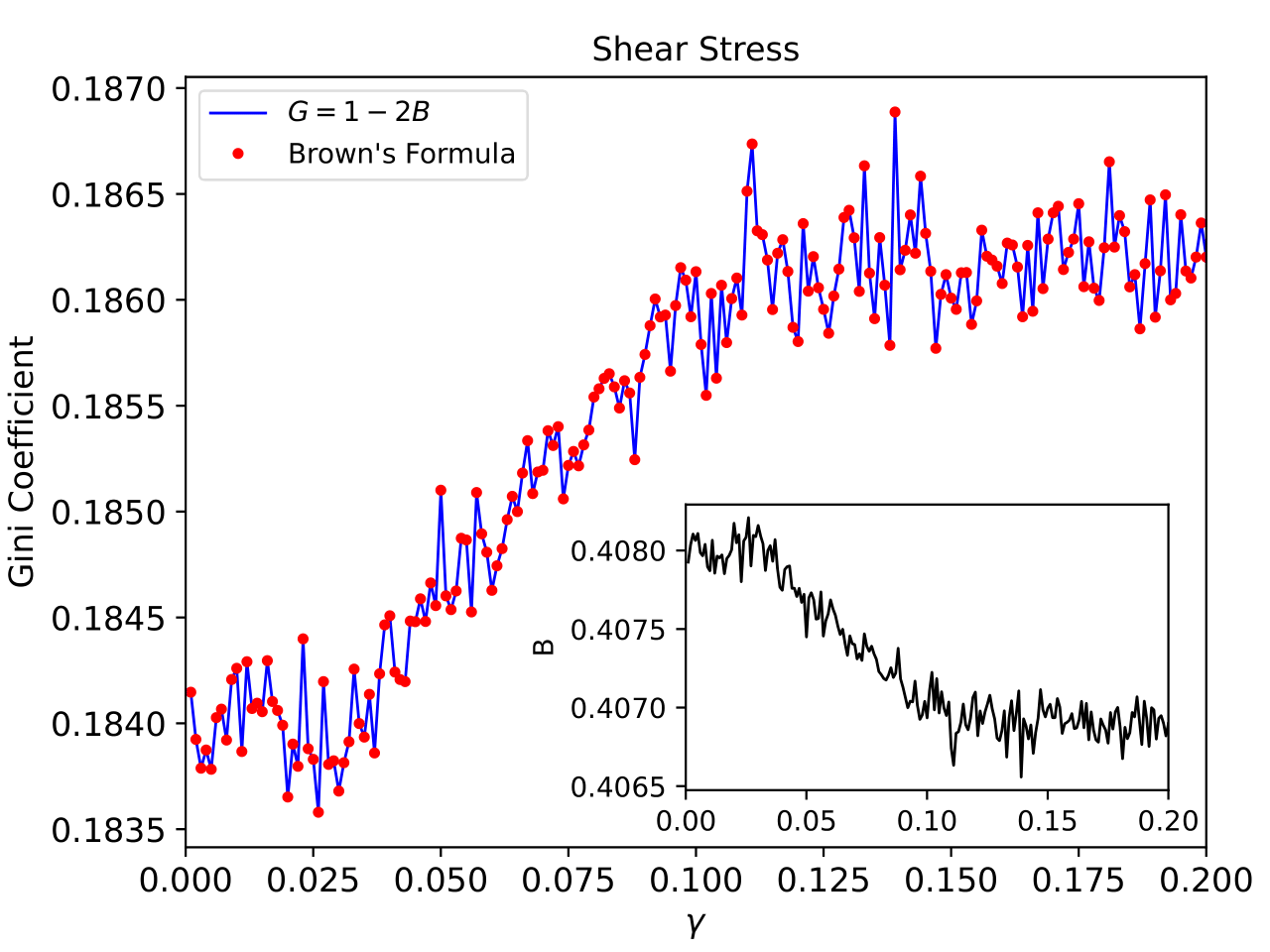}
                \includegraphics[scale=0.4]{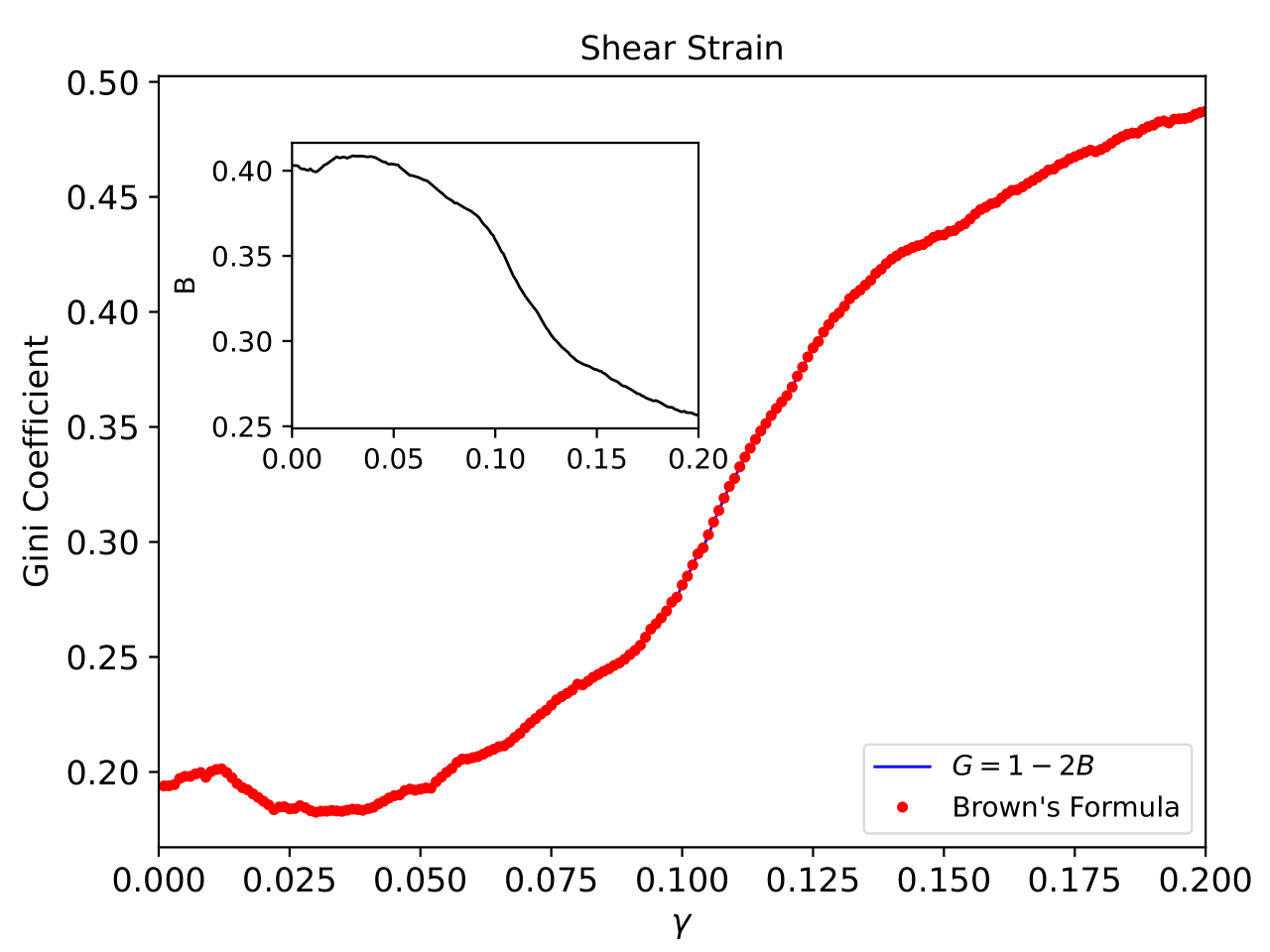}\\
                \includegraphics[scale=0.4]{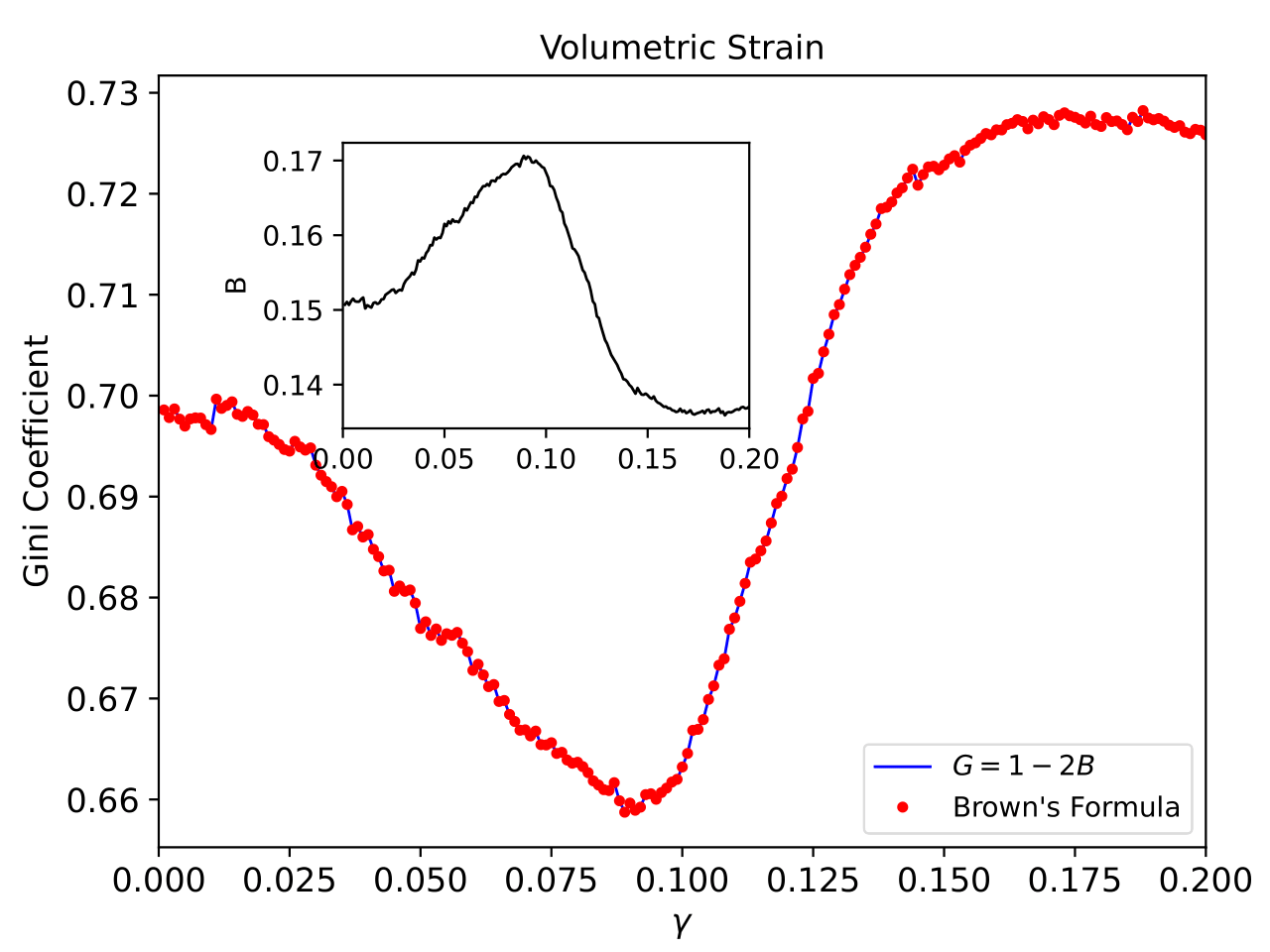}
                \includegraphics[scale=0.4]{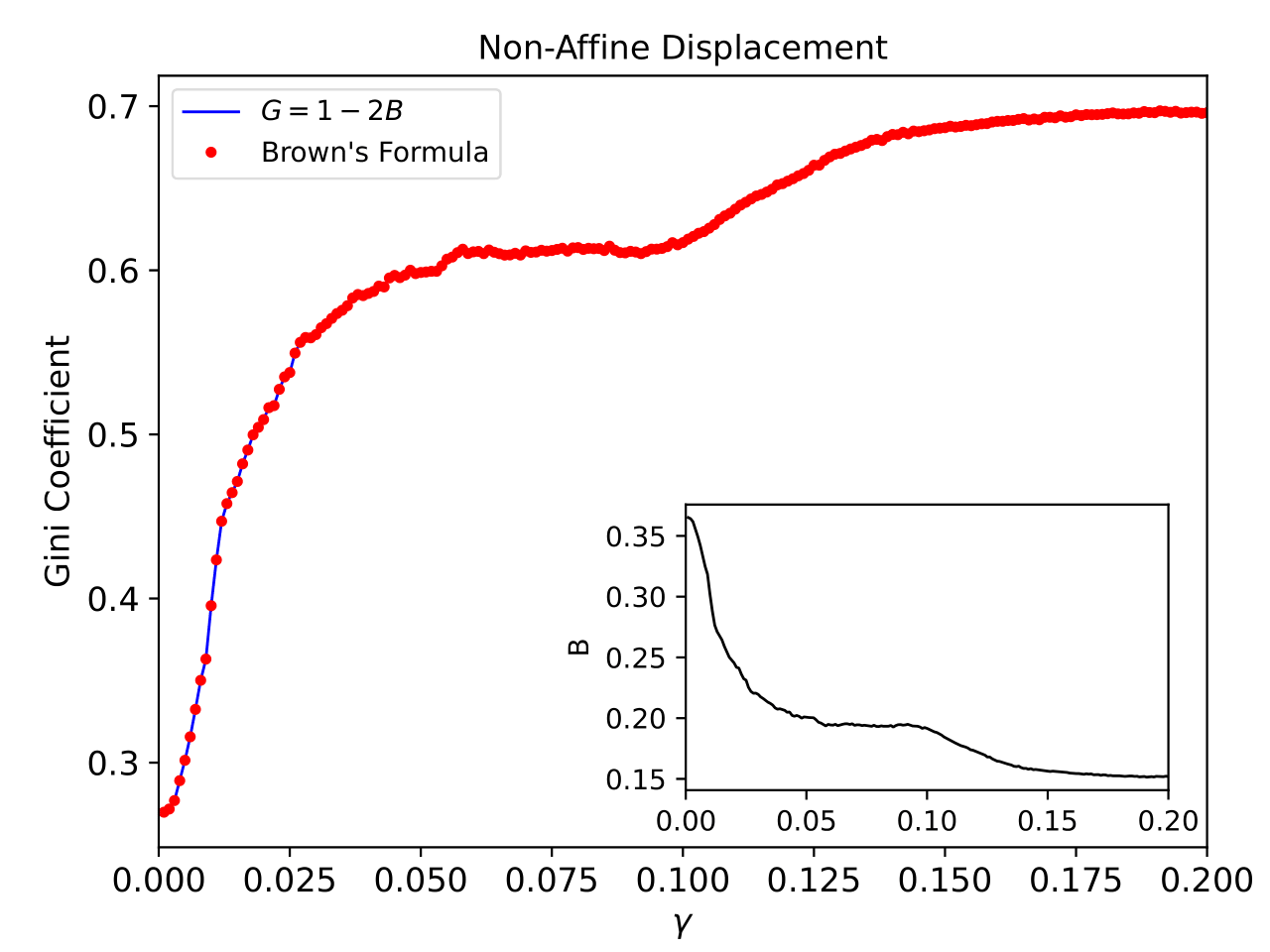}
                \caption{Gini coefficient as a function of strain for each of the distributions in Figure 
                (\ref{fig:pdf-descriptors}), computed using the area formula (numerical integration via the 
                trapezoidal method) and Brown's formula. The area under the Lorenz curves as a function of 
                strain is included in the inset. All the curves show a change as a function of the deformation 
                state of the system: elastic regime ($\gamma<0.45$), plastic regime ($0.45<\gamma<0.95$) and 
                plastic creep ($0.95<\gamma$).}
                \label{fig:gini-descriptors}
        \end{figure}

        \vs{0.2cm}
        \noindent The values that the Gini coefficient can take are between $0$ and $1$, where $0$ represents 
        perfect equality and $1$ maximum inequality. In relation to the analysis made on the Lorenz curves and 
        the results obtained in Figure (\ref{fig:gini-descriptors}), we see that the Gini coefficient gives us 
        better precision to evaluate inequalities. For example, in the case of $\sigma^{\text{Mises}}$, despite 
        the fact that its Lorenz curves reflected a relative equality in the data distribution, the Gini 
        coefficient shows how this relative equality is slowly lost when plastic deformations begin, saturating 
        until the SB formation process. Although the range of values that this coefficient takes for this 
        descriptor is $G\sim0.18$, it is interesting to note that information that is not possible to see with 
        all the previous results. On the other hand, for the descriptors $\eta^{\text{Mises}}$ and $\mathscr{D}^2$, 
        their Gini coefficient increases as a function of strain, registering variations up to $\Delta G\sim0.3$ for 
        the $\eta^{\text{Mises}}$ data and $\Delta G\sim0.45$ for the $\mathscr{D}^2$ data through the deformation    
        process. This positive variation indicates that the data for both descriptors are distributed 
        with less equality, which shows that there are atoms, or clusters of atoms that acquire even more atypical 
        values. For the $\eta^{\text{Mises}}$, the variations of inequality are recorded when the 
        plastic deformations $(\gamma>0.5)$ begin, and when the SB is located $(\gamma\sim0.95)$, unlike of the 
        $\mathscr{D}^2$, that such important variations are registered for the elastic deformations $(\gamma<0.5)$ 
        and when the SB is located. This result is an elegant way of verifying that the different mechanical states 
        that the system adopts under an external stress have multiple properties. On the one hand, the 
        \emph{Shear Strain} is relatively egalitarian in the elastic regime, until when the plastic regime begins, 
        select groups of atoms (apparently random), take a large part of the \emph{Shear Strain}, evidenced in the 
        formation of a SB in Figure (\ref{fig:spatial_distribution_descriptors}). And on the other hand, the 
        \emph{Non-Affine Displacement} evidences a growing inequality due to the underlying physical phenomena, 
        showing that a simple result of a minimization can reveal macroscopic properties.
        
        \vs{0.2cm}
        \noindent All the statistical analysis that we have presented in this section has allowed us to characterize 
        the deformation process in a microscopic way, identifying the different mechanical states that the system 
        adopts, the plastic events that are identified as the STZs and that give rise to the SBs, in order to finally, 
        have a macroscopic characterization of the material. In the following section we begin the presentation of our 
        methodology based on CNs for our research.


        \section{Complex network model}

        \noindent In general, a CN is a set of vertices and edges connecting them that 
        can be defined to represent a given physical system or process, by means of a suitable 
        definition for the vertices and edges. Such definitions depend on the specific problem that 
        we intend to model. As previously mentioned, CNs approaches have been successfully used to study 
        localized energy release processes such as seisms 
        \cite{Abe2011Universalities,Pasten2016Time,Martin2022Complex,Abe2011Finite}, solar flares 
        \cite{Gheibi2017Solar}, and sunspots \cite{Munoz2022Complex}. In these works, the vertices 
        correspond to a spatial region where an event occurs (seism, flare and/or sunspot), 
        and the connections are given by a temporal sequence. For a seismic catalog, only one event 
        occurs at a time, so the connections are between consecutive seisms. Thus, the spatial 
        information is contained in the vertices, and the temporal evolution in the edges of the 
        network. In the reference \cite{Munoz2022Complex}, the authors follow a similar strategy, 
        except that the vertices were defined using a certain threshold for the magnetic field, and 
        defining them as pixels in a solar magnetogram. The resulting network is a representation of 
        the spatiotemporal evolution of sunspots. All these works have shown that the topological 
        measures of the resulting network contain information about the underlying physical processes. 
        Our purpose then is to study whether a similar technique can be used to study the deformation 
        process in MG sample.

        \vs{0.2cm}
        \noindent In this study, we are interested in characterizing the atomic events that originate 
        in the MG during a shear deformation. For this, we have used only three descriptors, which are 
        $\alpha=\{\sigma^{\text{Mises}},\eta^{\text{Mises}},\mathscr{D}^2\}$. Then, following the 
        references \cite{Pasten2016Time,Martin2022Complex,Munoz2022Complex}, we have defined the vertices 
        as the atoms that satisfy the following conditions:

        \vs{-0.2cm}
        \begin{eqnarray}
        \sigma_0^{\text{Mises}} &<& \sigma^{\text{Mises}}_{\ell s} \label{mrc_shear_stress} \\
        \eta_0^{\text{Mises}}   &<& \eta^{\text{Mises}}_{\ell s}   \label{mrc_shear_strain} \\ 
        \mathscr{D}_0^2         &<& \mathscr{D}_{\ell s}^2         \label{mrc_nonaffinedisp}
        \end{eqnarray}

        \vs{0.2cm}
        \noindent where $\alpha_0=\{\sigma_0^{\text{Mises}},\eta_0^{\text{Mises}},\mathscr{D}_0^2\}$ are 
        defined as the vertices selector thresholds. On the other hand, the 
        $\alpha_{\ell s}=\{\sigma_{\ell s}^{\text{Mises}},\eta_{\ell s}^{\text{Mises}},\mathscr{D}_{\ell s}^2\}$ 
        are the values that each descriptor takes for the atoms $\ell$, monitored at time $t_s$ of the 
        simulation. Independently, if any of the above inequalities is satisfied, the atom $\ell$ at time 
        $t_s$ becomes a vertex of the network. Each descriptor is worked independently, this means that 
        if we are working with $\sigma^{\text{Mises}}$ descriptor, it is enough that (\ref{mrc_shear_stress}) 
        is fulfilled to generate the vertices. If we are working with $\eta^{\text{Mises}}$, it is enough 
        that (\ref{mrc_shear_strain}) is fulfilled and so on with the other. It is important to mention 
        that the choice of these thresholds determines the density of vertices that each CN will have.

        \vs{0.2cm}
        \noindent After having classified the atoms as vertices using any of the descriptors, we proceed 
        to establish the edges between them. For this particular model, the edges are established between 
        consecutive times of the simulation, that is, all the vertices of the time $t_{s}$ are connected 
        by an edge with all the vertices of the time $t_{s+1}$. This is a similar protocol used for solar 
        magnetograms \cite{Munoz2022Complex}. Note that this allows one atom to be connected to any other, 
        as long as (\ref{mrc_shear_stress}-\ref{mrc_nonaffinedisp}) is satisfied at consecutive times. 
        However, although edges do not imply causality, one could argue that the connections between atoms 
        are more meaningful if they are closer, given the locality of interactions leading to deformation. 
        Thus, we add a locality condition, which states that a connection between two 
        vertices only occurs if it is satisfied:

        \begin{equation}\label{locality_cond} 
        \sqrt{(x_{\ell}-x_{m})^2+(y_{\ell}-y_{m})^2+(z_{\ell}-z_{m})^2}\leq R_0.
        \end{equation}

        \vs{0.2cm}
        \noindent Here, $\n{r}_{\ell}=(x_{\ell}, y_{\ell}, z_{\ell})$ and $\n{r}_m=(x_m, y_m, z_m) $ are the 
        position vectors of the vertices $\ell,m$ respectively, and $R_0$ is the value of cutoff radius that 
        defines the neighborhood where connections are allowed for vertex $\ell$.

        \vs{0.2cm}
        \noindent Other rules that satisfy the connections in our model and that are consistent with previous 
        work \cite{Pasten2016Time,Martin2022Complex,Munoz2022Complex} are:

        \begin{enumerate}
                \item The graphs are undirected, i.e., the edges have no direction of connection.
                \item Vertices at the initial instant $t_0$ have no edges. The first edges appear 
                at the next time.
                \item Suppose we are working with the descriptor $\sigma^{\text{Mises}}$. A vertex 
                at $t_s$ will remain in the network along the simulation even if at later times 
                the condition $\sigma_0^{\text{Mises}} < \sigma^{\text{Mises}}_{\ell p}$ is not 
                satisfied with $t_s<t_p$. The only consequence of this is that said vertex will not 
                receive connections in $t_p$. The above also applies to the other descriptors.
                \item Multi-edge and self-connections are not allowed.
                \item The locality condition (\ref{locality_cond}) respects periodic boundary 
                conditions, i.e., it is possible for two vertices to be connected even though they 
                are at opposite ends of the cell.
        \end{enumerate}

        \vs{0.2cm}
        \noindent In this model we propose, each atomic configuration of the MG is mapped to a set of vertices, 
        providing instantaneous information about the system. The vertices are then connected as the simulation 
        progresses. When this is carried out during a certain time interval, the resulting network is expected 
        to contain relevant information about microscopic events via computation of topological metrics.

        \vs{0.2cm}
        \noindent Interested in the study of plasticity and microscopic events generated in MG, we have developed 
        a shear deformation up to $20$\% with a strain rate of $\dot{\gamma}=5\times10^8$ s $^{-1}$. This implies, 
        that we must simulate for $0.4$ ns (400,000 time steps). Every $2$ ps (2000 time steps), we monitor the 
        positions, velocities, and tensors associated with the deformation of all the atoms, to update the growth 
        of the network. This implies that the times in which the network will receive new vertices and edges are 
        $t_{s}=\{0,2,4,6,...,400\}$ ps with $s=\{0,1 ,2, ...,200\}$ and with its corresponding state/strain value 
        $\gamma_s=\{0,0.001,0.002,...0.2\}$.

        \begin{figure}[ht!]
                \centering
                \includegraphics[scale=0.6]{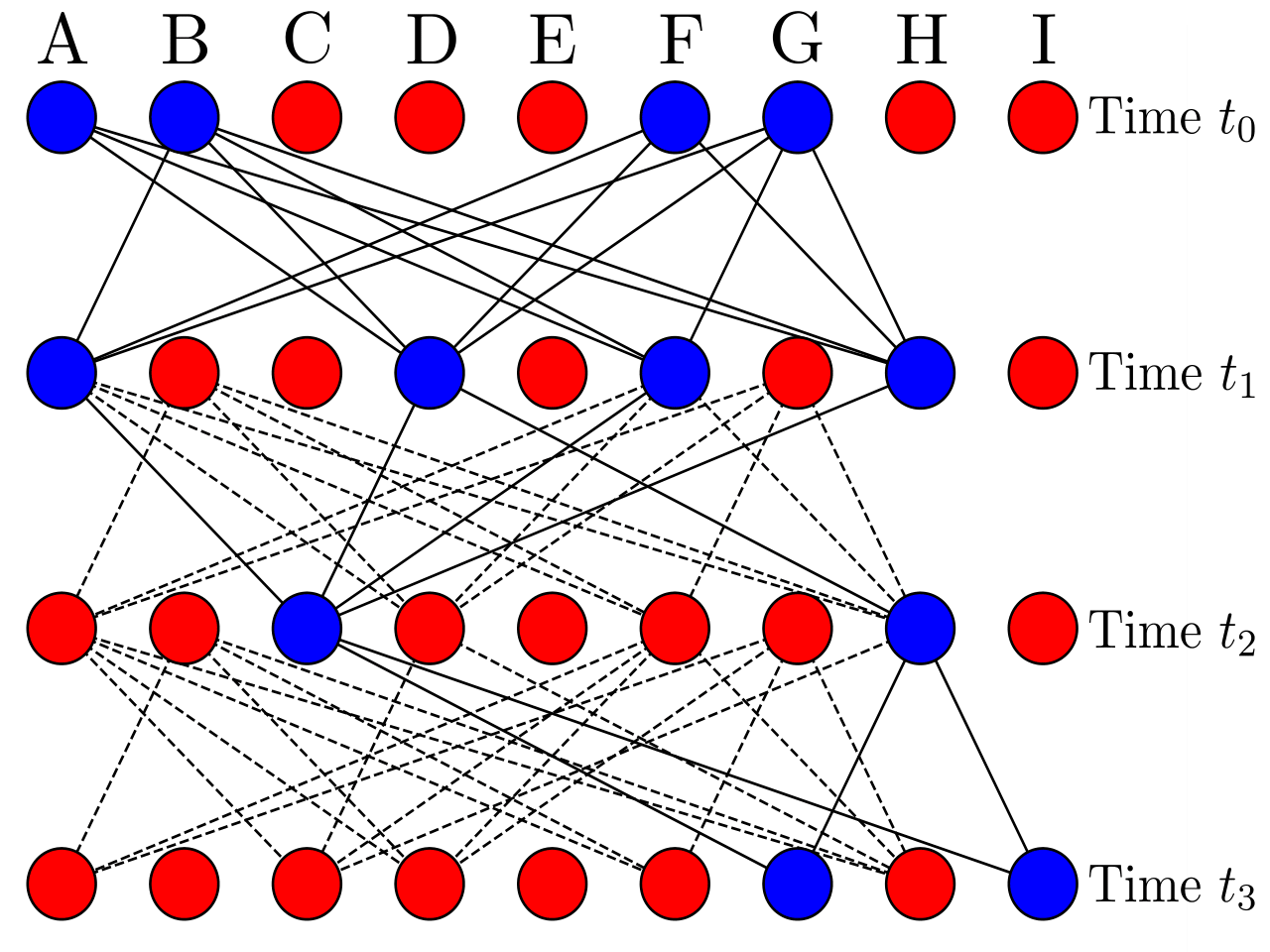}
                \caption{Representation of the CN model that we propose. (a) Diagram that illustrates the protocol 
                to set the edges between pairs of vertices while the MD performs shear deformation. Four consecutive 
                times and a nine-atoms system have been considered. The blue atoms are identified as the 
                network vertices and the red ones are the atoms that do not satisfy the threshold condition. Vertices at 
                time $t_s$ are connected to all vertices at time $t_{s+1}$. The graph grows in time, so the number 
                of edges of each vertice changes. Solid and dashed lines represent new and past edges, respectively. 
                For this particular case, we have selected a neighborhood $R_0$ large enough to allow all connections 
                between the vertices from time $t_s$ to $t_{s+1}$.}
                \label{fig:network_example}
        \end{figure}

        \vs{0.2cm}
        \noindent Figure (\ref{fig:network_example}) illustrates the network construction process, showing how 
        the edges connect pairs of vertices at consecutive times of the simulation. In particular and without loss 
        of generality, we have considered the first four times (6 ps equals 6000 time steps). Initially, there are 
        only vertices, always preserving the previous connections through the dashed lines.

        \subsubsection{Topological metrics}

                \noindent To study the dynamical and structural properties of CN, we compute topological metrics 
                such as degree, clustering coefficient, and centrality, which provide a characterization of the 
                network as the strain increases.

                \vs{0.2cm}
                \noindent The degree ($k_{\ell}$) is the number of connections of the vertex indexed by $\ell$. In our model we work with undirected networks, therefore, we do not have the concept of incoming and outgoing degree. Thus, the average degree $\langle k\rangle$ of the network is defined as:

                \begin{equation}\label{average_degree}
                \langle k \rangle_{\gamma_s} = \frac{1}{n(\nu_s)}\sum_{\ell\in\nu_s}^{n(\nu_s)}k_{\ell},
                \end{equation}

                \vs{0.2cm}
                \noindent where $\nu_{s}$ is the set of vertices, $n(\nu_s)$ is the number of vertices, 
                $k_{\ell}$ is the degree, which is computed by adding all the edges that the 
                vertex $\ell$ has in $\nu_s$, and $\langle k \rangle_{\gamma_s}$ is the average degree 
                for a strain $\gamma_s$ at time $t_s$.

                \vs{0.2cm}
                \noindent Based on this metric, one way to obtain a more complete characterization of the network 
                is by computing the degree probability distribution. This distribution contains information about 
                the nature of the network and the underlying physical processes involved in its construction 
                \cite{Albert2002Statistical}, being able to distinguish between purely random processes if the 
                distribution is Poisson, and processes under preferential attachment growth, if the distribution is 
                a power law \cite{Barabasi1999Emergence}.

                \vs{0.2cm}
                \noindent Another useful metric is the clustering coefficient 
                \cite{Albert2002Statistical,Newman2006Structure,Watts1998Collective}, which quantifies the number of 
                connections between the neighbors of a given vertex. Thus, this provides information on the density of 
                connections in the network. If all the neighbors of a vertex $\ell$ are connected to each other, then 
                the clustering coefficient of $\ell$ takes the maximum value $1$, whereas if they are not connected, 
                the clustering coefficient would be $0$. For undirected networks, we compute the clustering coefficient 
                of vertex $\ell$ at time $t_s$ and its average over the network as:

                \begin{equation} \label{average_clustering}
                C_{\ell s} = \sum_{j,k}\frac{2e_{jk}}{k_{\ell}(k_{\ell}-1)}  
                \quad \;\;\;\;\text{and}\;\;\;\quad \langle C \rangle_{\gamma_s}  
                = \frac{1}{n(\nu_s)}\sum_{\ell\in\nu_s}^{n(\nu_s)}C_{\ell s},
                \end{equation}

                \vs{0.2cm}
                \noindent where the indices $j$ and $k$ denote the neighbors of vertex $\ell$. The terms $e_{jk}$ are 
                the coefficients of the adjacency matrix of the network and can take the values $1$ or $0$, indicating 
                if the neighbors $j$ and $k$ are connected or disconnected respectively. The average clustering 
                coefficient $\langle C \rangle_{\gamma_s}$ can provide information about the global structure, existence 
                of communities, connectivity and how robust or vulnerable is a CN.

                \vs{0.2cm}
                \noindent To study the centrality of a graph and evaluate the importance of the vertices, metrics such 
                as betweeness centrality, closeness centrality, among others, are used. Each of them quantifies the 
                importance of the vertices in a different way, helping to give different interpretations to the network 
                structure. For our study we use betweeness centrality, a metric that measures the importance of a vertex 
                in terms of how many times it bridges along geodesics between other vertices. It is defined as the sum 
                of the fraction of all pairs of shortest paths through which a vertex $\ell$ passes. Thus, 

                \begin{equation}\label{betweeness_centrality}
                C^B_{\ell s} = \frac{2}{\big(n(\nu_s)-1\big)\big(n(\nu_s)-2\big)}\sum_{i,j\in \nu_s} \frac{\sigma(i,j|\ell)}{\sigma(i,j)} 
                \quad \;\;\;\;\text{and}\;\;\;\quad \langle C^B \rangle_{\gamma_s}  
                = \frac{1}{n(\nu_s)}\sum_{\ell\in\nu_s}^{n(\nu_s)}C^B_{\ell s},
                \end{equation}

                \vs{0.2cm}
                \noindent where $\sigma(i,j)$ is the number of shortest paths joining the vertices $i$ and $j$, and 
                $\sigma(i,j|\ell)$ is the number of shortest paths through which the vertex $\ell$ passes and that join 
                the vertices $i$ and $j$. If $i=j$, then $\sigma(i,j)=1$, and if $\ell=\{i,j\}$ then $\sigma(i,j|\ell)=0$. 
                The term that precedes the sum is the factor that normalizes the metric, in our case, for an undirected 
                graph. The average betweenness centrality $\langle C^B \rangle_{\gamma_s}$ is a very useful metric to 
                have information about the entire network. For example, it is a good indicator of vulnerability and 
                resilience. Vulnerability is understood as a network that is sensitive to being disconnected if a vertex 
                with high betweenness is eliminated, while a resilient network is a more robust network, i.e., it 
                preserves its properties even if it loses important vertices. Also, the distribution of this metric allows 
                us to analyze whether there are hierarchies in CNs.

                \vs{0.2cm}
                \noindent The other centrality measure we use is closeness centrality, a metric that measures the average 
                distance from a given vertex to all other vertices in terms of geodesics. The closeness is defined as the 
                multiplicative inverse of the distance between two vertices. Mathematically

                \begin{equation}\label{closeness_centrality}
                C^C_{\ell s} = \frac{n(\nu_s)-1}{\sum_{i\in \nu_s} d(\ell,i)} 
                \quad \;\;\;\;\text{and}\;\;\;\quad \langle C^C \rangle_{\gamma_s}  
                = \frac{1}{n(\nu_s)}\sum_{\ell\in\nu_s}^{n(\nu_s)}C^C_{\ell s},
                \end{equation}

                \vs{0.2cm}
                \noindent where $d(\ell,i)$ is the geodesic distance between vertices $\ell$ and $i$, and $n(\nu_s)-1$ are 
                all possible connections that vertex $\ell$ can have and that is used to normalize the metric.

                \vs{0.2cm}
                \noindent Graph theory provides a wide variety of metrics and topological measures to characterize the 
                structure of networks. In this study we use the degree, the clustering coefficient and centrality to 
                characterize the resulting networks from the mapping of each atomic configuration product of the deformation 
                of the MG. The objective is to carry out a microscopic analysis of the deformation, and to observe the 
                macroscopic physical phenomena by means our CN methodology.


        \section{Topological metric calculations}

        \noindent We start the construction of the CN by randomly selecting different values 
        for the vertex selection threshold for the three descriptors. We remember that the 
        selection of these thresholds cannot be arbitrary since the density of vertices of the graphs 
        depends precisely on this value. For very low thresholds, the density of vertices increases, 
        making the numerical computation of CN untenable, but for higher thresholds, the density of 
        vertices decreases, and therefore we will have unpopulated networks for the study. Another 
        parameter that we must select is the cutoff radius that defines the neighborhood where 
        connections are allowed for a vertex. The selection of this parameter influences the 
        density of connections that the CN will have, so and in view of the dimensions of the 
        simulation cell, for our study we have selected the radii $R_0=\{10,20,30,40, 50\}$ \AA.

        \vs{0.2cm}
        \noindent Before proceeding with the construction of the networks and the calculation of 
        the metrics, we present the data of all the thresholds that we have used to apply our 
        methodology. The objective is to carry out a preliminary analysis to decide which threshold(s) 
        is(are) most suitable for the study.

        \begin{table}[h!]
                \centering
                \begin{tabular}{|cccc|l|cccc|}
                \cline{1-4} \cline{6-9}
                \multicolumn{4}{|c|}{\textbf{Shear Stress} [GPa$\cdot\text{A}^3/V$]} & \hs{0.5cm} & \multicolumn{4}{c|}{\textbf{Shear Strain}}                               \\ \cline{1-4} \cline{6-9} 
                \multicolumn{1}{|c|}{$[\sigma_0^{\text{Mises}}]_1$} & \multicolumn{1}{c|}{$2.25\times10^6$} & \multicolumn{1}{c|}{$[\sigma_0^{\text{Mises}}]_{11}$} & \multicolumn{1}{c|}{$2.56\times10^6$} &  & \multicolumn{1}{c|}{$[\eta_0^{\text{Mises}}]_1$} & \multicolumn{1}{c|}{$0.5649$} & \multicolumn{1}{c|}{$[\eta_0^{\text{Mises}}]_{11}$} & \multicolumn{1}{c|}{$0.7524$} \\ \cline{1-4} \cline{6-9} 
                
                \multicolumn{1}{|c|}{$[\sigma_0^{\text{Mises}}]_2$} & \multicolumn{1}{c|}{$2.28\times10^6$} & \multicolumn{1}{c|}{$[\sigma_0^{\text{Mises}}]_{12}$} & \multicolumn{1}{c|}{$2.59\times10^6$} &  & \multicolumn{1}{c|}{$[\eta_0^{\text{Mises}}]_2$} & \multicolumn{1}{c|}{$0.5837$} & \multicolumn{1}{c|}{$[\eta_0^{\text{Mises}}]_{12}$} & \multicolumn{1}{c|}{$0.7711$} \\ \cline{1-4} \cline{6-9} 
                
                \multicolumn{1}{|c|}{$[\sigma_0^{\text{Mises}}]_3$} & \multicolumn{1}{c|}{$2.31\times10^6$} & \multicolumn{1}{c|}{$[\sigma_0^{\text{Mises}}]_{13}$} & \multicolumn{1}{c|}{$2.62\times10^6$} &  & \multicolumn{1}{c|}{$[\eta_0^{\text{Mises}}]_3$} & \multicolumn{1}{c|}{$0.6024$} & \multicolumn{1}{c|}{$[\eta_0^{\text{Mises}}]_{13}$} & \multicolumn{1}{c|}{$0.7899$} \\ \cline{1-4} \cline{6-9} 
                
                \multicolumn{1}{|c|}{$[\sigma_0^{\text{Mises}}]_4$} & \multicolumn{1}{c|}{$2.34\times10^6$} & \multicolumn{1}{c|}{$[\sigma_0^{\text{Mises}}]_{14}$} & \multicolumn{1}{c|}{$2.65\times10^6$} &  & \multicolumn{1}{c|}{$[\eta_0^{\text{Mises}}]_4$} & \multicolumn{1}{c|}{$0.6212$} & \multicolumn{1}{c|}{$[\eta_0^{\text{Mises}}]_{14}$} & \multicolumn{1}{c|}{$0.8086$} \\ \cline{1-4} \cline{6-9} 
                
                \multicolumn{1}{|c|}{$[\sigma_0^{\text{Mises}}]_5$} & \multicolumn{1}{c|}{$2.37\times10^6$} & \multicolumn{1}{c|}{$[\sigma_0^{\text{Mises}}]_{15}$} & \multicolumn{1}{c|}{$2.69\times10^6$} &  & \multicolumn{1}{c|}{$[\eta_0^{\text{Mises}}]_5$} & \multicolumn{1}{c|}{$0.6399$} & \multicolumn{1}{c|}{$[\eta_0^{\text{Mises}}]_{15}$} & \multicolumn{1}{c|}{$0.8273$} \\ \cline{1-4} \cline{6-9} 
                
                \multicolumn{1}{|c|}{$[\sigma_0^{\text{Mises}}]_6$} & \multicolumn{1}{c|}{$2.40\times10^6$} & \multicolumn{1}{c|}{$[\sigma_0^{\text{Mises}}]_{16}$} & \multicolumn{1}{c|}{$2.72\times10^6$} &  & \multicolumn{1}{c|}{$[\eta_0^{\text{Mises}}]_6$} & \multicolumn{1}{c|}{$0.6587$} & \multicolumn{1}{c|}{$[\eta_0^{\text{Mises}}]_{16}$} & \multicolumn{1}{c|}{$0.8461$} \\ \cline{1-4} \cline{6-9} 
                
                \multicolumn{1}{|c|}{$[\sigma_0^{\text{Mises}}]_7$} & \multicolumn{1}{c|}{$2.43\times10^6$} & \multicolumn{1}{c|}{$[\sigma_0^{\text{Mises}}]_{17}$} & \multicolumn{1}{c|}{$2.75\times10^6$} &  & \multicolumn{1}{c|}{$[\eta_0^{\text{Mises}}]_7$} & \multicolumn{1}{c|}{$0.6774$} & \multicolumn{1}{c|}{$[\eta_0^{\text{Mises}}]_{17}$} & \multicolumn{1}{c|}{$0.8648$} \\ \cline{1-4} \cline{6-9} 
                
                \multicolumn{1}{|c|}{$[\sigma_0^{\text{Mises}}]_8$} & \multicolumn{1}{c|}{$2.47\times10^6$} & \multicolumn{1}{c|}{$[\sigma_0^{\text{Mises}}]_{18}$} & \multicolumn{1}{c|}{$2.78\times10^6$} &  & \multicolumn{1}{c|}{$[\eta_0^{\text{Mises}}]_8$} & \multicolumn{1}{c|}{$0.6961$} & \multicolumn{1}{c|}{$[\eta_0^{\text{Mises}}]_{18}$} & \multicolumn{1}{c|}{$0.8836$} \\ \cline{1-4} \cline{6-9} 
                
                \multicolumn{1}{|c|}{$[\sigma_0^{\text{Mises}}]_9$} & \multicolumn{1}{c|}{$2.50\times10^6$} & \multicolumn{1}{c|}{$[\sigma_0^{\text{Mises}}]_{19}$} & \multicolumn{1}{c|}{$2.81\times10^6$} &  & \multicolumn{1}{c|}{$[\eta_0^{\text{Mises}}]_9$} & \multicolumn{1}{c|}{$0.7149$} & \multicolumn{1}{c|}{$[\eta_0^{\text{Mises}}]_{19}$} & \multicolumn{1}{c|}{$0.9023$} \\ \cline{1-4} \cline{6-9} 
                
                \multicolumn{1}{|c|}{$[\sigma_0^{\text{Mises}}]_{10}$} & \multicolumn{1}{c|}{$2.53\times10^6$} & \multicolumn{1}{c|}{$[\sigma_0^{\text{Mises}}]_{20}$} & \multicolumn{1}{c|}{$2.84\times10^6$} &  & \multicolumn{1}{c|}{$[\eta_0^{\text{Mises}}]_{10}$} & \multicolumn{1}{c|}{$0.7336$} & \multicolumn{1}{c|}{$[\eta_0^{\text{Mises}}]_{20}$} & \multicolumn{1}{c|}{$0.9211$} \\ \cline{1-4} \cline{6-9} 
                \end{tabular}
        \end{table}
        \begin{table}[h!]
                \centering
                \begin{tabular}{|llllllll|}
                \hline
                \multicolumn{8}{|c|}{\textbf{Non-Affine Displacement}} \\ \hline
                \multicolumn{1}{|c|}{$[\mathscr{D}^2_0]_1$}    & \multicolumn{1}{c|}{$632$} & \multicolumn{1}{c|}{$[\mathscr{D}^2_0]_{6}$}  & \multicolumn{1}{c|}{$791$} & \multicolumn{1}{c|}{$[\mathscr{D}^2_0]_{11}$} & \multicolumn{1}{c|}{$949$} & \multicolumn{1}{c|}{$[\mathscr{D}^2_0]_{16}$} & \multicolumn{1}{c|}{$1107$}   \\ \hline
                \multicolumn{1}{|c|}{$[\mathscr{D}^2_0]_2$}    & \multicolumn{1}{l|}{$664$} & \multicolumn{1}{c|}{$[\mathscr{D}^2_0]_{7}$}  & \multicolumn{1}{c|}{$822$} & \multicolumn{1}{c|}{$[\mathscr{D}^2_0]_{12}$} & \multicolumn{1}{c|}{$980$} & \multicolumn{1}{c|}{$[\mathscr{D}^2_0]_{17}$} & \multicolumn{1}{c|}{$1139$}   \\ \hline
                \multicolumn{1}{|c|}{$[\mathscr{D}^2_0]_3$}    & \multicolumn{1}{l|}{$696$} & \multicolumn{1}{c|}{$[\mathscr{D}^2_0]_{8}$}  & \multicolumn{1}{c|}{$854$} & \multicolumn{1}{c|}{$[\mathscr{D}^2_0]_{13}$} & \multicolumn{1}{c|}{$1012$} & \multicolumn{1}{c|}{$[\mathscr{D}^2_0]_{18}$} & \multicolumn{1}{c|}{$1170$}   \\ \hline
                \multicolumn{1}{|c|}{$[\mathscr{D}^2_0]_4$}    & \multicolumn{1}{l|}{$727$} & \multicolumn{1}{c|}{$[\mathscr{D}^2_0]_{9}$}  & \multicolumn{1}{c|}{$885$} & \multicolumn{1}{c|}{$[\mathscr{D}^2_0]_{14}$} & \multicolumn{1}{c|}{$1044$} & \multicolumn{1}{c|}{$[\mathscr{D}^2_0]_{19}$} & \multicolumn{1}{c|}{$1202$}   \\ \hline
                \multicolumn{1}{|c|}{$[\mathscr{D}^2_0]_5$}    & \multicolumn{1}{l|}{$759$} & \multicolumn{1}{c|}{$[\mathscr{D}^2_0]_{10}$} & \multicolumn{1}{c|}{$917$} & \multicolumn{1}{c|}{$[\mathscr{D}^2_0]_{15}$} & \multicolumn{1}{c|}{$1075$} & \multicolumn{1}{c|}{$[\mathscr{D}^2_0]_{20}$} & \multicolumn{1}{c|}{$1234$}   \\ \hline
                \end{tabular}
                \caption{Thresholds values that map an atom to a vertex for the descriptors 
                $\alpha=\{\sigma^{\text{Mises}},\eta^{\text{Mises}},\mathscr{D}^2\}$. 20 thresholds 
                for each descriptor are chosen and used for the construction of the networks.}
                \label{table:thresholds_values}
        \end{table}

        \vs{0.2cm}
        \noindent The previous study consisted in building the CN using each of the thresholds of the 
        Table (\ref{table:thresholds_values}) and the set of cutoff radii $R_0$. The parameters set 
        $(\alpha,[\alpha_0]_{i},R_0)$, where 
        $[\alpha_0]_{i}=\Big\{[\sigma_0^{\text{Mises}}],[\eta_0^{\text{Mises}}],[\mathscr{D}^2_0]\Big\}$ 
        and $i=\{1,2,3,...,19,20\}$, define the construction of a network based on the $\alpha$ descriptor. 
        For each strain value $\gamma$, the atomic configuration of the system is mapped to a graph, 
        where we have monitored the number of vertices and edges based on their growth. This study has 
        allowed us to review the density of vertices and edges that our graphs reach as a function of 
        the parameters with the purpose of selecting the most suitable parameters for the analysis of 
        topological metrics.

        \vs{0.2cm}
        \noindent The first metric we study is the average degree of the network as a function of strain. 
        The parameters we have selected to do the calculations are: For the \emph{Shear Stress}, all 
        $[\sigma_0^{\text{Mises}}]_i$ with $i=\{1,2,3,4,5,6,7,8,9,10\}$, for the \emph{Shear Strain}, all 
        $[\eta_0^{\text{Mises}}]_j$ with $j=\{1 ,2,3,4,5,6,7,8,9,10\}$ and for the \emph{Non-Affine Displacement}, 
        all $[\mathscr{D}^2_0]_k$ with $k =\{1,2,3,4,5,6,7,8,9,10\}$.

        \begin{figure}[ht!]
                \centering
                \includegraphics[scale=0.28]{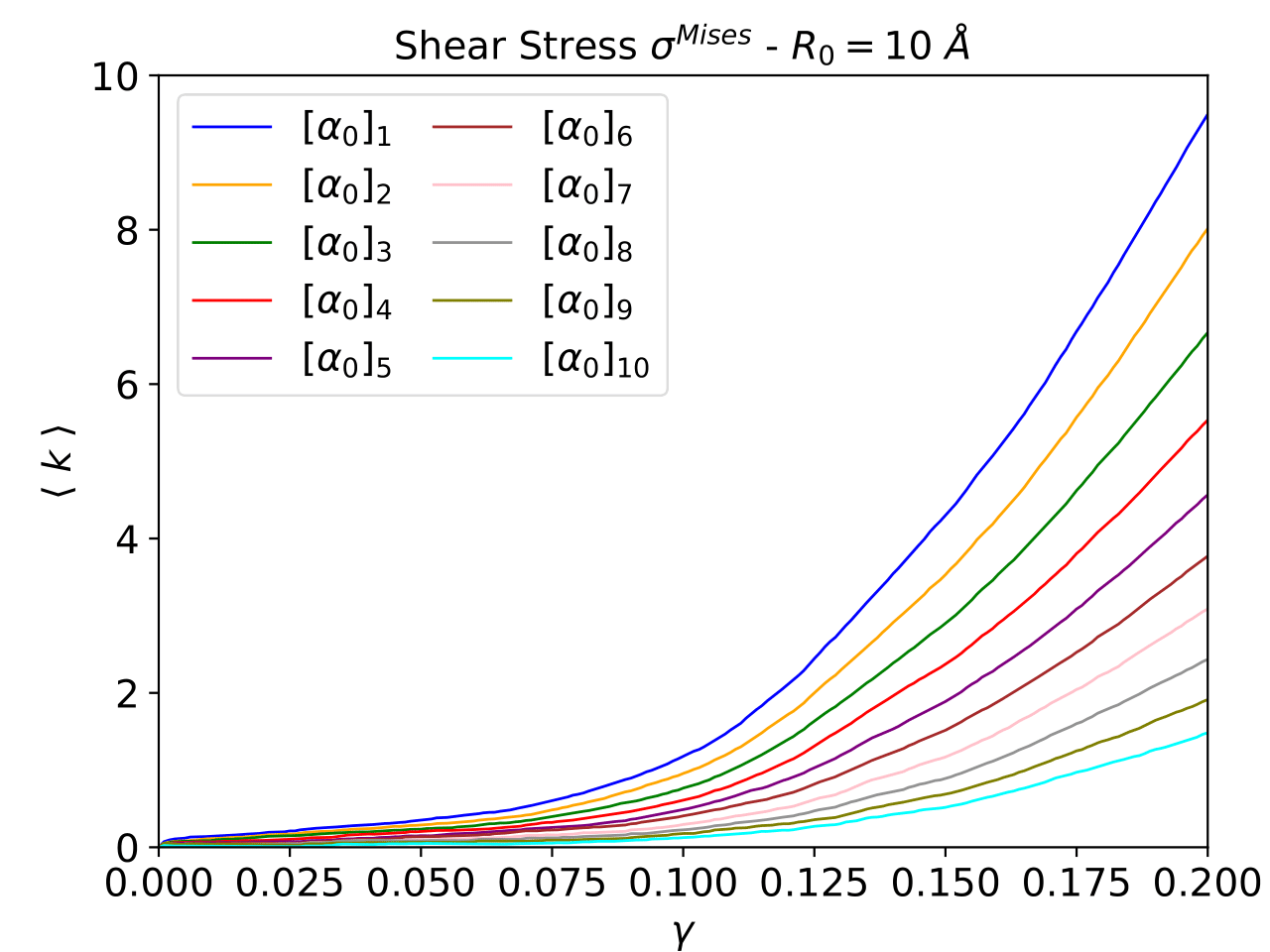}
                \includegraphics[scale=0.28]{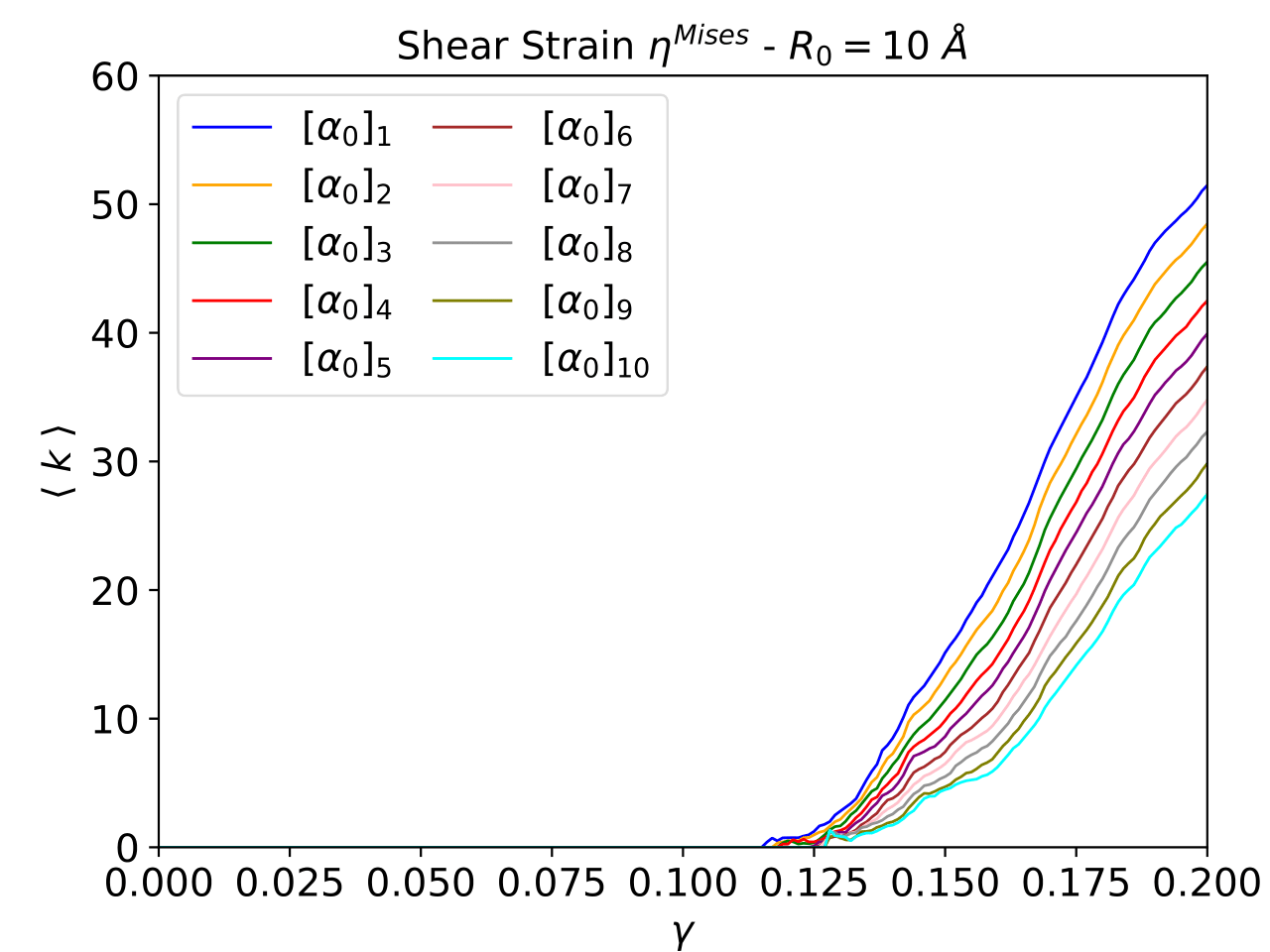}
                \includegraphics[scale=0.28]{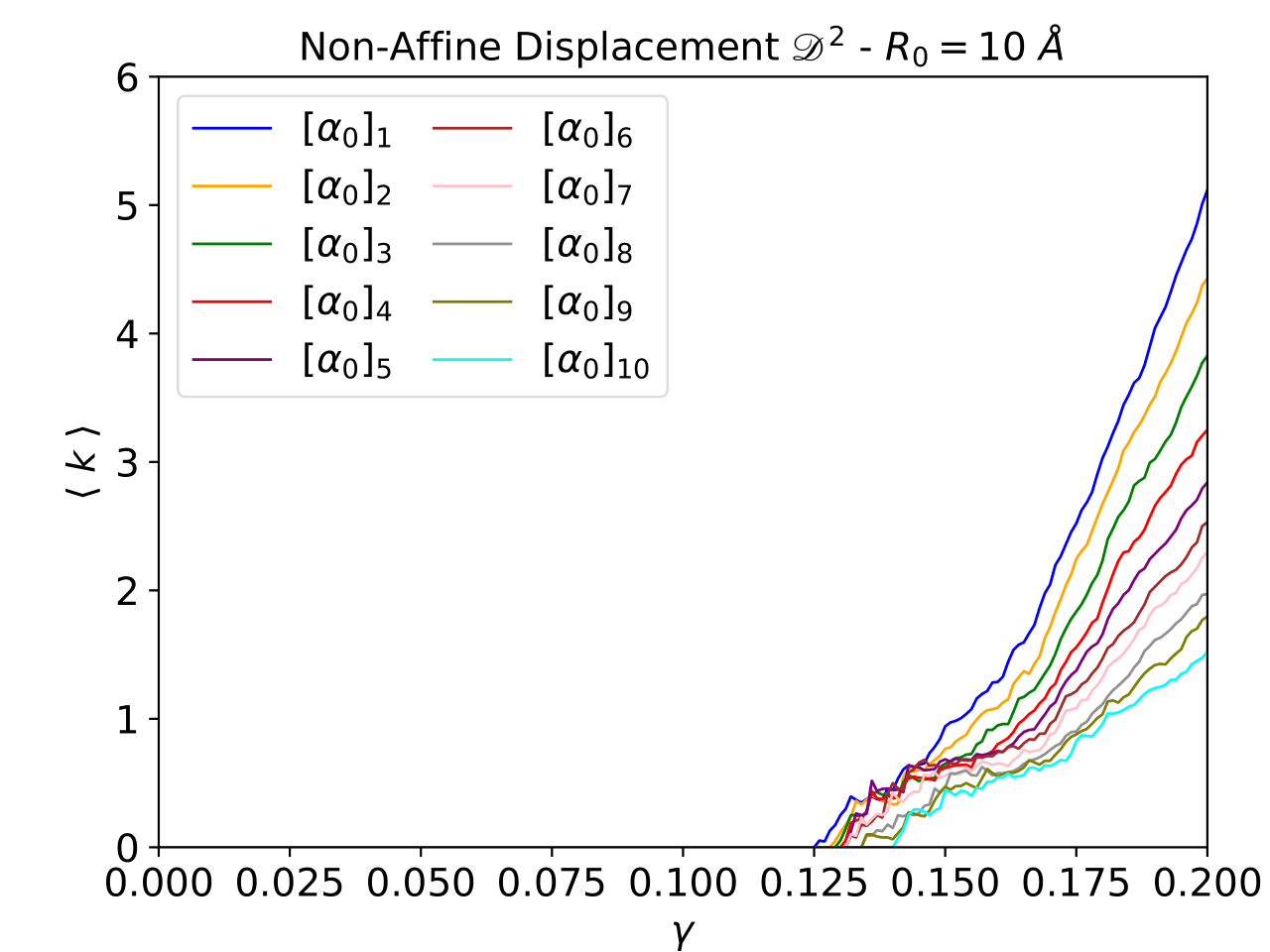}\\
                \includegraphics[scale=0.28]{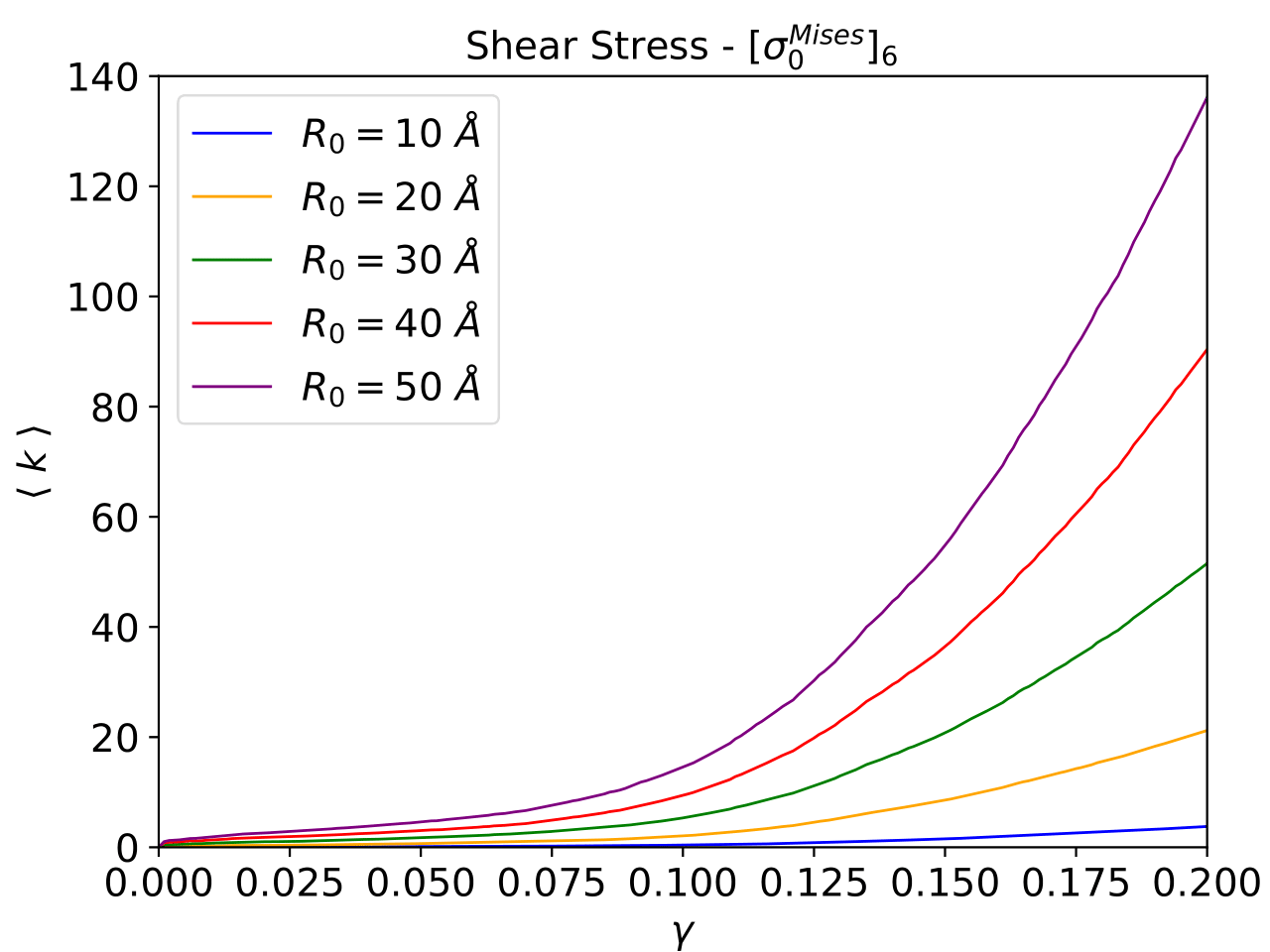}
                \includegraphics[scale=0.28]{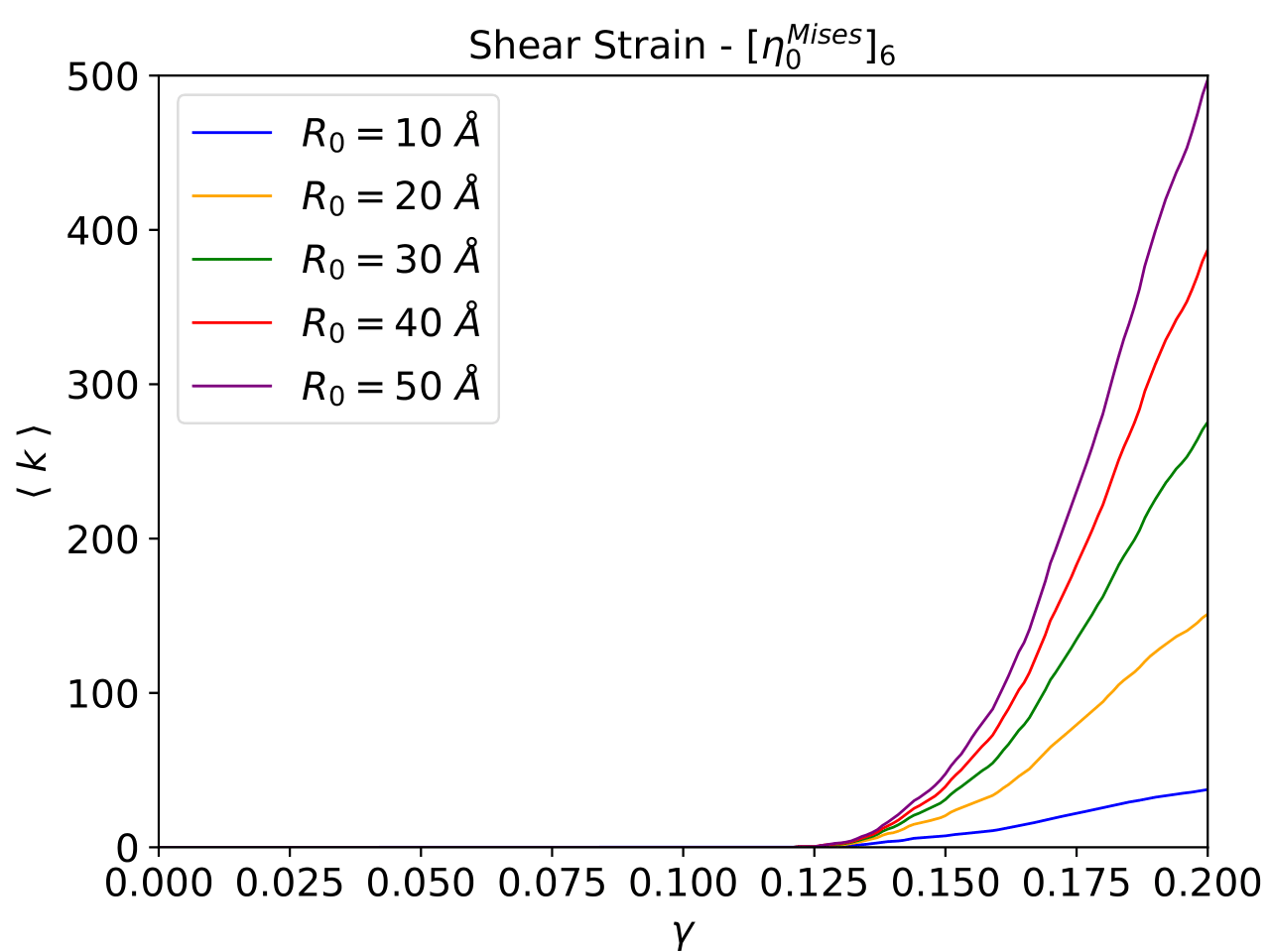}
                \includegraphics[scale=0.28]{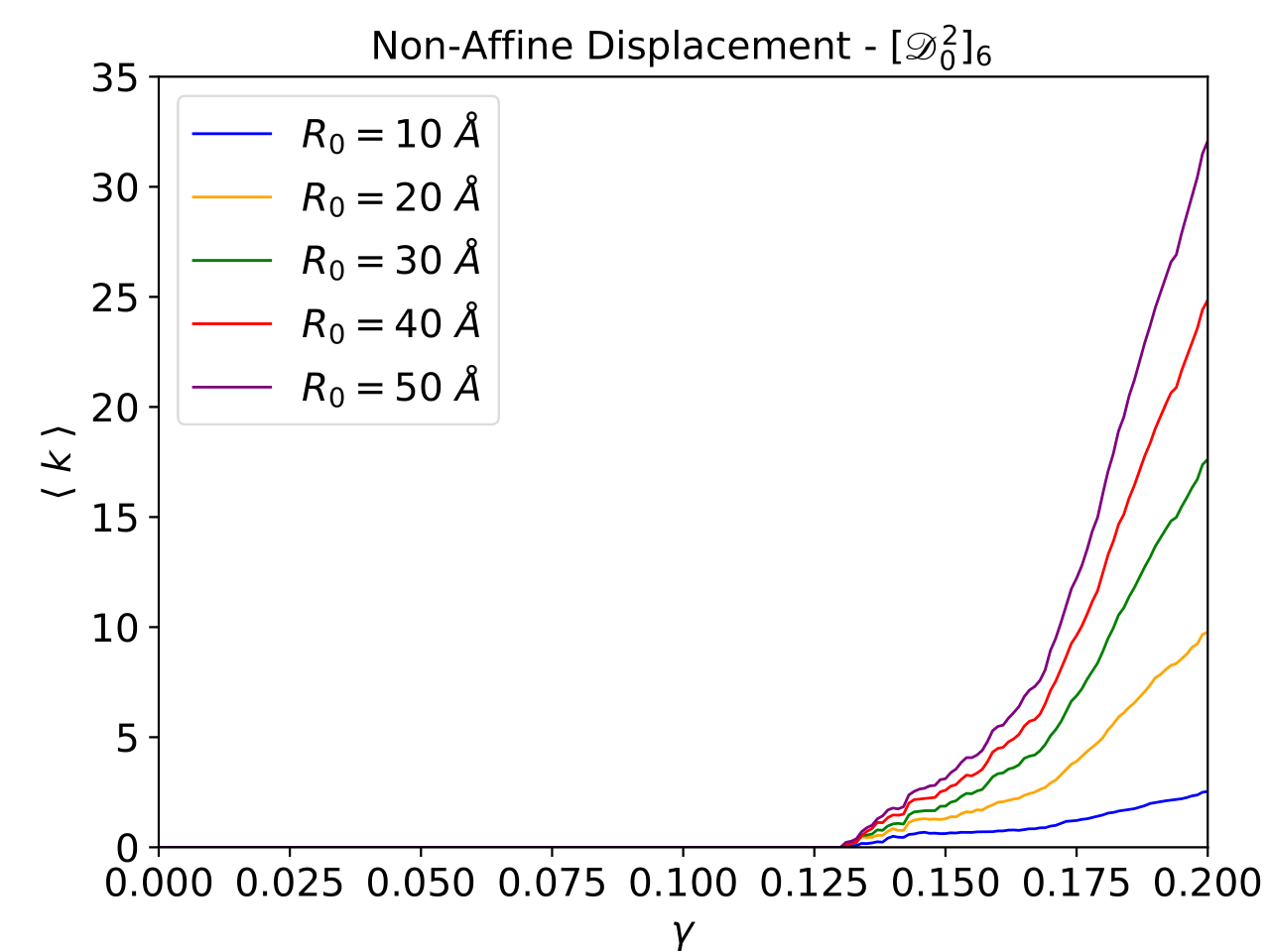}
                \caption{Average degree $\langle k \rangle$ as a function of strain for different physical 
                descriptors, thresholds, and cutoff radii. Metric computed for CNs built based on the descriptors 
                $\alpha=\{\sigma^{\text{Mises}},\eta^{\text{Mises}},\mathscr{D}^2\}$. Plots in the first row show a 
                comparison of the $\langle k \rangle$ calculated using different thresholds at a fixed cutoff 
                radius $R_0=10$ \AA. For all cases, to a higher threshold value, lower is the network average degree, 
                since the population of vertices and edges decrease. Plots in the second row show a comparison of 
                the $\langle k \rangle$ calculated using different cutoff radii with a fixed threshold (depending 
                on the descriptor). For all cases, using a larger cutoff radius increases the degreee since 
                connections increase.}
                \label{fig:degree_descriptors}
        \end{figure}

        \vs{0.2cm}
        \noindent Numerically, the average degree has been the easiest metric to calculate, since we only 
        need to count the number of vertices and edges of the networks constructed for each strain value 
        $\gamma$. As seen in Figure (\ref{fig:degree_descriptors}), this metric reports different results 
        depending on the descriptor used. For example, for networks with $\alpha=\sigma^{\text{Mises}}$, 
        the degree exhibits a monotonically increasing behavior for all deformation regimen, increasing its 
        slope rate when plastic events give rise to the SB. This microscopic phenomenon of energy dissipation 
        due to irreversible deformations is captured by the degree, a simple and purely topological element, 
        which regardless of the structure of the network, tells us about a macroscopic response suffered by 
        the system. For this type of network, the increase of average degree can be interpreted as an increase 
        in the internal energy of the system due to atomic displacements. For elastic deformations, the atoms 
        oscillate around their equilibrium point (low degree), while for plastic deformations, the atoms 
        become progressively disordered, and some migrate due to the failure caused by the STZs, increasing 
        internal energy (increasing degree).

        \vs{0.2cm}
        \noindent Similar to networks with $\alpha=\eta^{\text{Mises}}$ and $\alpha=\mathscr{D}^2$, the 
        degree increases but only when the system has developed the SBs and the creep zone. This has happened 
        because the thresholds used have not been low enough to capture atoms in another deformation level. 
        In other words, for the elasto-plastic regime, the atoms do not show a \emph{Shear Strain} or a 
        \emph{Non-Affine Displacement} comparable to the considered thresholds. Basically, the network built 
        for these descriptors is formed by the atoms that live in the SBs. Therefore, the degree for these 
        cases quantifies the increase of irreversible deformations as shown in the maps of Figure 
        (\ref{fig:spatial_distribution_descriptors}).

        \vs{0.2cm}
        \noindent In Figure (\ref{fig:degree_descriptors}) we also show an example of how the degree changes 
        depending on the different cutoff radii for a fixed threshold. We see that regardless of the radius, 
        the degree always tends to increase, but with different slope rates, which tells us that for this 
        methodology, physical phenomena in the order between 10-50 \AA, do not show a difference beyond values 
        that the degree takes. However, we have observed that for smaller radii and thresholds even larger 
        than $[\sigma^{\text{Mises}}_0]_{16}$ (only for $\alpha=\sigma^{\text{ Mises}}$), the degree is no 
        longer monotonically increasing, but fluctuates.

        \begin{figure}[ht!]
                \centering
                \includegraphics[scale=0.43]{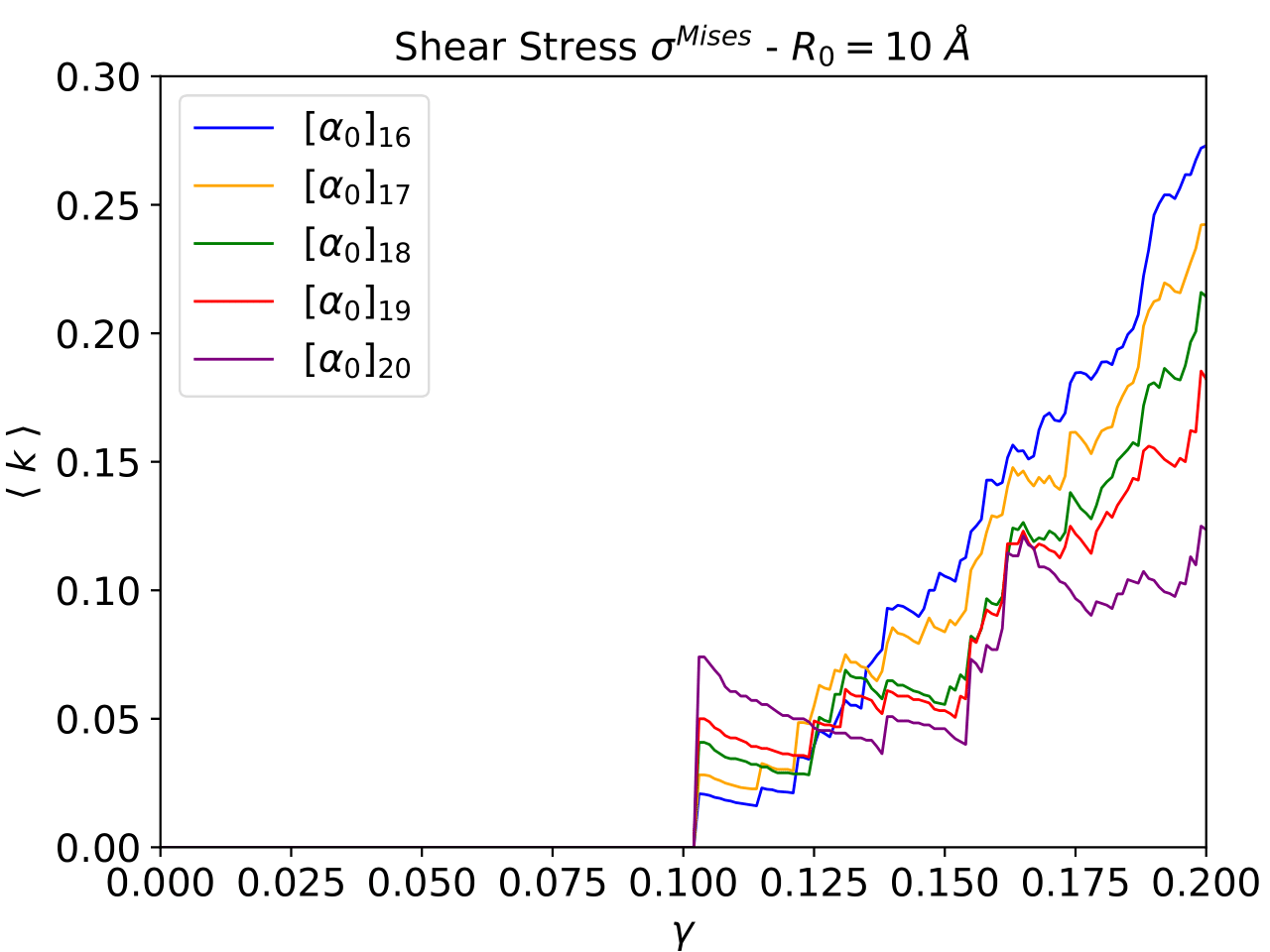}
                \includegraphics[scale=0.43]{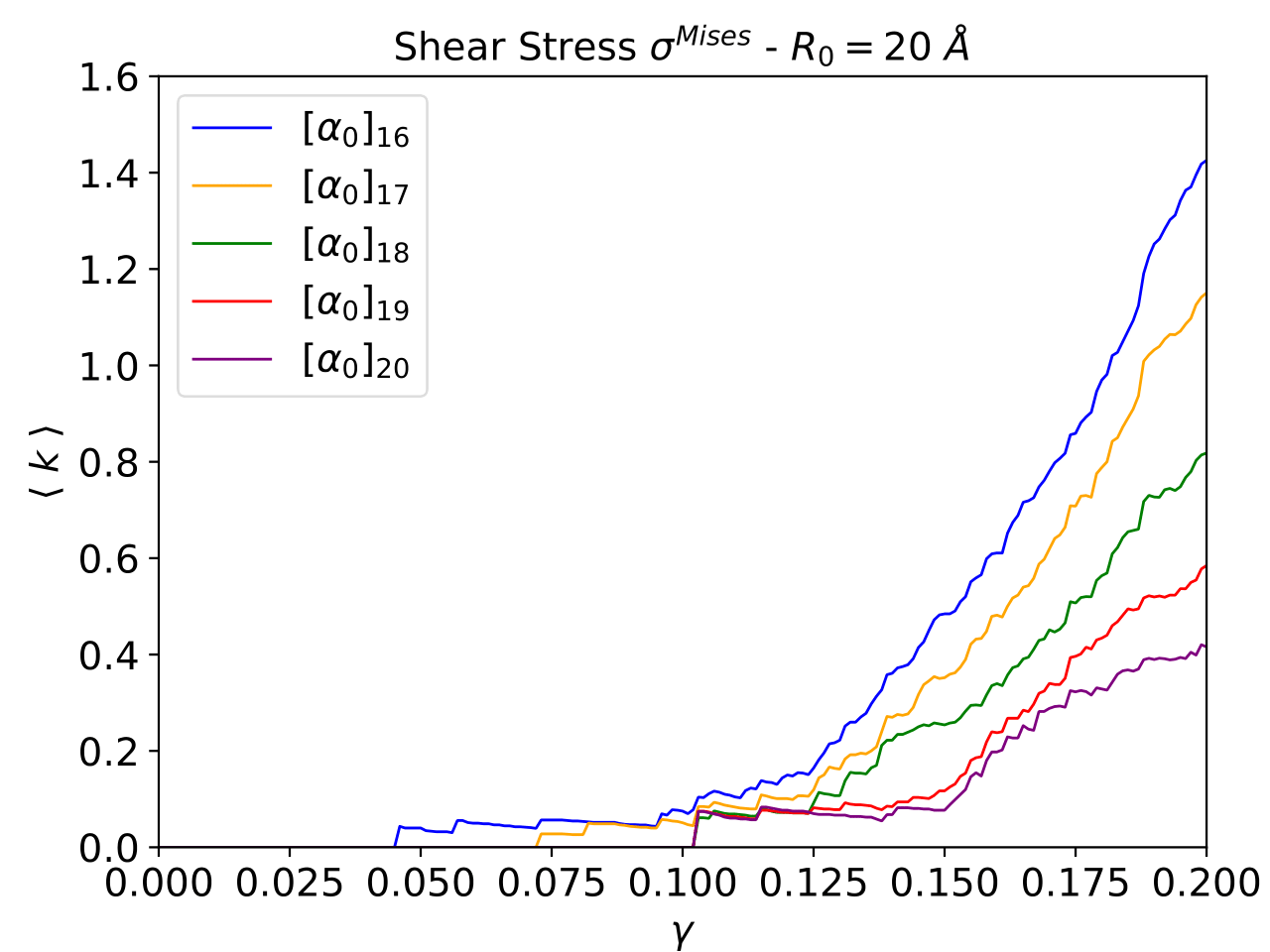}
                \caption{Average degree $\langle k \rangle$ as a function of strain for a network with 
                $\alpha=\sigma^{\text{Mises}}$ for different thresholds and cutoff radii $R_0=10\text{ \AA}$ 
                and $R_0=20\text{ \AA}$.}
                \label{fig:degree_shearstress}
        \end{figure}

        \vs{0.2cm}
        \noindent The average degree for a network with $\sigma^{\text{Mises}}$ in Figure 
        (\ref{fig:degree_descriptors}), increases as strain increases regardless of the threshold used. These 
        curves are smooth with a certain slope rate that could be calculated with a numerical derivative. 
        However, if we build these networks with even higher thresholds, we obtain an interesting result to 
        analyze. In Figure (\ref{fig:degree_shearstress}) the average degree calculated with a $R_0=10\text{ \AA}$ 
        shows less smoothness than the previous results and a variable slope rate, but this it is only a 
        consequence of the physics that the metric is representing. With these parameters, the network begins to 
        grow once the SB has been located ($\gamma\sim0.1$), revealing that there are highly stressed atoms 
        involved in the construction of the graph. According to our previous interpretation, the average degree is 
        like an internal energy of the system, being in this case, the energy that these highly stressed atoms 
        contribute, but the interesting result is when that energy is released via decrease of the degree. For the 
        different thresholds that have been used in Figure (\ref{fig:degree_shearstress}), they all exhibit degree drops, which 
        we physically interpret as a release of internal energy, causing an increase of temperature. This result 
        proves the existence of nanofractures/avalanches precisely due to those highly stressed atoms.

        \vs{0.2cm}
        \noindent The results obtained for the case $R_0=20\text{ \AA}$ still show some degree drops, but they are 
        less and less noticeable. This is already a sign that the network begins to hide some details that occur 
        locally and only show the progressive increase in degree tending to smoothness and a constant slope rate. 
        The average degree for networks with $\alpha=\eta^{\text{Mises}}$ and $\alpha=\mathscr{D}^2$ constructed 
        for higher thresholds has also been revised, but does not reveal interesting results. given the lack of 
        vertices. Physically, there are not enough highly deformed atoms to build a network.

        \vs{0.2cm}
        \noindent The next metric that we study are the clustering coefficient of the network as a function of strain. 
        The parameters we have selected for this case are: For the \emph{Shear Stress}, $[\sigma_0^{\text{Mises}}]_{11}$ , 
        for the \emph{Shear Strain}, $[ \eta_0^{\text{Mises}}]_{15}$ and for the \emph{Non-Affine Displacement}, 
        $[\mathscr{D}^2]_6$.

        \begin{figure}[ht!]
                \centering
                \includegraphics[scale=0.28]{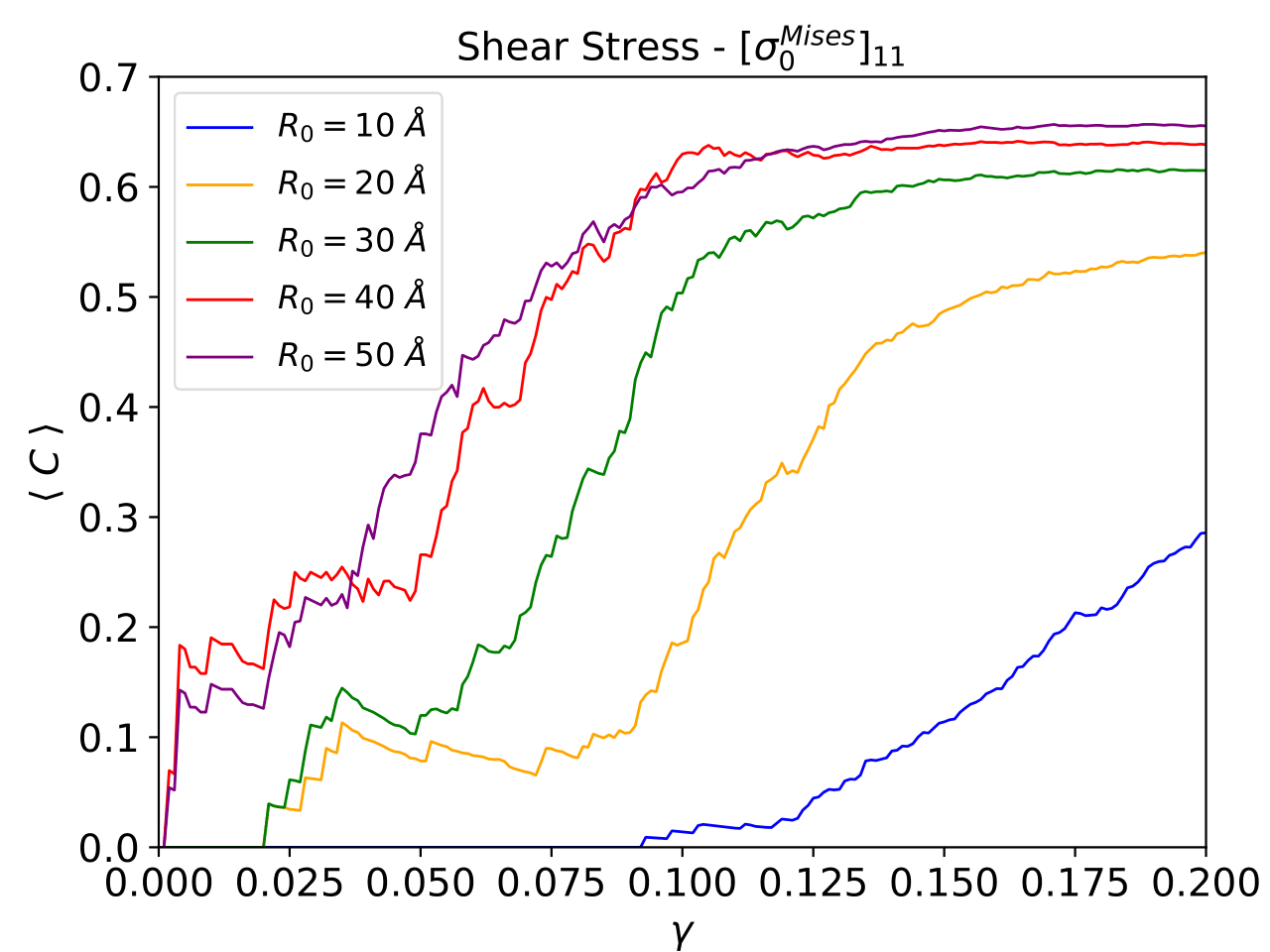}
                \includegraphics[scale=0.28]{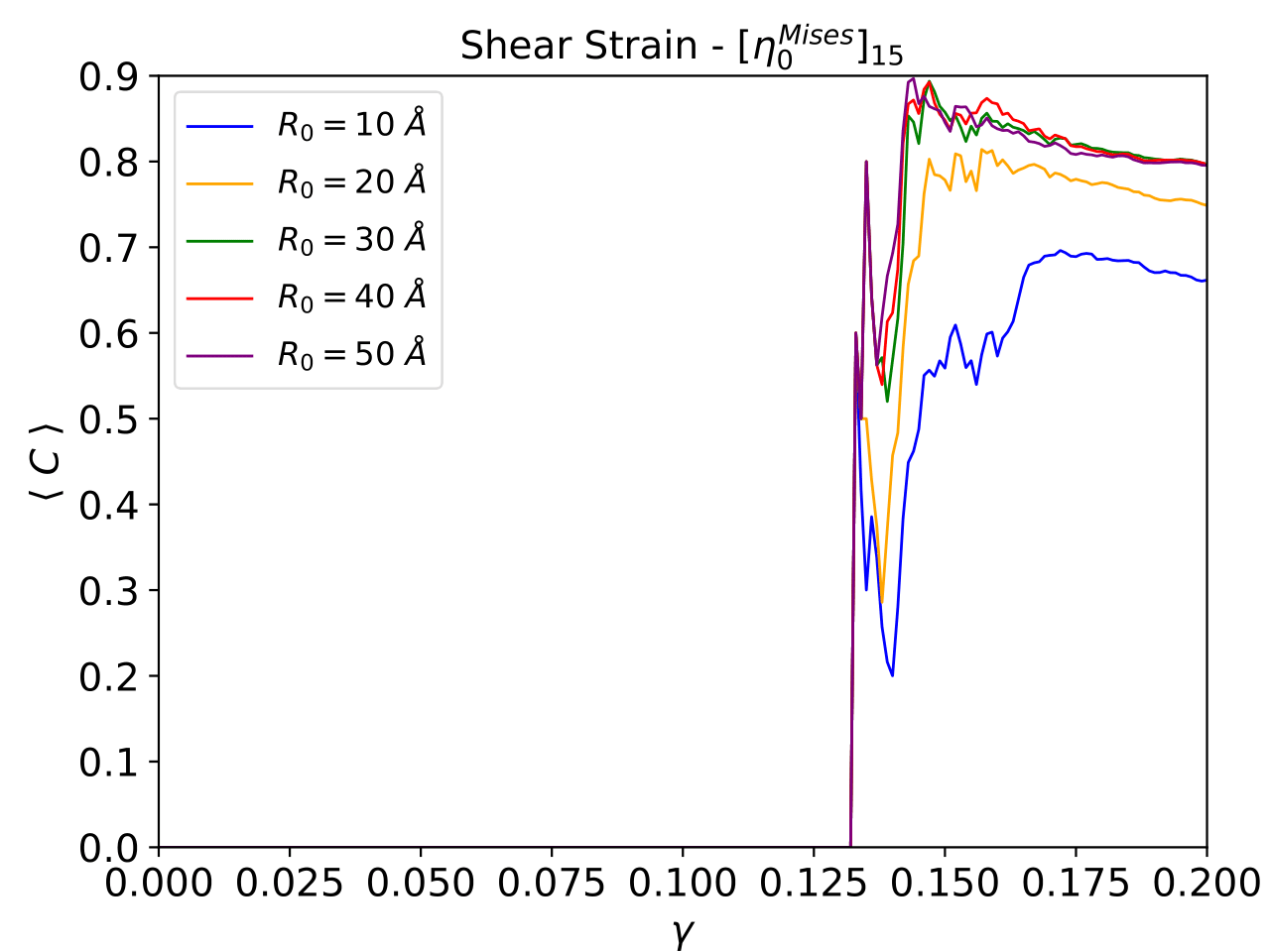}
                \includegraphics[scale=0.28]{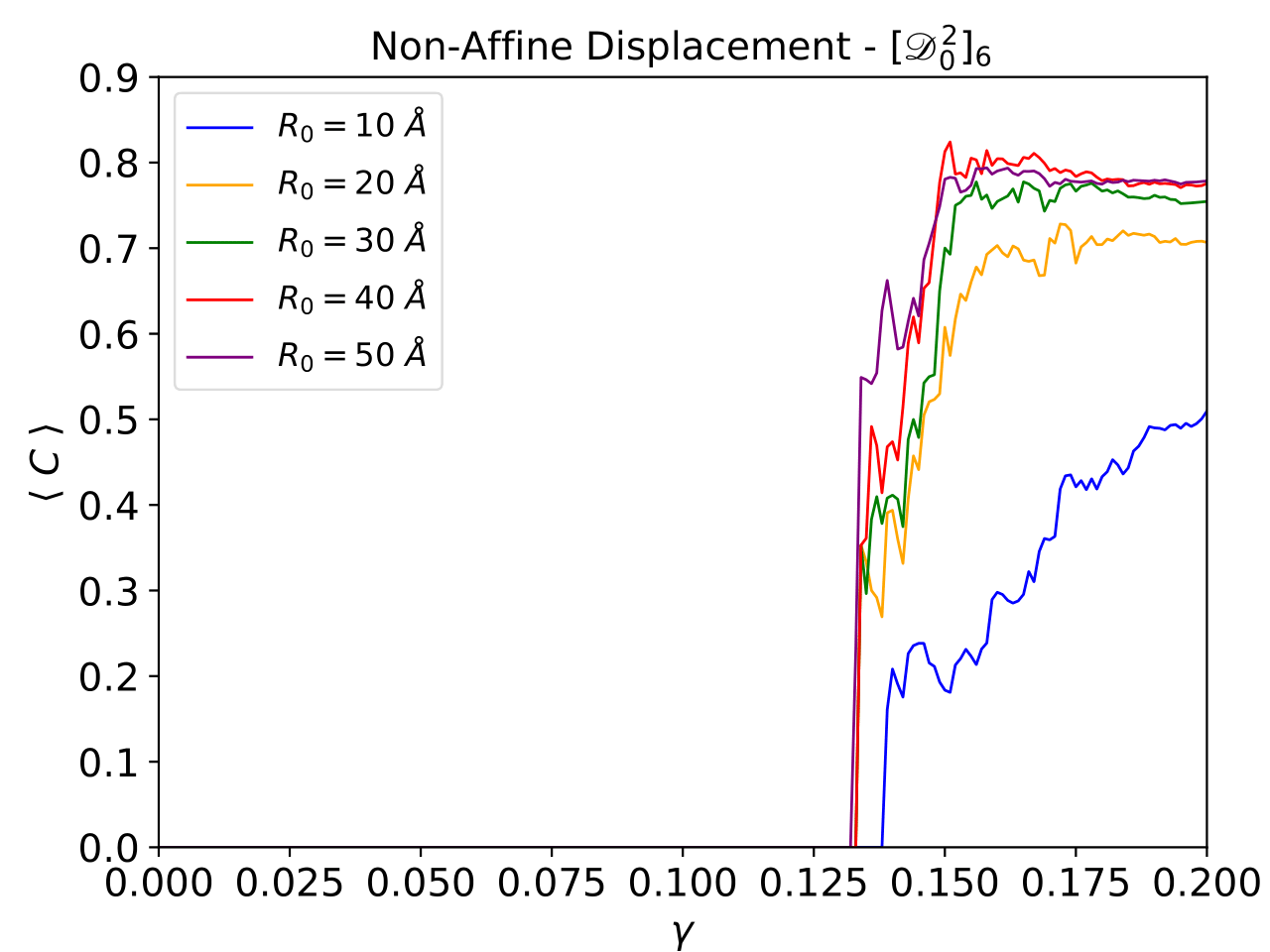}
                \caption{Average clustering coefficient $\langle C \rangle$ as a function of strain for different 
                cutoff radii and physical descriptors. Metric calculated for CNs built based on the descriptors 
                $\alpha=\{\sigma^{\text{Mises}},\eta^{\text{Mises}},\mathscr{D}^2\}$. Comparison of the average 
                clustering coefficients calculated for different cutoff radii with a fixed threshold (depending on 
                the descriptor).}
                \label{fig:clustering_descriptors}
        \end{figure}

        \vs{0.2cm}
        \noindent In this case, we compute the clustering coefficient of the network as the sum of the clustering 
        coefficient of each vertex averaged by the number of vertices for each strain value $\gamma$. This 
        coefficient is a measure of information about the local structure of the network and how the vertices are 
        connected to each other. In particular, it measures the probability that the neighbors of a vertex are also 
        neighbors of each other, so the range of values it takes is between $0$ and $1$. Therefore, the average 
        over the network of this measure provides information about the global structure, and a characterization 
        of how robust is the network against threats such as vertex and/or edge extraction.

        \vs{0.2cm}
        \noindent In Figure (\ref{fig:clustering_descriptors}) we present the results for the average clustering 
        coefficient as a function of strain for the three descriptors. For networks with 
        $\alpha=\sigma^{\text{Mises}}$, clustering of vertices takes place from the start, indicating an increase as 
        strain increases. Although the curves are not smooth, it is possible to identify some important variations 
        depending on the different deformation regimes. For the cutoff radius $R_0=20\text{ \AA}$ we see this 
        metric suffers a considerable increase during the SB location, while for $R_0>20\text{ \AA}$, the 
        clustering saturates to $\langle C \rangle\sim0.6$. This occurs because the clustering coefficient intensifies 
        after the elastic limit in response to the formation of the STZs and SBs.

        \vs{0.2cm}
        \noindent For the other descriptors, the average clustering undergoes a drastic increase since the first 
        vertices and edges appear in the graph. We remember that the atoms that map to these vertices are those 
        that live in the SB, therefore the interactions in time are more robust, which increases the clustering up 
        to a value of $\langle C \rangle\sim0.9 -0.7$. By means of this metric, these networks perfectly 
        represent the SB showing how strong is the interaction between its atoms. The above invites us to think 
        about how robust are these networks and how they will respond to attacks on the vertices or edges. For 
        example, what happens to the clustering coefficient if we remove any vertex or edge? What happens to the 
        connectivity if we attack the network in this way? Physically, removing vertices/atoms from the network/cell 
        will prevent the location of SB and/or fracture of the material?. These are some questions that would be 
        interesting to address.

        \vs{0.2cm}
        \noindent The last metrics we study are the betweenness centrality and the closeness centrality of the network 
        as a function of strain. The parameters that we have selected for this case are: For the \emph{Shear Stress}, 
        $[\sigma_0^{\text{Mises}}]_{8}$, for the \emph{Shear Strain}, $ [\eta_0^{\text{Mises}}]_{20}$ and for the 
        \emph{Non-Affine Displacement}, $[\mathscr{D}^2]_{6}$.

        \begin{figure}[ht!]
                \centering
                \includegraphics[scale=0.28]{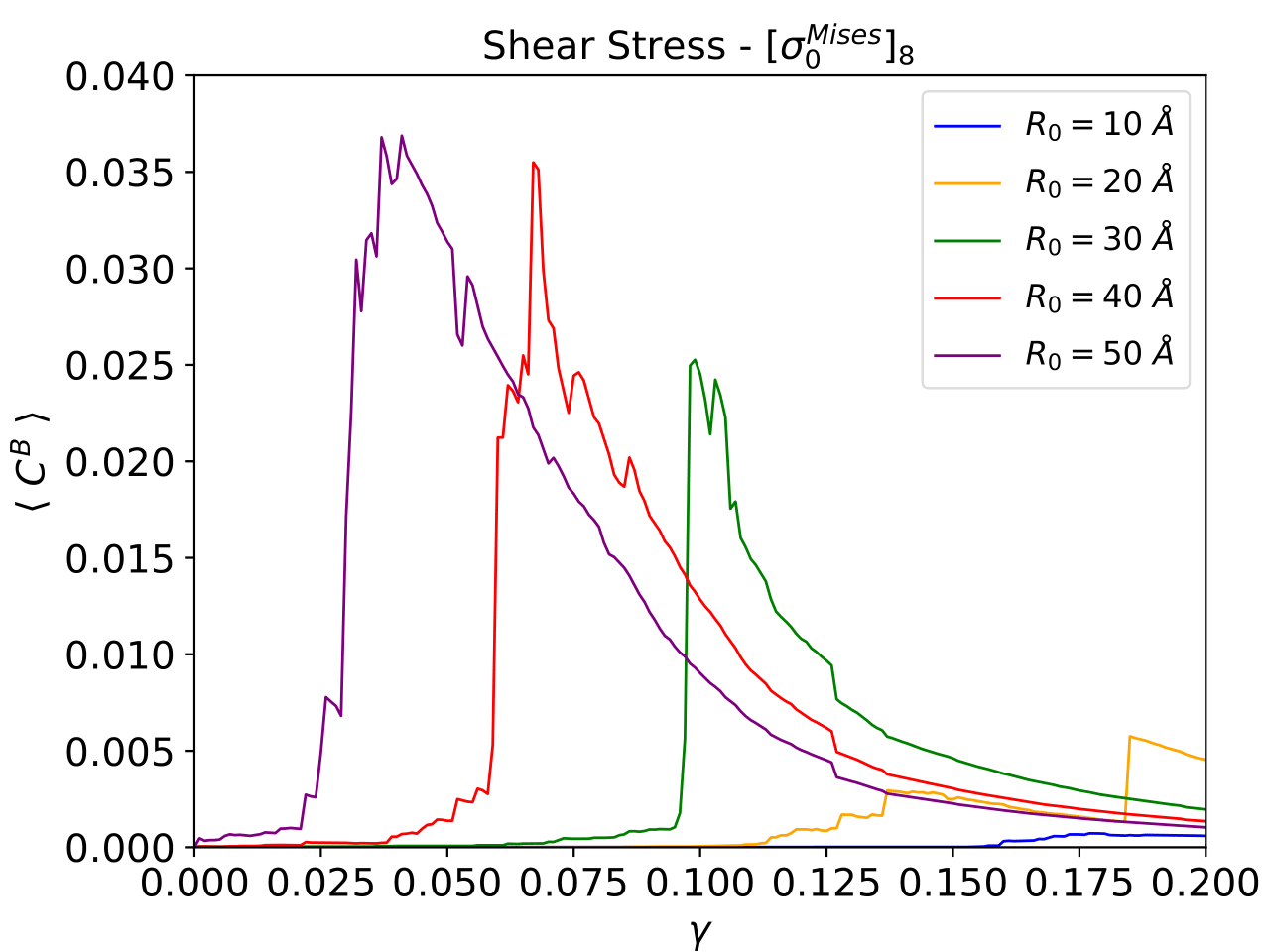}
                \includegraphics[scale=0.28]{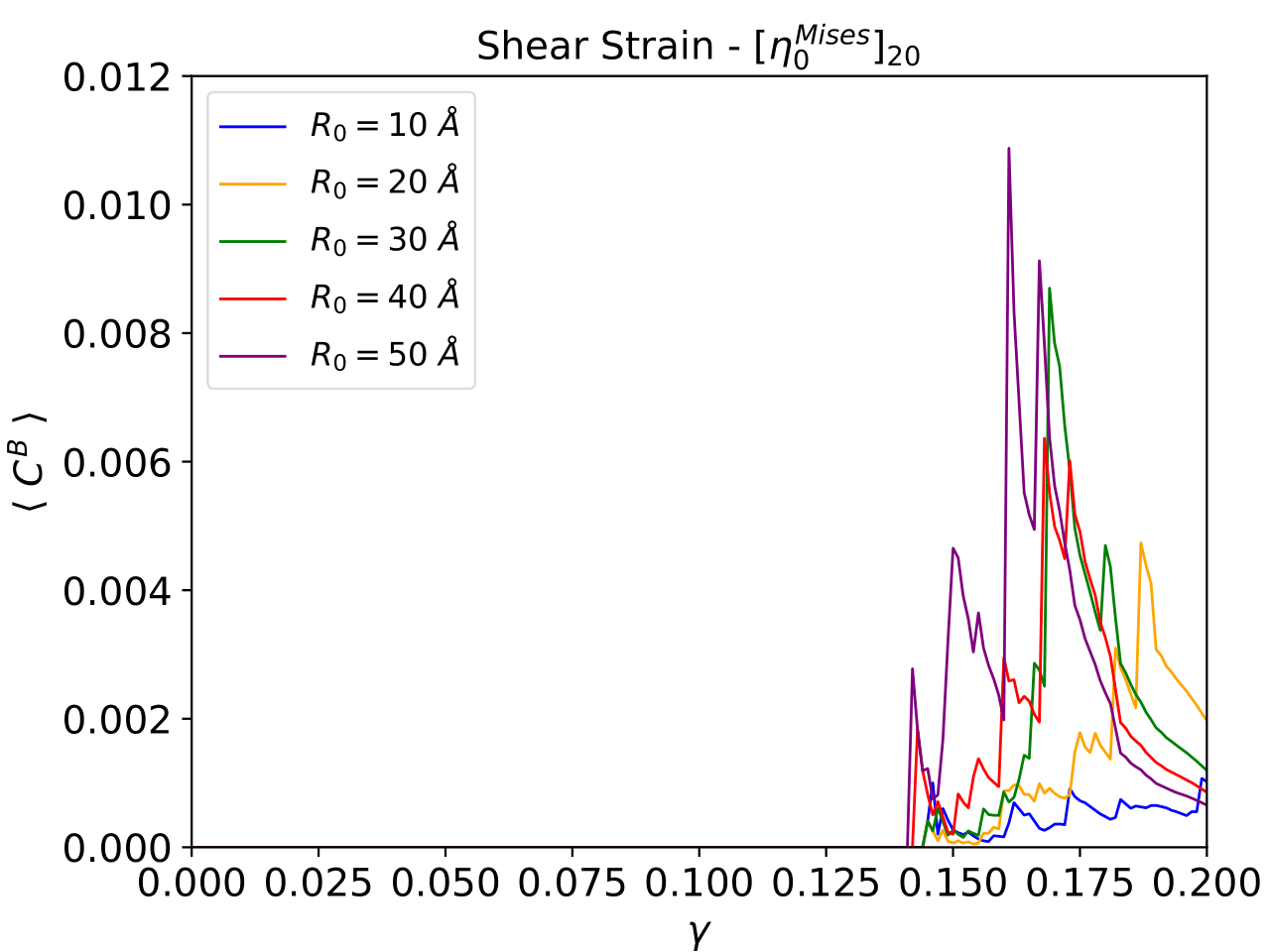}
                \includegraphics[scale=0.28]{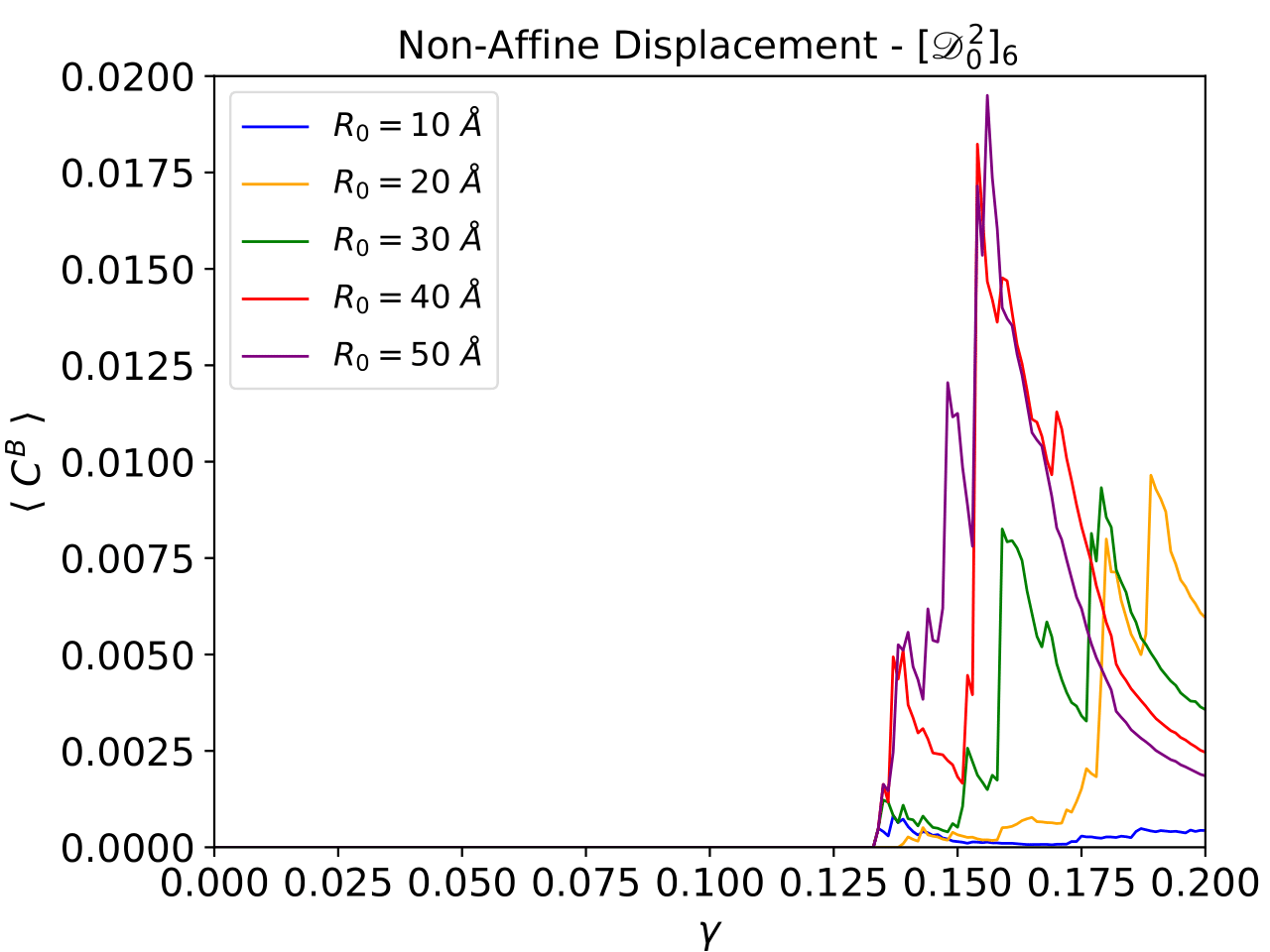}\\
                \includegraphics[scale=0.28]{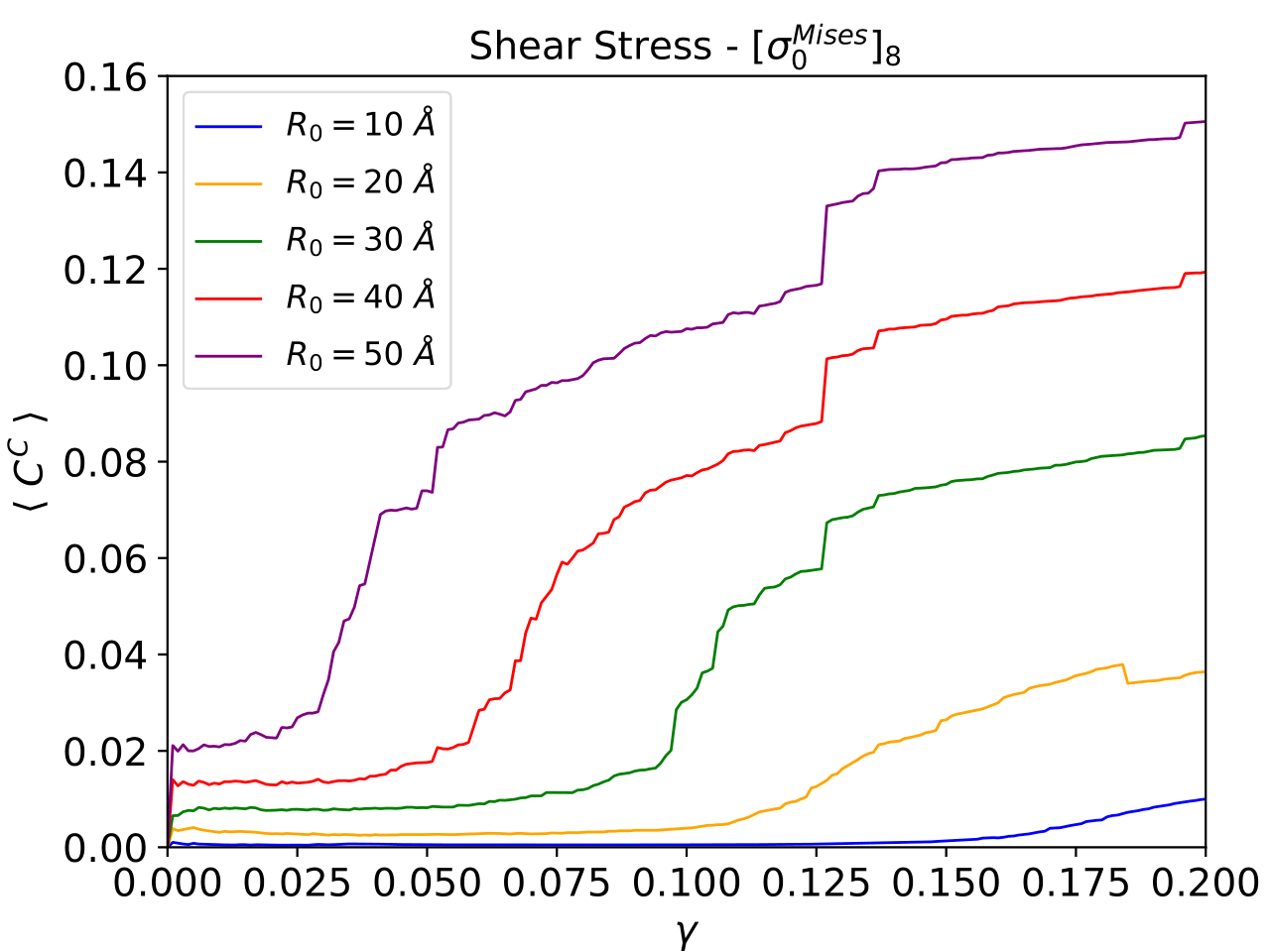}
                \includegraphics[scale=0.28]{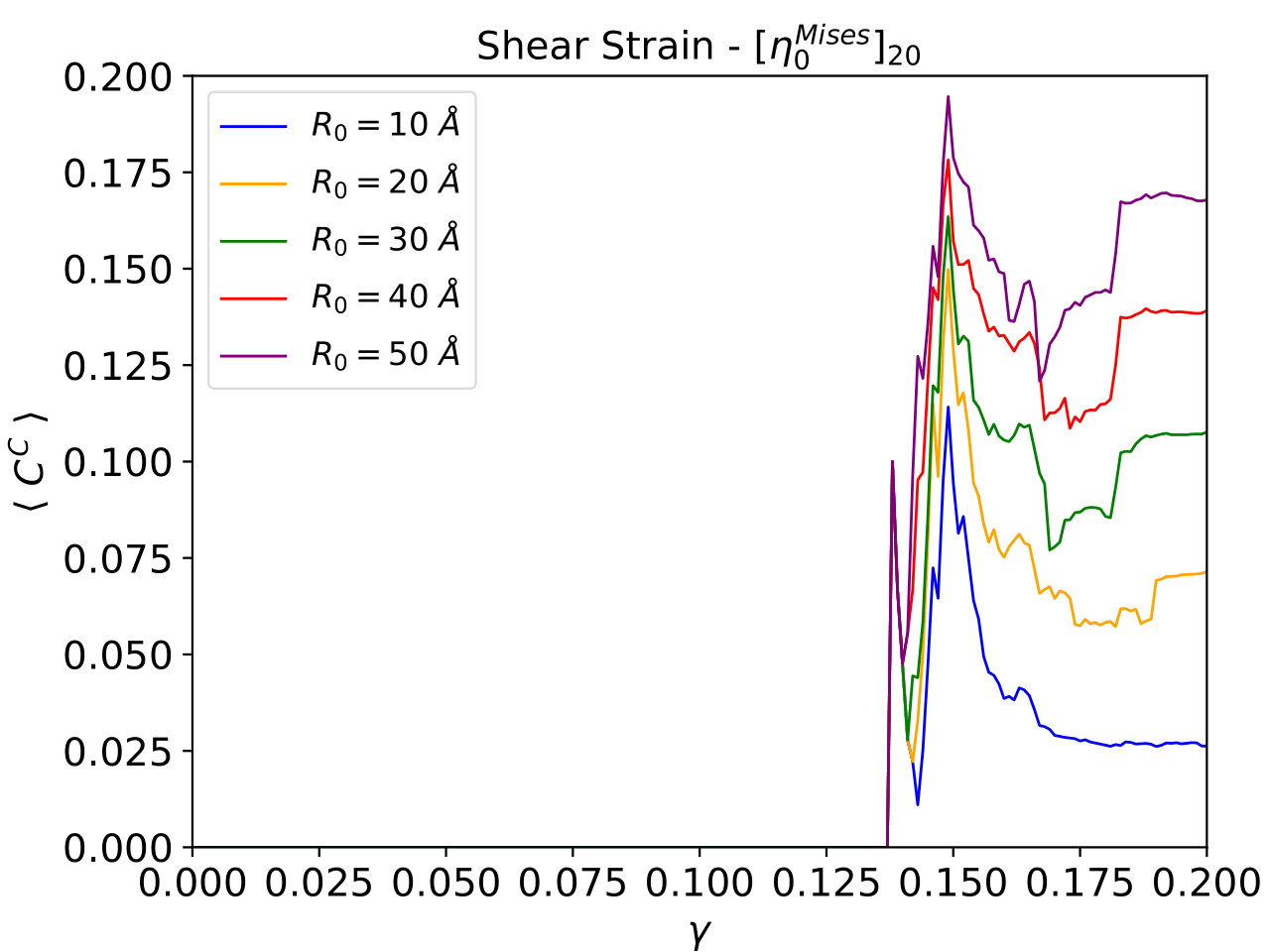}
                \includegraphics[scale=0.28]{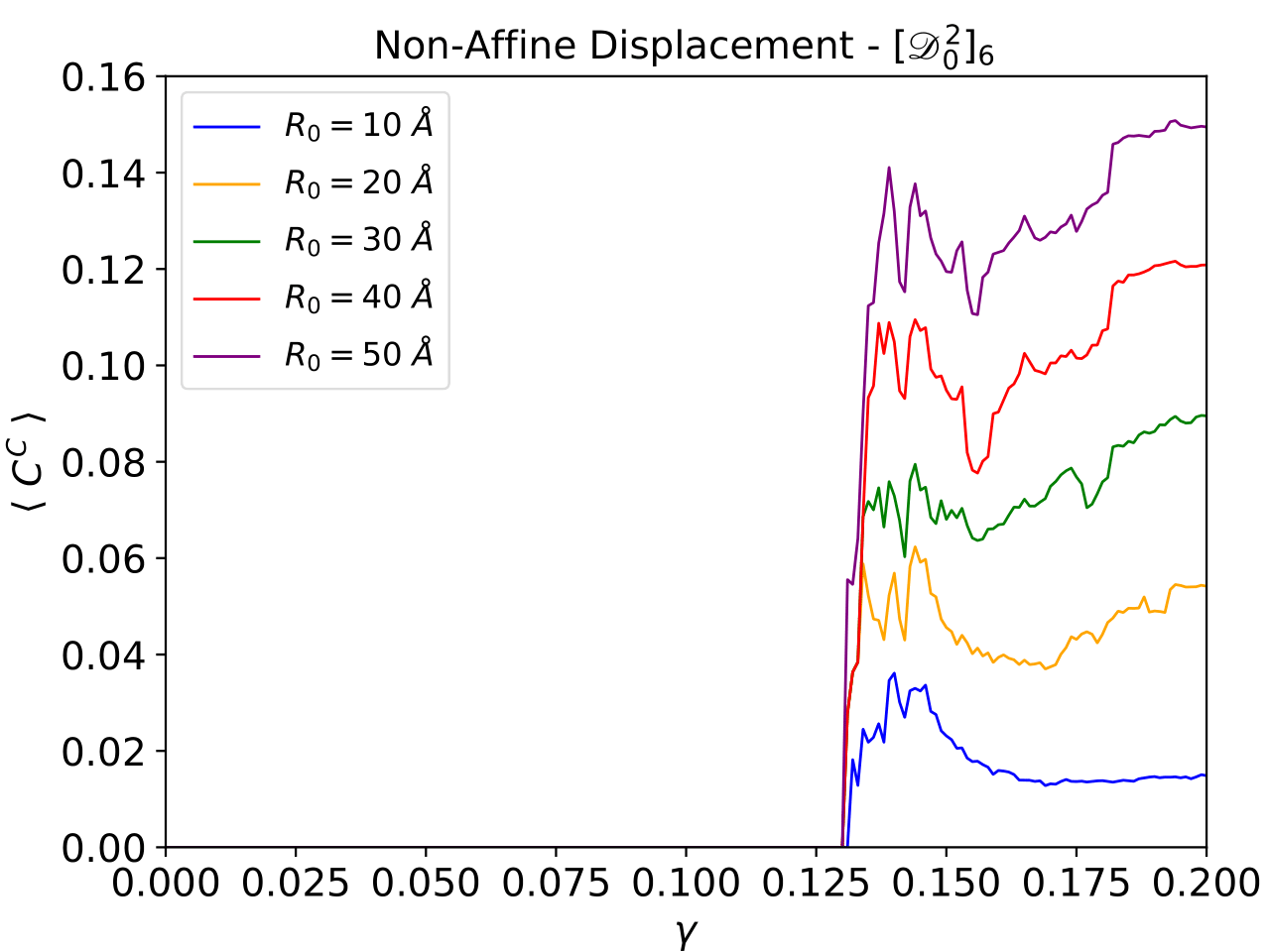}
                \caption{Average betweenness $\langle C^B\rangle$ and closeness $\langle C^C\rangle$ centrality as a function 
                of strain for different cutoff radii and physical descriptors. Metric calculated for CNs built based on the 
                descriptors $\alpha=\{\sigma^{\text{Mises}},\eta^{\text{Mises}},\mathscr{D}^2\}$. Plots in the first row show 
                a comparison of the average betweeness centrality calculated for different cutoff radii at a fixed threshold 
                (depending on the descriptor). Plots in the second row show the same as above but now for the average closeness 
                centrality.}
                \label{fig:centrality_descriptors}
        \end{figure}

        \vs{0.2cm}
        \noindent We have calculated the average betweenness and closeness centrality of the network as the independent 
        sum of the betweenness and closeness of each vertex averaged by the number of vertices for each strain value 
        $\gamma$. We know that these metrics quantify the importance of the vertices, providing a local characterization 
        of the network. Despite this, the average, which indicates a global property, allows us to obtaine information 
        about the vulnerability and robustness of the network, which is what interests us most in this context. In Figure 
        (\ref{fig:centrality_descriptors}) we have a very interesting result for networks with 
        $\alpha=\sigma^{\text{Mises}}$ and cutoff radii $R_=\{30,40,50 \}$ \AA. We see that the average betweennes 
        reaches a maximum and then decreases, while the closeness only increases for the entire strain interval. A valid 
        interpretation for this decrease in the betweenness of the network is that from a moment on, important vertices 
        that fulfill the role of bridges stop appearing, the network becomes denser, but not necessarily with edges that 
        are what give it the quality of importance to the vertices. Even so, this does not imply that the network is more 
        vulnerable. Direct evidence of this is the result that we obtain for the closeness of the network. The increase 
        in this metric is proof that shorter paths appear that give importance to the vertices, making the network more 
        robust. With this result we are in the presence of the fact that there are and exist vertices/atoms that mediate 
        the interaction of other groups of atoms/subnetworks or communities, which is physically the topological 
        representation of the interaction between different STZs and that finally collapse in a SB.

        \vs{0.2cm}
        \noindent In the case of networks with $\alpha=\eta^{\text{Mises}}$ and $\alpha=\mathscr{D}^2$, it is not 
        possible to capture a clear result. Given that the networks begin their growth once the material has fractured, 
        we cannot draw any conclusions about what precedes this phenomenon, therefore the fluctuations of these metrics 
        are only a response to an uncontrolled growth of vertices and edges as a result of the high strains achieved.

        
        \subsection{Complex networks and spatial distribution of topological metrics}

                \noindent Until now we have only presented the compute of topological metrics as a function of strain, 
                analyzing each curve with respect to the physical phenomena involved, but to have a more complete study, 
                we present a visual representation of these CNs once the simulation is finished. For this, we have used 
                the Python NetworkX module \cite{Hagberg2008Exploring} to graph these networks.

                \begin{figure}[ht!]
                        \centering
                        \includegraphics[scale=0.85]{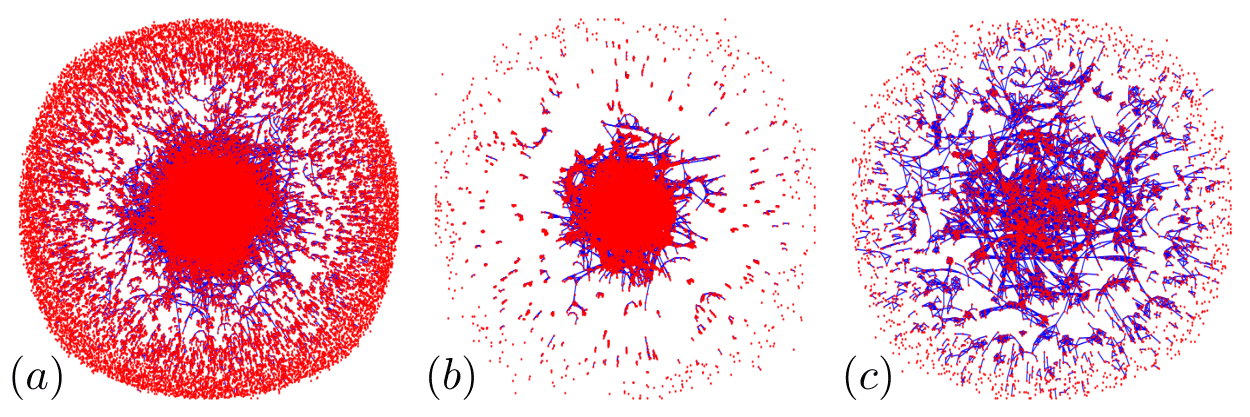}
                        \caption{Undirected CNs resulting at the end of the deformation process. The networks were built 
                        with the parameter set (a) $\{\sigma^{\text{Mises}}, [\sigma^{\text{Mises}}_0]_{6}, 
                        10\text{ \AA }\}$, (b) $\{\eta^{\text{Mises}}, [\eta^{\text{Mises}}_0]_{6}, 10\text{ \AA}\}$ and 
                        (c) $\{\mathscr{D}^2, [\mathscr{D}^2_0]_{6}, 10\text{ \AA}\}$. The red points are the vertices 
                        and the blue lines are the edges.}
                        \label{fig:complex_networks_1}
                \end{figure}

                \vs{0.2cm}
                \noindent 
                In Figure 
                (\ref{fig:complex_networks_1}) we show an example of how looks like each network built for the three $\alpha$ 
                descriptors once the deformation process is completed. Each network acquires a large population of vertices 
                and edges, where we see that the representation algorithm places vertices in the center and the perimeter, 
                depending on the degree of interaction it has with its neighbors. The center of each network is where the 
                highest density of dominant edges and vertices is located. A cluster of vertices and edges is the only thing 
                that highlights for networks with $\alpha=\{\sigma^{\text{Mises}},\eta^{\text{Mises}}\}$. Due to the high number of 
                interactions, it is visually difficult to identify if there are preferential connections or some degree of 
                randomness in the edge distribution, but the most interesting case is the network with $\alpha=\mathscr{D}^2$, 
                since thanks to its lower vertex density (compared to the other two cases), we can identify that there are 
                sub-networks or small communities that have their own local interaction, and even community interaction.

                \vs{0.2cm}
                \noindent The interesting about this is that through a mathematical methodology based on graph theory, it is 
                possible to characterize microscopic and macroscopic properties of a system subject to deformation. Basically, 
                we are describing the physical phenomena involved in the elasto-plastic regime using the abstraction of 
                topological metrics and a representation technique via CNs. We well know that these networks are topological 
                structures and that their properties do not depend on their representation, therefore, we present the same 
                previous graphs but now with the vertices located in relation to their respective atom in the simulation cell. 
                In these cases, we have eliminated the edges since the vertices are colored according to the corresponding 
                topological metrics that we have studied.

                \begin{figure}[ht!]
                        \centering
                        \includegraphics[scale=0.28]{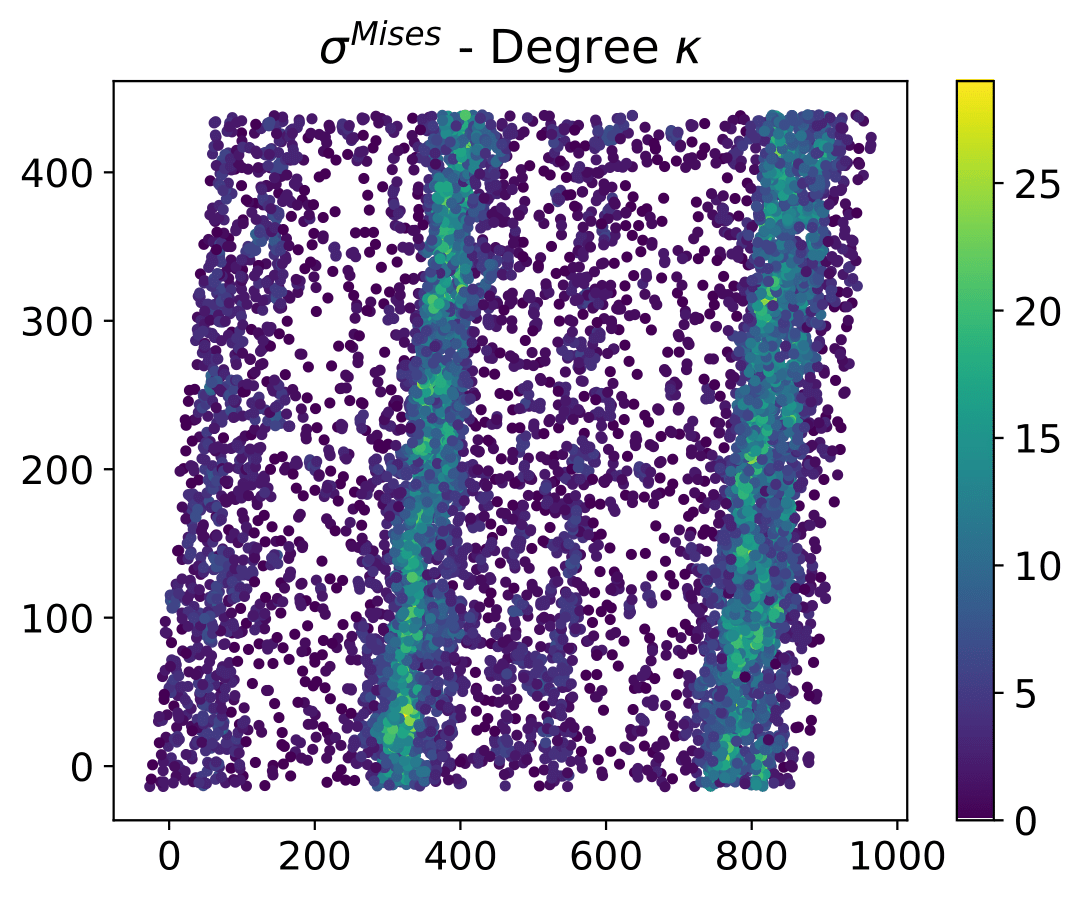}
                        \includegraphics[scale=0.28]{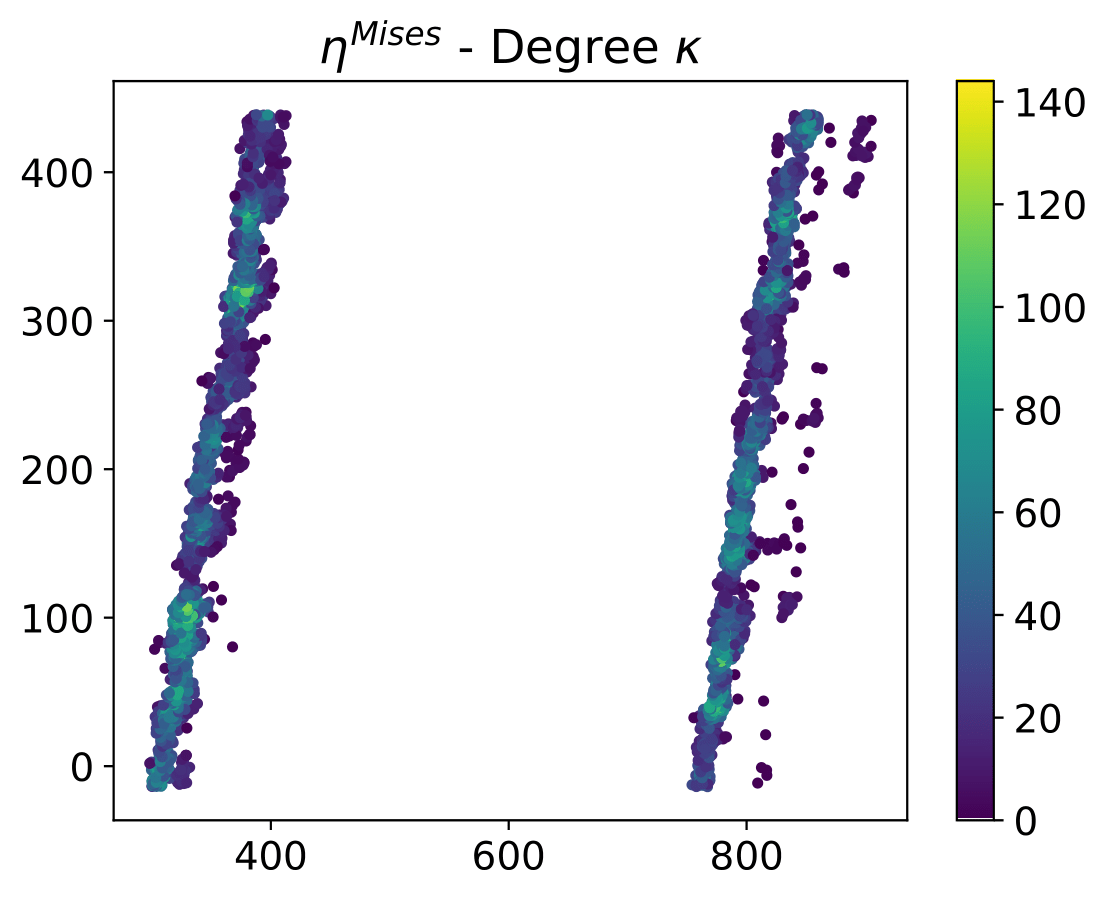}
                        \includegraphics[scale=0.28]{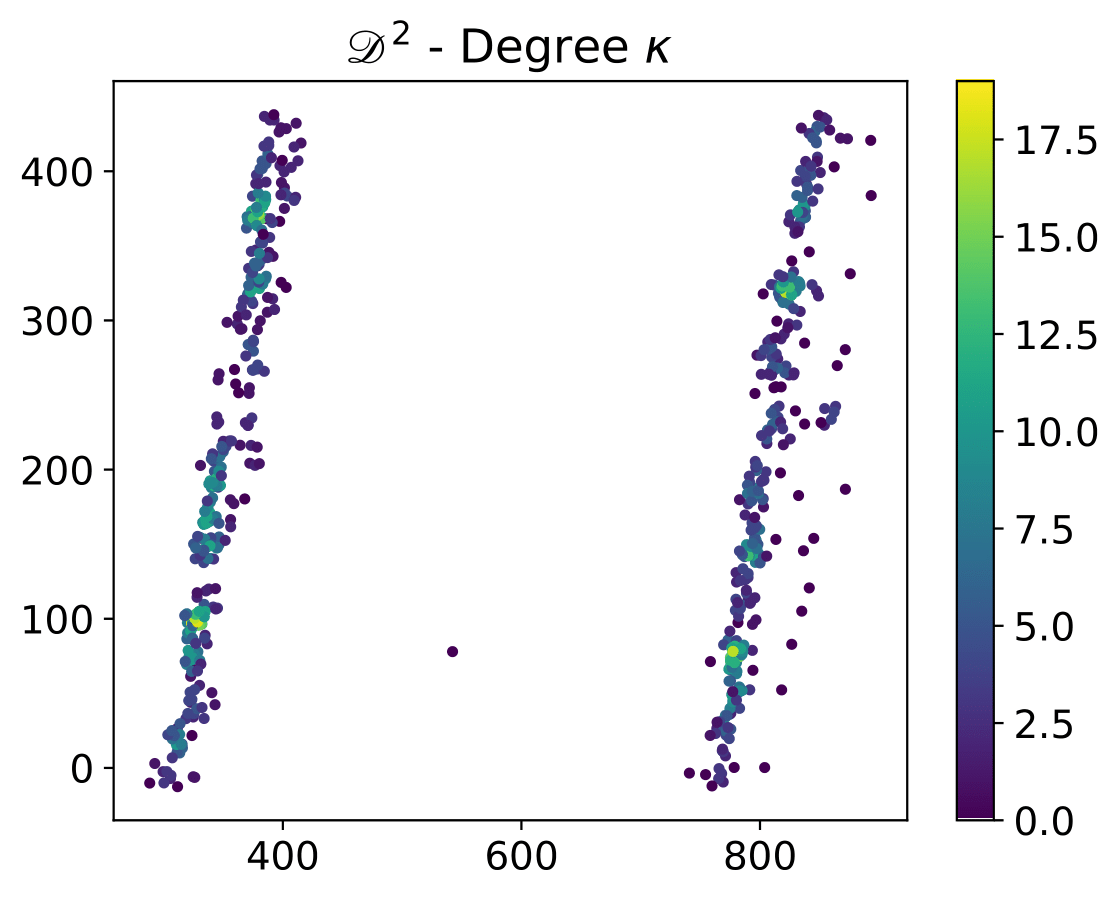}\\
                        \includegraphics[scale=0.28]{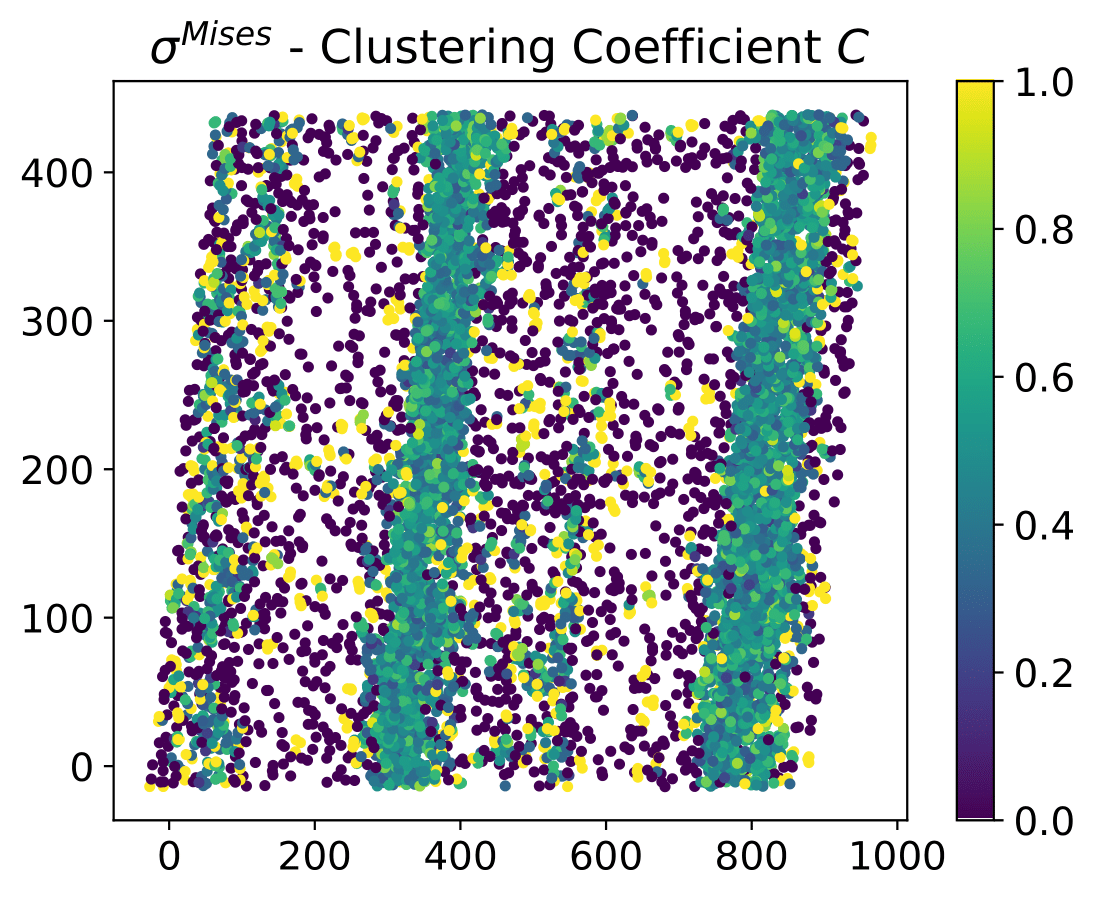}
                        \includegraphics[scale=0.28]{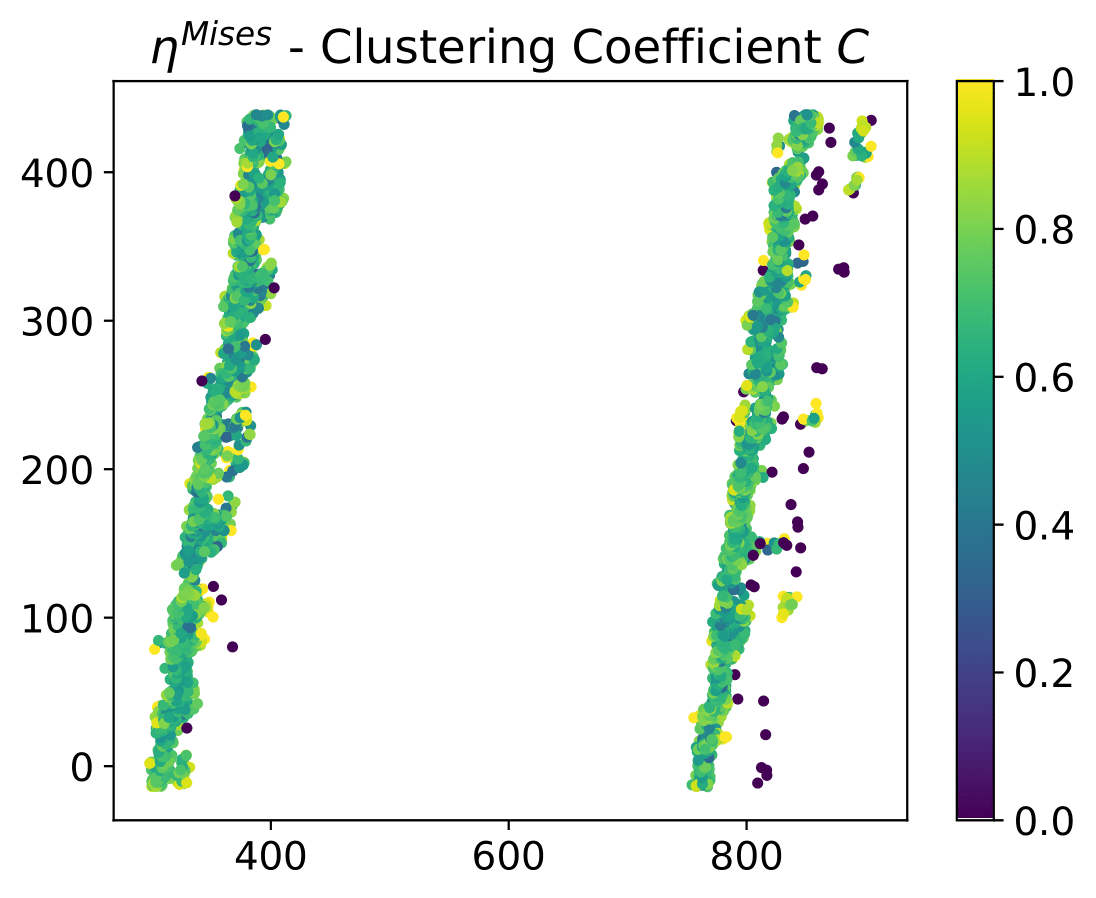}
                        \includegraphics[scale=0.28]{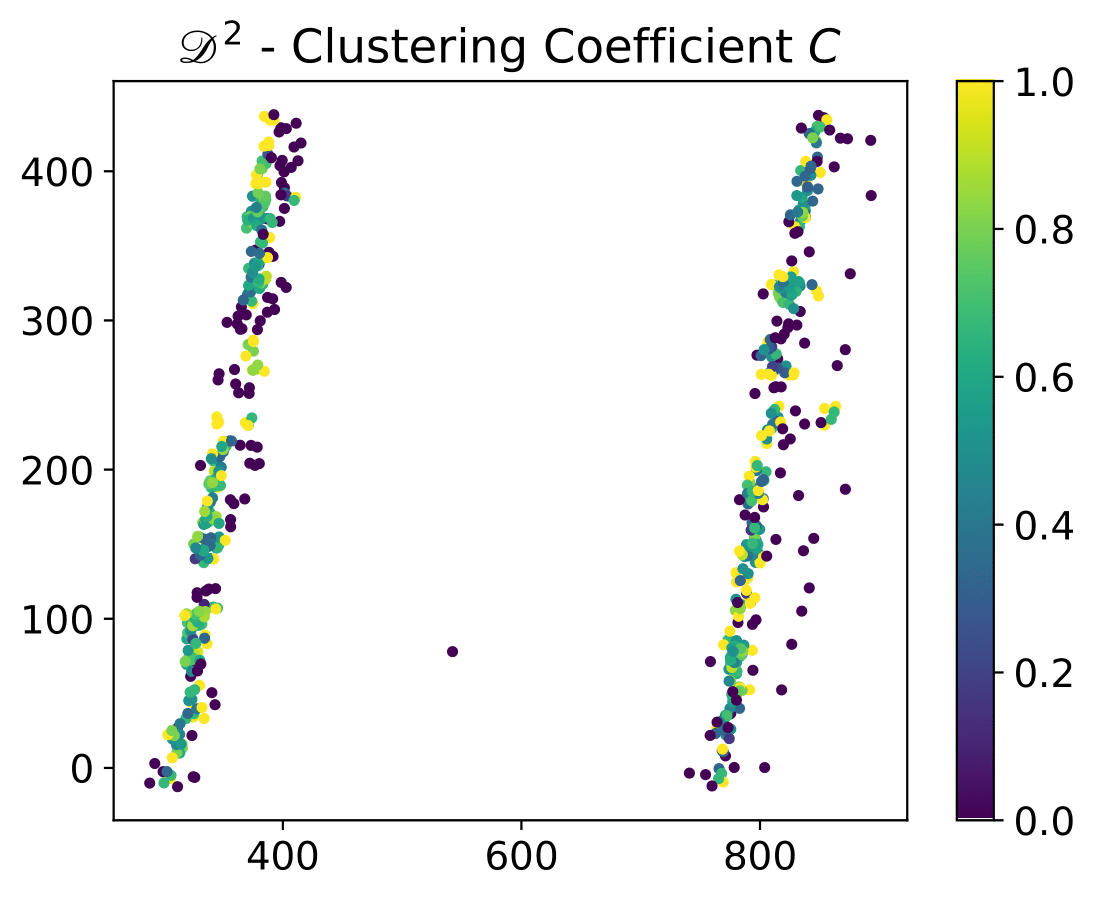}\\
                        \includegraphics[scale=0.28]{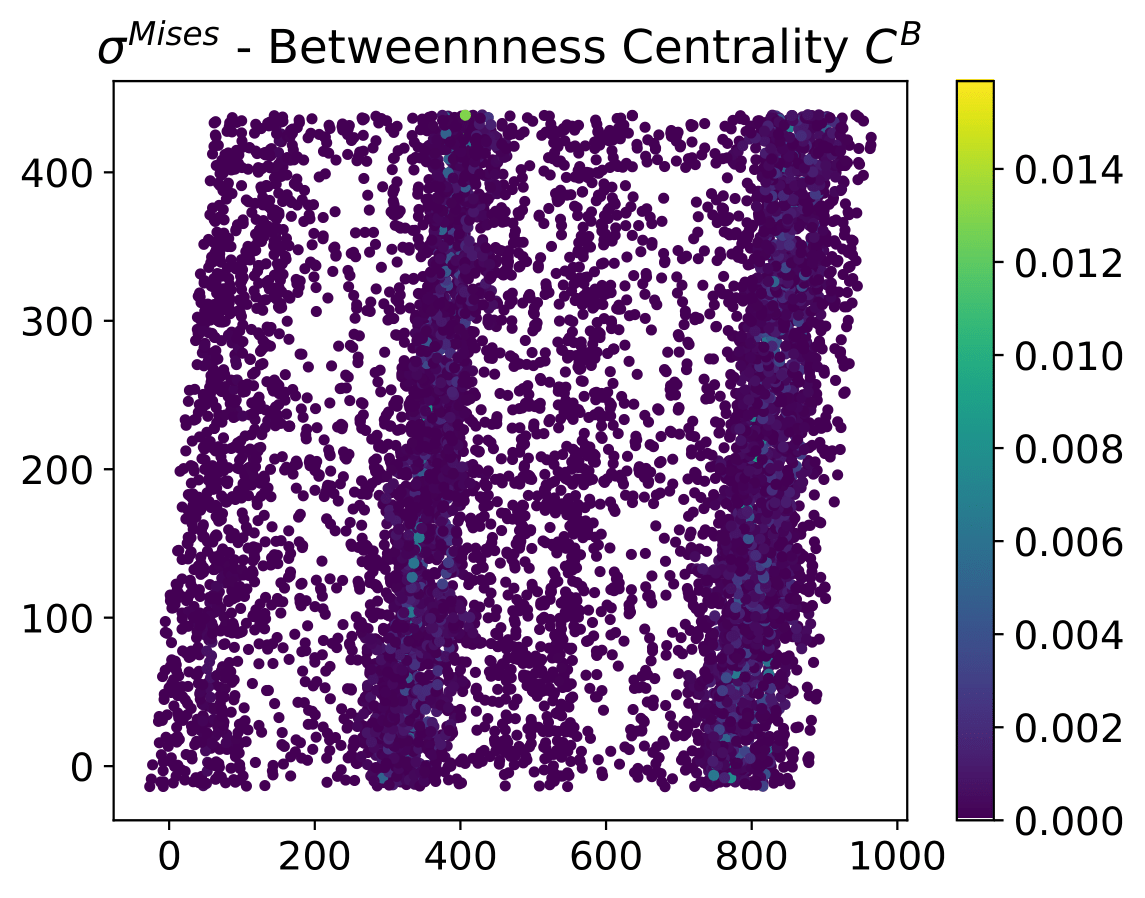}
                        \includegraphics[scale=0.28]{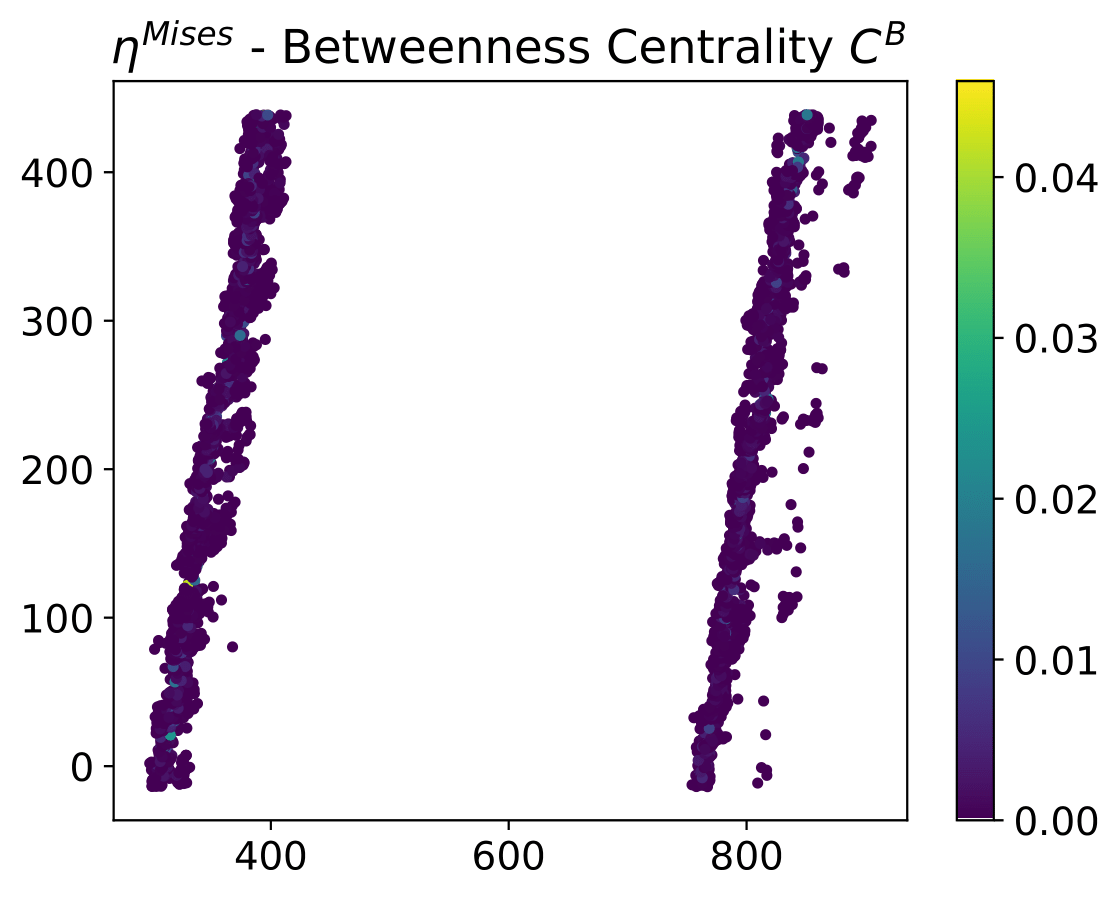}
                        \includegraphics[scale=0.28]{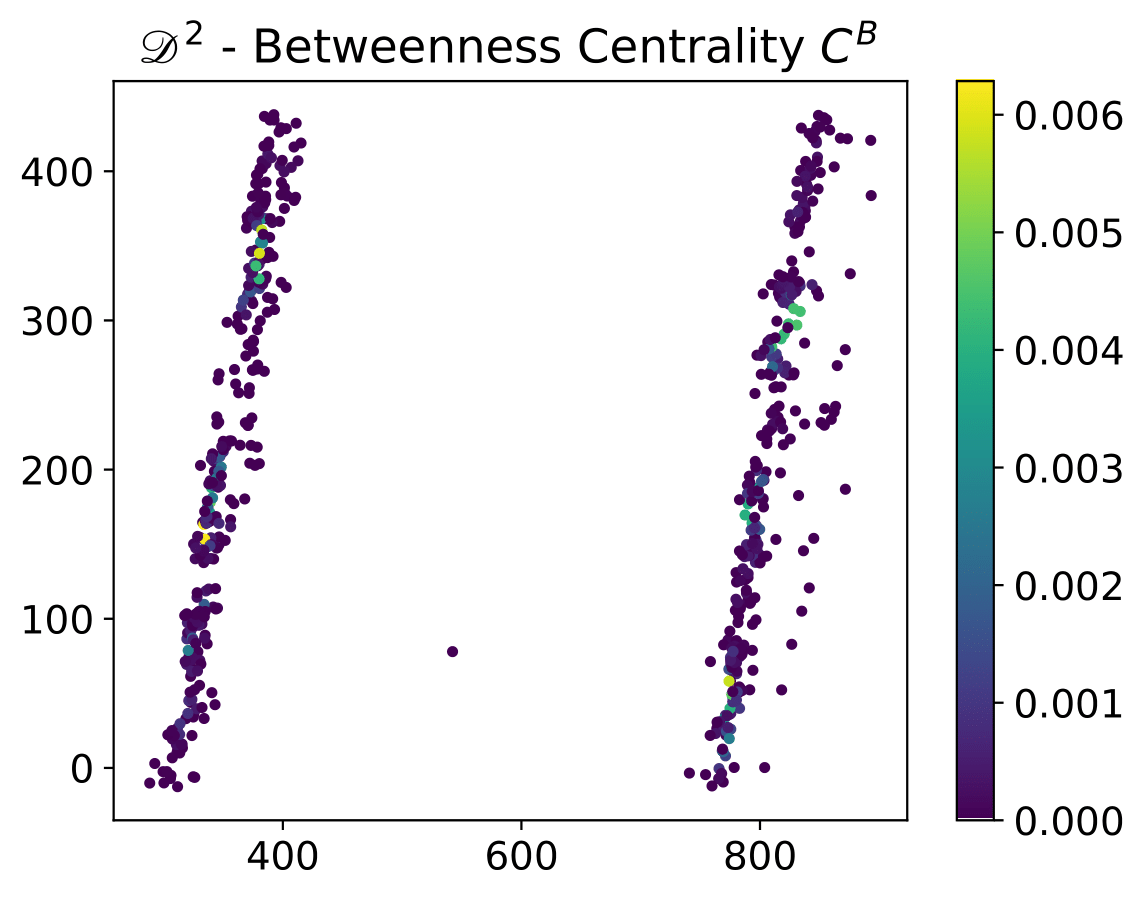}\\
                        \includegraphics[scale=0.28]{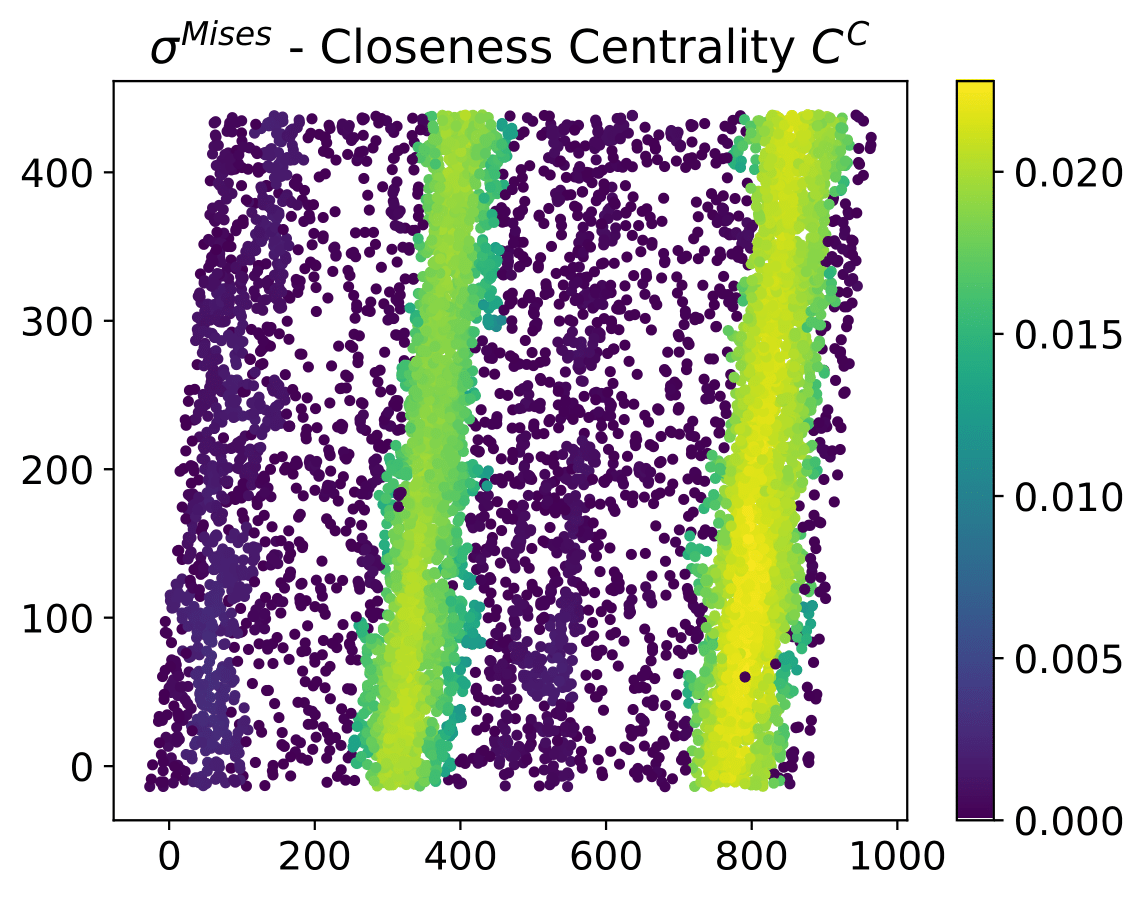}
                        \includegraphics[scale=0.28]{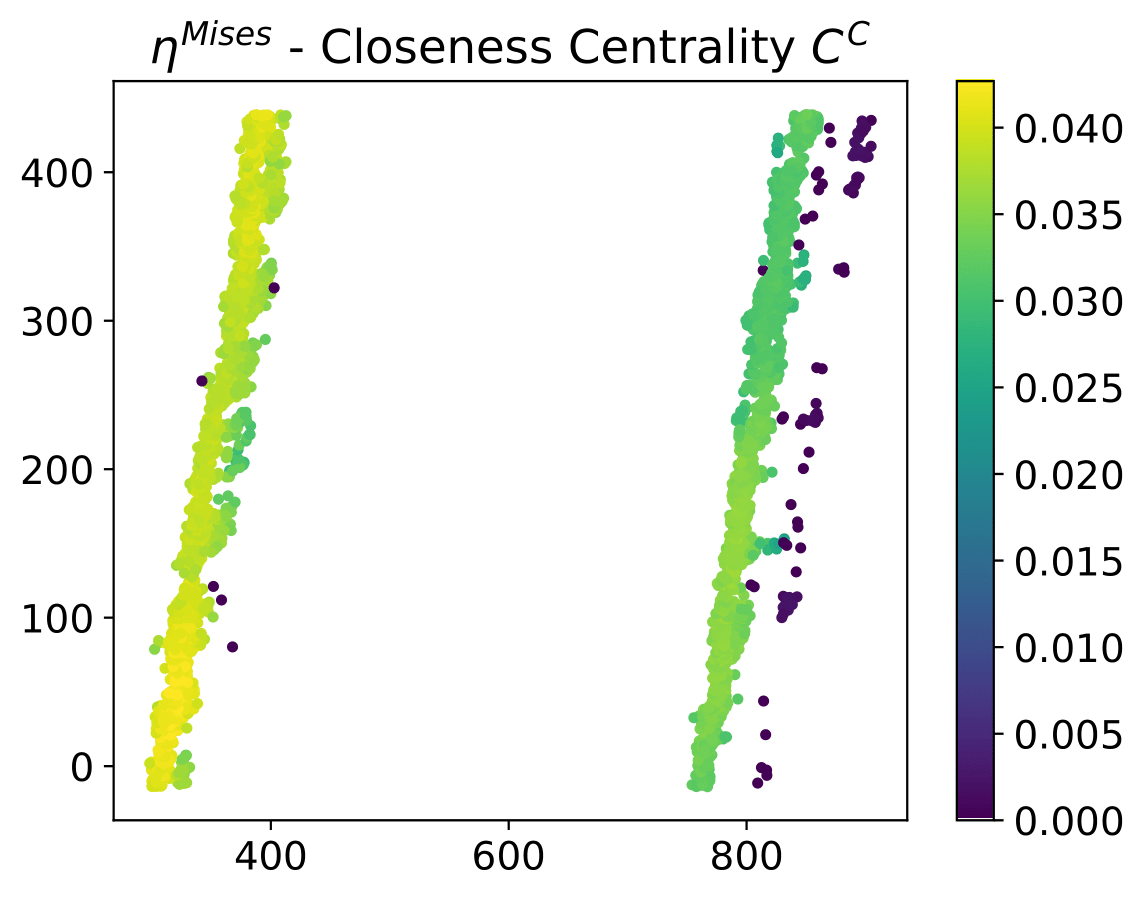}
                        \includegraphics[scale=0.28]{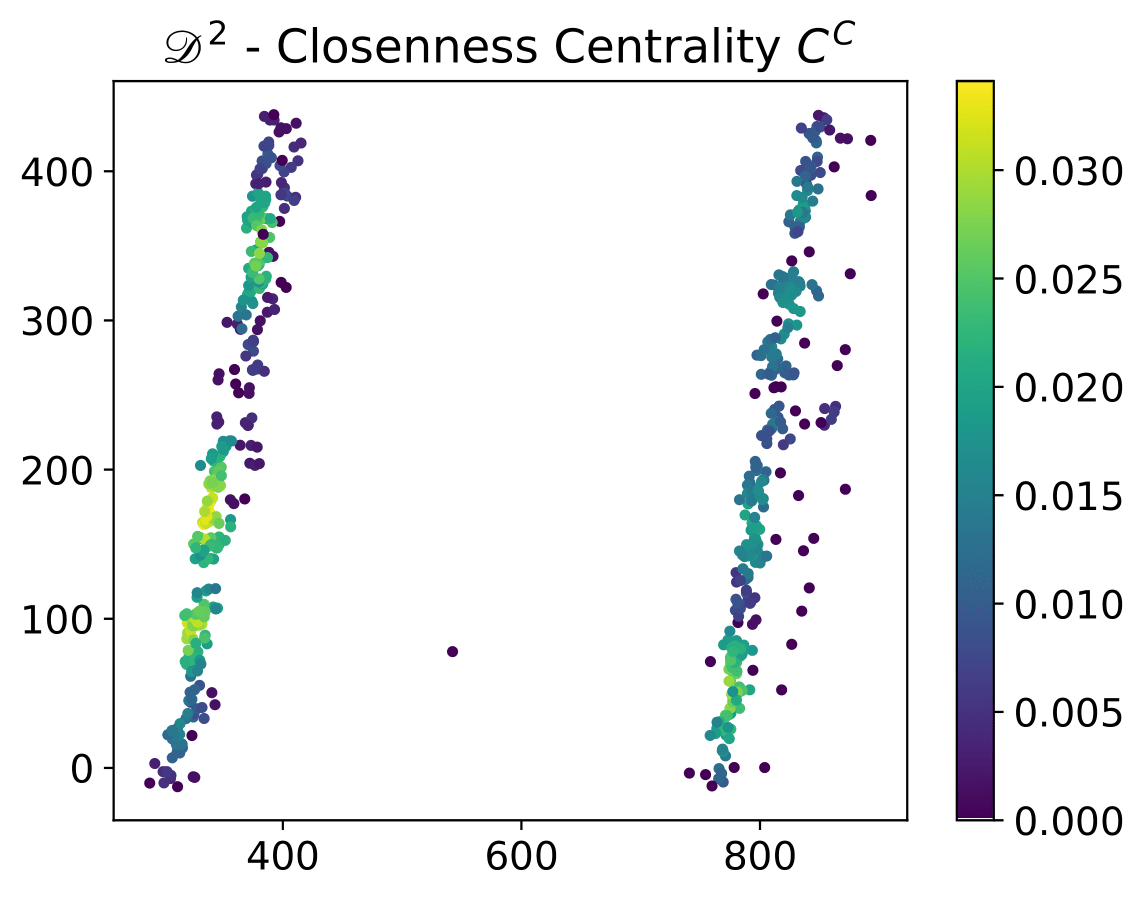}
                        \caption{Spatial distribution (CN without edges) of the metrics calculated for the networks of Figure 
                        (\ref{fig:complex_networks_1}) under the topological representation where the vertices are located in 
                        the position of their corresponding atom for the instant that the system reaches the deformation state 
                        $\gamma=0.20$. A color bar is included to record the range of values that each metric takes.}
                        \label{fig:complex_networks_2}
                \end{figure}

                \vs{0.2cm}
                \noindent The results that we have obtained for the spatial distribution of the metrics show that our methodology, 
                based on graph theory and CNs, has allowed a microscopic characterization of the deformation and 
                identification of plastic events that give rise to the SB and eventual material fracture. Figure 
                (\ref{fig:complex_networks_2}) shows the networks for the three descriptors and for each one, the atoms are colored 
                according to the degree, the clustering coefficient, the betweenness centrality and the closeness centrality, where 
                each metric shows that there is a area of the material where the interactions are more intense and complex, either 
                due to the levels of energy released by the plastic deformation (degree), or due to the regrouping and packing by 
                irreversible movements of the atoms (clustering), or due to the interaction collective group of atoms (centrality) 
                known as STZs.

                \vs{0.2cm}
                \noindent The relevant result here is that such metrics have been able to signal the location of the SB generated 
                by the deformation process, suggesting this description via CNs to study the phenomenon of plasticity in amorphous 
                materials.


       \section{Conclusions}

        \noindent In present work we have performed a shear deformation of the Cu$_{50}$Zr$_{50}$ metallic glass using molecular
        dynamic simulations and by means statistical analysis and application of a complex network model, we have studied 
        the plasticity in this type of amorphous material. Our results suggest the complex network description may be a useful 
        tool for the study of material properties and description of physical phenomena that occur at the microscopic level.

        \vs{0.2cm}
        \noindent The statistical analysis of the physical descriptors has allowed a better understanding of the physics of 
        the system under a deformation process. The probability density functions of these descriptors were calculated to gain 
        insight into the randomness of the parameters. However, the most notable result is the time series of the gini coefficient. 
        This measure together with the Lorenz curves show considerable changes precisely in regimes where a physical phenomenon 
        occurs. For example, in the elasto-plastic transition $\gamma\sim0.045$, in the plastic regime $0.045<\gamma$, in the location 
        of the SB $\gamma\sim0.095$ and eventual fracture of the material $\gamma\sim0 .130$. Basically, the microscopic 
        statistical analysis of the descriptors has predicted what happens at the macroscopic level as a consequence of the 
        deformation.

        \vs{0.2cm}
        \noindent The methodology based on complex networks, where the atomic configurations of the system are mapped to a graph that 
        increases in number of vertices and edges, added to the calculation of topological metrics, has turned out to be a very useful 
        mathematical tool for the microscopic characterization of deformation. In this research, we develop the calculation of the degree, 
        the clustering coefficient and measures of centrality such as betweennes and closeness for the characterization of the plastic 
        events that transit during the deformation. The topological structure of the networks and the metrics have shown interesting 
        results as proof of the existence of energy dissipation due to irreversible movements in the plastic regime and, as a consequence, 
        the increase in temperature (increase in degree), as well as the evidence of communities or groups of atoms that interact among 
        themselves and other groups, providing a new justification for the hypothesis of shear transformation zones as a fundamental 
        element in the theory of plasticity in amorphous materials.

        \vs{0.2cm}
        \noindent Because of amorphous systems exhibit greater complexity at the atomic configuration level, efforts to understand 
        their properties using condensed matter physics theories is still a challenge. Therefore, the complex network techniques 
        presented here represents an alternative, strong technique, to study the physical properties of these systems. 


        \section*{Acknowledgments}

        This project was supported financially by the National Agency for Research and Development ANID-PFCHA/Doctorado 
        Nacional/2020-2121991 (F.C.) and  by FONDECyT under contract No. 1201967 (V.M.). F.C. is grateful to Denisse 
        Pastén for useful discussion and suggestions.

        \section*{CRediT authorship contribution statement }

        {\bf F. Corvacho:         } Conceptualization, Methodology, Software, Formal analysis, Data Curation, Writing (Original Draft), Visualization.
        {\bf V. Muñoz:            } Conceptualization, Methodology, Formal analysis, Writing (Review \& Editing), Supervision, Project administration.
        {\bf M. Sepúlveda-Macías: } Software, Resources, Writing (Review \& Editing).
        {\bf G. Gutiérrez:        } Writing (Review \& Editing), Supervision.

        \section*{Declaration of Competing Interest}

        The authors declare that they have no known competing financial interests 
        or personal relationships that could have appeared to influence the work 
        reported in this paper.

        \section*{Declaration of Generative AI and AI-assisted technologies in the writing process}

        The autors declare that they have not used any type of AI technologies in this paper.

        \section*{Additional Information}
        {\bf Correspondence} and requests for materials should be addressed to Fernando Corvacho.


\begin{thebibliography}{10}

\bibitem{Wang2009Bulk}
W.H. Wang.
\newblock \href{}{Bulk metallic glasses with functional physical properties}.
\newblock {\em Advanced Materials}, 21(45):4524--4544, 2009.

\bibitem{Tiberto2007Magnetic}
P.~Tiberto, M.~Baricco, E.~Olivetti, and R.~Piccin.
\newblock \href{}{Magnetic properties of bulk metallic glasses}.
\newblock {\em Advanced Engineering Materials}, 9(6):468--474, 2007.

\bibitem{Greer1995Metallic}
A.L. Greer.
\newblock \href{}{Metallic glasses}.
\newblock {\em Science}, 267(5206):1947--1953, 1995.

\bibitem{Anantharaman1984Metallic}
T.R. Anantharaman.
\newblock {\em \href{}{Metallic glasses: production, properties and
  applications}}.
\newblock Trans Tech Publ, 1984.

\bibitem{Ashby2006Metallic}
M.F. Ashby and A.L. Greer.
\newblock \href{}{Metallic glasses as structural materials}.
\newblock {\em Scripta Materialia}, 54(3):321--326, 2006.

\bibitem{Klement1960Non}
W.~Klement, R.H. Willens, and P.~Duwez.
\newblock \href{}{Non-crystalline structure in solidified gold--silicon
  alloys}.
\newblock {\em Nature}, 187(4740):869--870, 1960.

\bibitem{Peker1993Highly}
A.~Peker and W.L. Johnson.
\newblock \href{}{A highly processable metallic glass:
  $\text{Zr}_{41,2}\text{Ti}_{13,8}\text{Cu}_{12,5}\text{Ni}_{10.0}\text{Be}_{22,5}$}.
\newblock {\em Applied Physics Letters}, 63(17):2342--2344, 1993.

\bibitem{Inoue1993New}
A.~Inoue, Y.~Horio, and T.~Masumoto.
\newblock \href{}{New amorphous $\text{Al}$--$\text{Ni}$--$\text{Fe}$ and
  $\text{Al}$--$\text{Ni}$--$\text{Co}$ alloys}.
\newblock {\em Materials Transactions, JIM}, 34(1):85--88, 1993.

\bibitem{Suryanarayana2017Bulk}
C.~Suryanarayana and A.~Inoue.
\newblock {\em \href{}{Bulk metallic glasses}}.
\newblock CRC Press, 2017.

\bibitem{Trexler2010Mechanical}
M.M. Trexler and N.N. Thadhani.
\newblock \href{}{Mechanical properties of bulk metallic glasses}.
\newblock {\em Progress in Materials Science}, 55(8):759--839, 2010.

\bibitem{Chen2008Mechanical}
M.~Chen.
\newblock \href{}{Mechanical behavior of metallic glasses: $\text{M}$icroscopic
  understanding of strength and ductility}.
\newblock {\em Annu. Rev. Mater. Res.}, 38:445--469, 2008.

\bibitem{Schuh2007Mechanical}
C.A. Schuh, T.C. Hufnagel, and U.~Ramamurty.
\newblock \href{}{Mechanical behavior of amorphous alloys}.
\newblock {\em Acta Materialia}, 55(12):4067--4109, 2007.

\bibitem{Plummer2015Metallic}
J.~Plummer.
\newblock \href{}{Is metallic glass poised to come of age?}
\newblock {\em Nature Materials}, 14(6):553--555, 2015.

\bibitem{Argon1979Plastic}
A.S. Argon.
\newblock \href{}{Plastic deformation in metallic glasses}.
\newblock {\em Acta Metall.}, 27(1):47--58, 1979.

\bibitem{Falk1998Dynamics}
M.L. Falk and J.S. Langer.
\newblock \href{}{Dynamics of viscoplastic deformation in amorphous solids}.
\newblock {\em Physical Review E}, 57(6):7192, 1998.

\bibitem{Shimizu2007Theory}
F.~Shimizu, S.~Ogata, and J.~Li.
\newblock \href{}{Theory of shear banding in metallic glasses and molecular
  dynamics calculations}.
\newblock {\em Materials Transactions}, 2007.

\bibitem{Greer2013Shear}
A.L. Greer, Y.Q. Cheng, and E.~Ma.
\newblock \href{}{Shear bands in metallic glasses}.
\newblock {\em Materials Science and Engineering: R: Reports}, 74(4):71--132,
  2013.

\bibitem{Rezaei2017Ductility}
R.~Rezaei, C.~Deng, M.~Shariati, and H.~Tavakoli-Anbaran.
\newblock \href{}{The ductility and toughness improvement in metallic glass
  through the dual effects of graphene interface}.
\newblock {\em Journal of Materials Research}, 32(2):392, 2017.

\bibitem{Sepulveda2018Tensile}
M.~Sepulveda-Macias, N.~Amigo, and G.~Gutierrez.
\newblock \href{}{Tensile behavior of $\text{Cu}_{50}\text{Zr}_{50}$ metallic
  glass nanowire with a $\text{B}$2 crystalline precipitate}.
\newblock {\em Physica B: Condensed Matter}, 531:64--69, 2018.

\bibitem{Albert2002Statistical}
R.~Albert and A.-L. Barab{\'a}si.
\newblock \href{}{Statistical mechanics of complex networks}.
\newblock {\em Rev. Mod. Phys.}, 74(1):47, 2002.

\bibitem{Newman2006Structure}
M.E. Newman, A.-L. Barab{\'a}si, and D.J. Watts.
\newblock {\em \href{}{The structure and dynamics of networks}}.
\newblock Princeton University Press, 2006.

\bibitem{Gheibi2017Solar}
A.~Gheibi, H.~Safari, and M.~Javaherian.
\newblock \href{}{The solar flare complex network}.
\newblock {\em The Astrophysical Journal}, 847(2):115, 2017.

\bibitem{Gosak2018Network}
M.~Gosak, R.~Markovi{\v{c}}, J.~Dolen{\v{s}}ek, M.S. Rupnik, M.~Marhl,
  A.~Sto{\v{z}}er, and M.~Perc.
\newblock \href{}{Network science of biological systems at different scales: A
  review}.
\newblock {\em Physics of Life Reviews}, 24:118--135, 2018.

\bibitem{Wasserman1994Social}
S.~Wasserman, K.~Faust, et~al.
\newblock {\em \href{}{Social network analysis: Methods and applications}},
  volume~8.
\newblock Cambridge University Press, 1994.

\bibitem{Baiesi2005Complex}
M.~Baiesi and M.~Paczuski.
\newblock \href{}{Complex networks of earthquakes and aftershocks}.
\newblock {\em Nonlinear Processes in Geophysics}, 12(1):1--11, 2005.

\bibitem{Papadopoulos2018Network}
L.~Papadopoulos, M.A. Porter, K.E. Daniels, and D.S. Bassett.
\newblock \href{}{Network analysis of particles and grains}.
\newblock {\em Journal of Complex Networks}, 6(4):485--565, 2018.

\bibitem{Kiv2023Complex}
A.~Kiv, A.~Bryukhanov, V.~Soloviev, A.~Bielinskyi, T.~Kavetskyy, D.~Dyachok,
  I.~Donchev, and V.~Lukashin.
\newblock \href{}{Complex network methods for plastic deformation dynamics in
  metals}.
\newblock {\em Dynamics}, 3(1):34--59, 2023.

\bibitem{Kui2007Correlation}
X.~kui Xi, L.~long Li, B.~Zhang, W.~hua Wang, and Y.~Wu.
\newblock \href{}{Correlation of atomic cluster symmetry and glass-forming
  ability of metallic glass}.
\newblock {\em Physical Review Letters}, 99(9), 2007.

\bibitem{Lee2011Networked}
M.~Lee, C.-M. Lee, K.-R. Lee, E.~Ma, and J.-C. Lee.
\newblock \href{}{Networked interpenetrating connections of icosahedra:
  $\text{E}$ffects on shear transformations in metallic glass}.
\newblock {\em Acta Materialia}, 59(1):159--170, 2011.

\bibitem{Foroughi2018Medium}
A.~Foroughi, R.~Tavakoli, and H.~Aashuri.
\newblock \href{}{Medium range order evolution in pressurized sub-$\text{T}_g$
  annealing of $\text{Cu}_{64}\text{Zr}_{36}$ metallic glass}.
\newblock {\em Journal of Non-Crystalline Solids}, 481, 2018.

\bibitem{Abe2011Universalities}
S.~Abe, D.~Past{\'e}n, V.~Mu{\~n}oz, and N.~Suzuki.
\newblock \href{}{Universalities of earthquake-network characteristics}.
\newblock {\em Chinese Science Bulletin}, 56(34):3697--3701, 2011.

\bibitem{Vogel2017Time}
E.E. Vogel, G.~Saravia, D.~Past{\'e}n, and V.~Mu{\~n}oz.
\newblock \href{}{Time-series analysis of earthquake sequences by means of
  information recognizer}.
\newblock {\em Tectonophysics}, 712:723--728, 2017.

\bibitem{Pasten2016Time}
D.~Past{\'e}n, F.~Torres, B.~Toledo, V.~Mu{\~n}oz, J.~Rogan, and J.A. Valdivia.
\newblock \href{}{Time-based network analysis before and after the Mw 8.3
  $\text{I}$llapel earthquake 2015 $\text{C}$hile}.
\newblock {\em Pure and Applied Geophys}, 173:2267--2275, 2016.

\bibitem{Martin2022Complex}
F.A. Mart{\'\i}n and D.~Past{\'e}n.
\newblock \href{}{Complex networks and the b-value relationship using the
  degree probability distribution: $\text{T}$he case of three mega-earthquakes
  in $\text{C}$hile in the last decade}.
\newblock {\em Entropy}, 24(3):337, 2022.

\bibitem{Munoz2022Complex}
V.~Mu{\~n}oz and E.~Fl{\'a}ndez.
\newblock \href{}{Complex network study of solar magnetograms}.
\newblock {\em Entropy}, 24(6):753, 2022.

\bibitem{Inoue1999Stabilization}
A.~Inoue.
\newblock \href{}{Stabilization and high strain-rate superplasticity of
  metallic supercooled liquid}.
\newblock {\em Materials Science and Engineering: A}, 267(2):171--183, 1999.

\bibitem{Cheng2011Atomic}
Y.Q. Cheng and E.~Ma.
\newblock \href{}{Atomic-level structure and structure--property relationship
  in metallic glasses}.
\newblock {\em Progress in Materials Science}, 56(4):379--473, 2011.

\bibitem{Plimpton1995Fast}
S.J. Plimpton.
\newblock \href{}{Fast parallel algorithms for short-range molecular dynamics}.
\newblock {\em Journal of Computational Physics}, 117(1):1--19, 1995.

\bibitem{Stukowski2009Visualization}
A.~Stukowski.
\newblock \href{}{Visualization and analysis of atomistic simulation data with
  $\text{OVITO}$--the $\text{O}$pen $\text{V}$isualization $\text{T}$ool}.
\newblock {\em Modelling and Simulation in Materials Science and Engineering},
  18(1):015012, 2009.

\bibitem{Wang2012Structural}
C.C. Wang and C.H. Wong.
\newblock \href{}{Structural properties of
  $\text{Zr}_{x}\text{Cu}_{90-x}\text{Al}_{10}$ metallic glasses investigated
  by molecular dynamics simulations}.
\newblock {\em Journal of Alloys and Compounds}, 510(1):107--113, 2012.

\bibitem{Sepulveda2020Precursors}
M.~Sep{\'u}lveda-Mac{\'\i}as, G.~Gutierrez, and F.~Lund.
\newblock \href{}{Precursors to plastic failure in a numerical simulation of
  $\text{CuZr}$ metallic glass}.
\newblock {\em Journal of Physics: Condensed Matter}, 32(17), 2020.

\bibitem{Sepulveda2016Onset}
M.~Sep{\'u}lveda-Mac{\'\i}as, N.~Amigo, and G.~Guti{\'e}rrez.
\newblock \href{}{Onset of plasticity and its relation to atomic structure in
  $\text{CuZr}$ metallic glass nanowire: A molecular dynamics study}.
\newblock {\em Journal of Alloys and Compounds}, 655:357--363, 2016.

\bibitem{Baro2013Statistical}
J.~Bar{\'o}, A.~Corral, X.~Illa, A.~Planes, E.K.H. Salje, W.~Schranz, D.E.
  Soto-Parra, and E.~Vives.
\newblock \href{}{Statistical similarity between the compression of a porous
  material and earthquakes}.
\newblock {\em Phys. Rev. Lett.}, 110(8):088702, 2013.

\bibitem{Abe2011Finite}
S.~Abe, D.~Past{\'e}n, and N.~Suzuki.
\newblock \href{}{Finite data-size scaling of clustering in earthquake
  networks}.
\newblock {\em Physica A: Statistical Mechanics and its Applications},
  390(7):1343--1349, 2011.

\bibitem{Barabasi1999Emergence}
A.-L. Barab{\'a}si and R.~Albert.
\newblock \href{}{Emergence of scaling in random networks}.
\newblock {\em Science}, 286(5439):509--512, 1999.

\bibitem{Watts1998Collective}
D.J. Watts and S.H. Strogatz.
\newblock \href{}{Collective dynamics of ‘small-world’ networks}.
\newblock {\em Nature}, 393(6684):440--442, 1998.

\bibitem{Hagberg2008Exploring}
A.~Hagberg, P.~Swart, and D.~Schult.
\newblock \href{}{Exploring network structure, dynamics, and function using
  $\text{N}$etworkX}.
\newblock In Ga\"el Varoquaux, Travis Vaught, and Jarrod Millman, editors, {\em
  Proceedings of the 7th Python in Science Conference}, pages 11 -- 15,
  Pasadena, CA USA, 2008.

\end{thebibliography}

\end{document}